\documentclass[twocolumn,twocolappendix, tighten]{aastex6}
\usepackage{graphicx}
\usepackage{subfigure}

\shorttitle{Radio Parameters for the COBRA Survey}

\shortauthors{Golden-Marx et al.}

\begin{document}

\title{The High-Redshift Clusters Occupied by Bent Radio AGN (COBRA) Survey: Radio Source Properties}

\author{Emmet Golden-Marx\altaffilmark{1}\altaffilmark{2}, E.\,L. Blanton\altaffilmark{2}, R. Paterno-Mahler\altaffilmark{3}, M. Brodwin\altaffilmark{4}, M.\,L.\,N. Ashby\altaffilmark{5}, E. Moravec\altaffilmark{6}, L. Shen\altaffilmark{7}\altaffilmark{8}, B.C. Lemaux\altaffilmark{9}, L.M. Lubin\altaffilmark{9}, R.R. Gal\altaffilmark{10}, A.R. Tomczak\altaffilmark{9}} 

\email{emmetgm@bu.edu}

\altaffiltext{1}{Department of Astronomy, Tsinghua University, Beijing 100084, China}
\altaffiltext{2}{Department of Astronomy and The Institute for Astrophysical Research, Boston University, 725 Commonwealth Avenue, Boston, MA 02215, USA}
\altaffiltext{3}{WM Keck Science Center, 925 N. Mills Avenue, Claremont, CA  91711, USA}
\altaffiltext{4}{Department of Physics \& Astronomy, University of Missouri-Kansas City, 5110 Rockhill Road, Kansas City, MO 64110, USA}
\altaffiltext{5}{Center for Astrophysics $|$ Harvard \& Smithsonian, 60 Garden Street, Cambridge, MA 02138, USA}
\altaffiltext{6}{Astronomical Institute of the Czech Academy of Sciences, Bo\v cn\'i II 1401/IA, 14000 Praha 4, Czech Republic}
\altaffiltext{7}{CAS Key Laboratory for Research in Galaxies and Cosmology, Department of Astronomy, University of Science and Technology of China, Hefei 230026, China}
\altaffiltext{8}{School of Astronomy and Space Sciences, University of Science and Technology of China, Hefei 230026, China}
\altaffiltext{9}{Department of Physics, University of California, Davis, One Shields Avenue, Davis, CA 95616, USA}
\altaffiltext{10}{University of Hawai'i, Institute for Astronomy, 2680 Woodlawn Drive, Honolulu, HI 96822, USA}

\begin{abstract}
The shape of bent, double-lobed radio sources requires a dense gaseous medium.  Bent sources can therefore be used to identify galaxy clusters and characterize their evolutionary history.  By combining radio observations from the Very Large Array Faint Images of the Radio Sky at Twenty centimeters (VLA FIRST) survey with optical and infrared imaging of 36 red sequence selected cluster candidates from the high-$z$ Clusters Occupied by Bent Radio AGN (COBRA) survey (0.35 $<$ $z$ $<$ 2.2), we find that radio sources with narrower opening angles reside in richer clusters, indicating that the cluster environment impacts radio morphology.  Within these clusters, we determine 55.5$\%$ of our radio host galaxies are brightest cluster galaxies (BCGs) and that the remainder are associated with other luminous galaxies.  The projected separations between the radio sources and cluster centers and the sizes of the opening angles of bent sources follow similar distributions for BCG and non-BCG host populations, suggesting that COBRA host galaxies are either BCGs or galaxies that may evolve into BCGs.  By measuring the orientation of the radio sources relative to the cluster centers, we find between 30$\%$ and 42$\%$ of COBRA bent sources are outgoing and have passed through the cluster center, while between 8$\%$ and 58$\%$ of COBRA bent sources are infalling.  Although these sources typically do not follow directly radial paths, the large population of outgoing sources contrasts what is observed in low-$z$ samples of bent sources and may indicate that the intracluster medium is less dense in these high-$z$ clusters.      
     
\end{abstract}

\keywords{galaxies: clusters: general - galaxies:evolution - galaxies:high-redshift - infrared:galaxies - radio continuum:galaxies}

\section{Introduction}\label{sect:intro}
Galaxy clusters are the largest gravitationally-bound structures in the universe.  Clusters are partially characterized observationally by their galaxy populations and hot, X-ray emitting gas in the form of the intracluster medium (ICM).  At low redshift, cluster galaxy populations are very well established \citep[e.g.,][]{Eisenhardt2007}, with most clusters hosting large, densely populated cluster cores of massive, quiescent, early-type galaxies.  Although most galaxies in low-$z$ clusters are quiescent, these galaxies were not always ``red and dead".  The epoch of cluster formation is predicted to be at high redshift ($z$ $>$ 2), but at these redshifts, we find large diversity among galaxy populations and the environments they reside in (e.g., denser galaxy clusters or large and diffuse protoclusters; \citealp[e.g.,][]{Hatch2011,Lemaux2014,Lemaux2018}).   

At $z$ $>$ 2, the galaxy populations in clusters have been observed to be wildly diverse.  Some clusters have large quiescent galaxy populations \citep[e.g.,][]{Strazzullo2016,Willis2020}, while other clusters host galaxies with large star-forming populations \citep[e.g.,][]{Wang2016,Miller2018}, and still others host populations of galaxies that strongly resemble the distribution of field galaxies \citep[e.g.,][]{Alcorn2019}.  It is at 0.5 $<$ $z$ $<$ 2.0 where diffuse protocluster structures begin to collapse and form the established clusters observed in the low-$z$ universe.  During this epoch, cluster-cluster mergers with hot ICMs are also expected to become more prevalent and the processes associated with the typical evolution of cluster galaxies (e.g., ram pressure stripping, harassment) begin to mature.  These processes, all occurring in tandem, make this redshift range a dynamic time to examine the components of clusters.  

Observationally, at 0.5 $<$ $z$ $<$ 2.0, we find many examples of clusters with strong red sequences similar to those of low-$z$ clusters \citep[e.g.,][]{Gladders2000,Andreon2014,Cooke2016,Cerulo2016}, although an increasing number of counter examples exist \citep[e.g.,][]{Brodwin2013,Hennig2017}. Additionally, the color-density and star formation rate-density relations, which characterize the rate of star formation in clusters, are observed to persist in similar values as in low-$z$ clusters out to $z$ $\approx$ 1.5 \citep[e.g.][]{Foltz2018,Tomczak2019,Lemaux2019}, although there are hints that these trends begin to break down at higher redshifts \citep[e.g.,][]{Nantais2016,Nantais2017}.  Because we see a large range in cluster characteristics at $z$ $>$ 0.5, there is a need for increased studies of high redshift clusters.  Recently, new cluster surveys, including the Massive and Distant Clusters of Wise Survey (MaDCoWS; e.g.,  \citealt{Stanford2014,Brodwin2015,Gonzalez2015,Mo2018,Gonzalez2019,Moravec2019,Decker2019,Moravec2020}), which identified 2863 of the most massive galaxy clusters at 0.7 $<$ $z$ $<$ 1.5, have greatly increased those numbers.

However, the vast majority of clusters are not massive Coma cluster progenitors.  Thus, it is important to identify high-$z$ galaxy clusters that might be the progenitors of less massive galaxy clusters and groups to fully trace the role of cluster mass in cluster evolution.  To that end, identifying strong tracers of galaxy cluster candidates of all masses is vitally important.  One accurate tracer of high-$z$ galaxy clusters is radio active galactic nuclei (AGNs).  These energetic radio sources have long been known to be associated with galaxy clusters and were used in some of the earliest detections of what were then considered high-$z$ galaxy clusters \citep{Minkowski1960}.  More recently, radio loud AGNs have been used to trace both low- and high-$z$ clusters and protoclusters \citep[e.g.][]{Blanton2003,Wing2011,Galametz2012,Wylezalek2013,Wylezalek2014,Castignani2014,Rigby2014,Blanton2015,Cooke2015,Cooke2016,Noirot2016,Paterno-Mahler2017,Shen2017,Noirot2018,Shimakawa2018,Croston2019,Garon2019,Moravec2019,Golden-Marx2019,Moravec2020}.  For example, the Clusters Around Radio Loud AGN (CARLA) survey found that $\approx$ 55$\%$ of their 387 radio loud AGNs at 1.3 $<$ $z$ $<$ 3.2 were in overdense environments at the level of 2$\sigma$, with 10$\%$ of sources in overdense environments at the 5$\sigma$ level when compared to a background field \citep{Wylezalek2013}.  

More importantly, radio AGNs trace clusters with strong red sequences as well as clusters hosting younger stellar populations and dusty, star-forming galaxies, making radio AGNs an excellent tracer of clusters of all types \citep{Cooke2016,Noirot2016,Noirot2018,Golden-Marx2019}. Although there are a number of studies of the environments of radio AGN at low redshift ($z$ $<$ 0.5; \citealt{Best2000,Wing2011,Croston2019,Garon2019}) and high redshift ($z$ $>$ 1.0; e.g., \citealt{Wylezalek2013,Moravec2019,Moravec2020}), there is need for a sample to both fill the gap between and overlap with these surveys to better trace the evolution of these systems.

In creating a large sample of radio AGNs used to search for galaxy clusters and look for potential evolutionary effects, examining radio source morphology is important.  Specifically, the types of radio AGNs found in clusters tends to differ greatly with the cluster's redshift.  At lower redshifts, most radio AGNs in rich cluster environments are characterized by a bright radio core and faint radio lobes \citep[e.g.,][]{Longair1979,Prestage1988,Ledlow1996,Miller1999,Wing2011,Gendre2013}.  These sources, classified as Fanaroff-Riley I (FRI) radio sources, generally have P$_{1.44}$ $<$ 10$^{25}$\,W\,Hz$^{-1}$ \citep[e.g.,][]{Fanaroff1974,Ledlow1996} and are also found in similarly rich environments at high redshift, which makes them important cluster tracers \citep[e.g.,][]{Hill1991,Zirbel1997,Stocke1999,Fujita2016,Shen2017}.  By contrast, radio AGNs characterized by brighter radio lobes and a fainter radio core with P$_{1.44}$ $>$ 10$^{25}$\,W\,Hz$^{-1}$ \citep[e.g.,][]{Fanaroff1974,Ledlow1996} are classified as Fanaroff-Riley II (FRII) radio sources.  Unlike FRI radio sources, FRII radio sources are generally found in poor cluster environments and groups at low redshift and in richer protocluster environments at high redshift \citep[e.g.,][]{Best2000}, although examples of FRII radio sources in richer low-$z$ galaxy clusters do exist \citep[e.g.,][]{Wing2011}.  Due to this dichotomy and our desire to build a sample of high-$z$ clusters, incorporating FRI and FRII sources yields added diversity among cluster populations.  

One unique type of radio source that allows for a cluster sample across all ranges of cluster mass and morphology are bent, double-lobed radio sources \citep[e.g.,][]{Blanton2000,Blanton2001,Blanton2003,Blanton2015,Paterno-Mahler2017,Silverstein2018,Garon2019,Golden-Marx2019}.  Unlike their straight counterparts, the ``C" shape of these radio sources hints at the presence of a dense, gaseous medium acting on the radio source.  Physically, the bent nature of these radio sources has been linked directly to ram pressure exerted by the ICM \citep[e.g.,][]{Owen1976,ODonoghue1993,Hardcastle2005,Morsony2013}, where the relative motion of the ICM with respect to the host galaxy bends the radio lobes.  Additionally, there exists a long history of using bent radio sources to probe the intragroup medium (IGM) density \citep[e.g.,][]{Ekers1978,Freeland2008,Freeland2011} and at least one bent source has been found in a large-scale filament \citep{Edwards2010}.  Furthermore, many bent AGNs are Wide Angle Tail (WAT) radio sources.  WATs are bent sources with opening angles greater than 90$^{\circ}$ that are commonly hosted by bright central elliptical galaxies \citep[e.g.,][]{Owen1976,Valentijn1979,ODea1985}.  Morphologically, WATs are typically FRI sources, but have radio powers near the FRI/FRII border (10$^{24.75}$\,W\,Hz$^{-1}$ $<$ P$_{1.44}$ $<$ 10$^{25.75}$\,W\,Hz$^{-1}$) \citep[e.g.,][]{Blanton2000,Blanton2001,Wing2011}.  This makes WATs an ideal tracer for clusters of all masses, assuming those clusters host a dense ICM.   

At low redshift, \citet{Blanton2001} examined a sample of 40 bent sources, finding 54$\%$ are in clusters, while the remaining 46$\%$ are in groups, some of which are poor.  \citet{Wing2011} built on this work using the Very Large Array Faint Images of the Radio Sky at Twenty centimeters (VLA FIRST) survey \citep{Becker1995} to identify thousands of bent radio AGNs.  As VLA FIRST was designed to cover the same area of the sky as the Palomar Sky Survey, which is the same region of the sky covered by the Sloan Digital Sky Survey (SDSS), \citet{Wing2011} cross-matched each bent radio AGN with SDSS galaxies with m$_{r}$ $<$ 22.0\,mag within a search radius designed to ensure 95\,$\%$ accuracy for real associations between host galaxies and radio sources.  \citet{Wing2011} found that 40 - 80$\%$ of bent AGNs are in clusters, depending on the richness criteria.  \citet{Garon2019} more recently used the Radio Galaxy Zoo to study bent radio sources and found that of their 4304 bent AGNs at 0.02 $<$ $z$ $<$ 0.8, 27 $\%$ of their radio sources are within 1\,Mpc of optically identified clusters and 87$\%$ are within 15\,Mpc of optically identified clusters. 

Bent radio AGNs are associated with a diverse cluster sample and have been observed in both major cluster mergers \citep[e.g.,][]{Burns1990,Roettiger1996,Burns1996,Douglass2011} and relaxed clusters \citep[e.g.,][]{Paterno-Mahler2013}.  In relaxed clusters, bent AGNs can form due to ``sloshing spirals" of hot ICM gas caused by minor mergers where the gravitational pull of a subcluster passing by the main cluster bends the lobes.  Furthermore, \citet{Wing2013} did an optical substructure analysis on a subset of their low-$z$ SDSS sample and found that bent radio sources are no more likely to be in mergers than their straight counterparts. Thus, this range of masses and morphologies make bent AGNs an ideal tracer for clusters.  

As the ICM is a dense gaseous medium, bent radio sources without low-$z$ host galaxies are excellent tracers of high-$z$ clusters when other detection methods fail \citep[e.g.,][]{Blanton2003,Blanton2015,Paterno-Mahler2017,Obrien2018,Silverstein2018,Obrien2018,Golden-Marx2019}.  To this end, the high-$z$ Clusters Occupied by Bent Radio AGN (COBRA) survey of 646 bent, double-lobed radio sources identified from the VLA FIRST survey was built (see \citealp{Blanton2015}, \citealp{Paterno-Mahler2017}, and \citealp{Golden-Marx2019} for a complete overview of the high-$z$ COBRA survey).  Using a subset of the high-$z$ COBRA survey, we investigate the properties of the radio galaxies (bending angle, radio luminosity, and projected physical size) with respect to the optical and infrared properties of their host clusters (richness, host magnitude, offset from the cluster center).  In Section\,\ref{sect:data}, we discuss our sample selection, observations, and measurements.  In Section\,\ref{sect:AGNcluster}, we discuss the properties of the radio sources with respect to their host clusters.  In Section\,\ref{sect:discussion}, we investigate how our results compare to similar studies of high-$z$ ($z$ $>$ 0.5) radio AGNs and to samples of bent radio AGNs.  Throughout this work, we adopt a flat $\Lambda$CDM cosmology, using $H_{0}$ = 70\,km\,s$^{-1}$\,Mpc$^{-1}$, $\Omega_{m}$ = 0.3, and $\Omega_{\Lambda}$ = 0.7.  Unless otherwise noted, all magnitudes are given in AB magnitudes.  Additionally, all distances related to either the offset of radio sources from the cluster center or the size of the radio sources are given as projected distances.      

\section{Data}\label{sect:data}
To accurately characterize the relationship between bent radio AGNs and the clusters they reside in, we require a large sample of bent sources as well as the necessary optical, infrared, and radio observations needed to characterize cluster galaxies.  To that end, we utilize the high-$z$ COBRA Survey (Section\,\ref{sect:COBRAsurvey}) and the follow-up optical observations that allowed \citet{Golden-Marx2019} to create the red sequence selected cluster sample (Sections\,\ref{sect:Optical} and \ref{sect:clustercandidates}).  From the identification of red sequence galaxies, we introduce our measurement of the red sequence cluster center in Section\,\ref{sect:clustercenter}.  To describe each radio source, we include our methodology for measuring the size, power, and opening angles of our bent, double-lobed radio sources in Section\,\ref{sect:RSparameters}.  To determine the dynamics of bent radio AGNs in these clusters, we present the offset between the bent radio AGNs and the cluster centers in Section\,\ref{sect:RSoffset}.  

\subsection{The High-$z$ COBRA Survey Sample \label{sect:COBRAsurvey}} 
The high-$z$ COBRA survey consists of 646 bent, double-lobed radio sources that were initially identified in \citet{Wing2011} using either visual selection or an automated selection process.  All of these sources lacked an SDSS identified host galaxy in \citet{Wing2011}, making them high-$z$ candidates.  However, not all 646 sources are found in galaxy clusters.  As presented in \citet{Paterno-Mahler2017}, each of the 646 sources in the high-$z$ COBRA survey were observed as part of a $Spitzer$ Snapshot program (PI: Blanton) in IRAC 3.6\,$\mu$m and 135 were observed in 4.5\,$\mu$m. Using $Spitzer$ IRAC 3.6\,$\mu$m imaging, \cite{Paterno-Mahler2017} identified 190 cluster candidates based on a 2$\sigma$ single-band overdensity when compared to a background field.  

To further verify which COBRA radio sources reside in clusters, \citet{Golden-Marx2019} analyzed 90 fields with optical and IR observations (see Table 1 in \citealp{Golden-Marx2019} for the complete list of all COBRA fields with optical follow-up).  These fields were chosen because they were the strongest cluster candidates from \citet{Paterno-Mahler2017}, the radio source was a quasar, or their radio morphology was particularly suggestive of a cluster (see Figure 1 in \citealp{Golden-Marx2019} for a breakdown of the fraction of overdense fields observed with optical follow-up).  By combining these datasets to measure photometric redshifts for the host galaxies of the bent radio AGNs, \citet{Golden-Marx2019} found that 39 fields have a red sequence overdensity at or above 2$\sigma$ when compared to background fields, bringing the total number of cluster candidates identified to 195 (see Section\,\ref{sect:clustercandidates} or \citet{Golden-Marx2019} for a complete description of red sequence overdensities).  As in \citet{Blanton2001}, the remaining sources are primarily in overdense regions but below the 2$\sigma$ threshold used to identify cluster candidates.

\subsection{The Optical Follow-up Campaign \label{sect:Optical}}
Summarized here, the optical follow-up consists of observations taken on the 4.3\,m Lowell Discovery Telescope\footnote{The Lowell Discovery Telescope was previously named the Discovery Channel Telescope (DCT) from 2003 - 2019.  The observations attributed to the LDT are the same observations \citet{Golden-Marx2019} referred to as being taken on the DCT.  Here, we refer to the telescope as the LDT to reflect the name change.} (LDT) at Lowell Observatory using the Large Monolithic Imager (LMI).  LMI has a 12$\farcm$3 $\times$ 12$\farcm$3 field of view (F.O.V), as compared to the $\approx$ 5$\arcmin$ $\times$ 5$\arcmin$ F.O.V. on $Spitzer$ IRAC, allowing us to observe both the cluster and the surrounding field.  Observations consist of 90 fields in $i$-band and 38 in $r$-band for either 3 $\times$ 600\,s or 3 $\times$ 900\,s.  \citet{Golden-Marx2019} measured a unique zero-point for each field using the SDSS catalog and found that the limiting magnitudes of our $i$-band imaging was 24.0\,mag for 3 $\times$ 600\,s exposures (24.5\,mag for 3 $\times$ 900\,s exposures) and for our $r$-band imaging was 24.5\,mag for 3 $\times$ 600\,s exposures (25.0\,mag for 3 $\times$ 900\,s exposures).  To measure the color of each detected galaxy, \citet{Golden-Marx2019} matched sources in the shorter waveband of the desired color to sources in the longer waveband within a 1$\arcsec$ search radius (e.g., \citealp{Golden-Marx2019} match $i$-band with 3.6\,$\mu$m, 3.6\,$\mu$m with 4.5\,$\mu$m, and $r$-band with $i$-band) across the entire shared F.O.V.  A complete description of the data reduction and SExtractor parameters used is found in \citet{Golden-Marx2019}.    

%start here
\subsection{The Red Sequence Cluster Candidate Sample}\label{sect:clustercandidates}
Of the 90 fields with optical observations, \citet{Golden-Marx2019} determined redshift estimates for 77 of them.  The redshift estimates come primarily from comparisons between the color of the host galaxy and EzGal \citep{Mancone2012} Spectral Energy Distribution (SED) models of early-type galaxies without an AGN component (see Section 3 in \citealp{Golden-Marx2019} for a full description of the redshift estimates and EzGal modeling), spectroscopic redshifts of detected quasars (of the six quasars discussed in this paper, five are SDSS-detected broadline quasars, while COBRA130729.2+274659 lacks further classification on its quasar identification on the SDSS archive), or SDSS photometric redshifts.  

To determine which fields host cluster candidates with a strong red sequence, \citet{Golden-Marx2019} measured the overdensity of red sequence galaxies in each field relative to a background field.  \citet{Golden-Marx2019} defined the red sequence for fields with $i - [3.6]$ and $r - i$ colors as within $\pm$0.15\,mag of the color of the host galaxy, in agreement with earlier studies \citep[e.g.,][]{Blakeslee2003,Mei2006,Mei2009,Snyder2012,Lemaux2012,Cerulo2016}.  Because bent, double-lobed radio sources are not necessarily centrally located \citep[e.g.,][]{Sakelliou2000,Garon2019}, \citet{Golden-Marx2019} used this definition of red sequence galaxies to measure the surface density of all red sequence galaxies across the shared LDT LMI - $Spitzer$ IRAC F.O.V. and locate the cluster center (see Section\,\ref{sect:clustercenter} or \citet{Golden-Marx2019} for a complete description of how the cluster center is selected). Red sequence cluster candidates have an overdensity of red sequence galaxies above 2$\sigma$ within 1$\arcmin$ of either the AGN or the red sequence determined cluster center as compared to various background fields with unique magnitude limits set for each field (due to the small expected background contamination in the $i - [3.6]$ color, \citealp{Golden-Marx2019} also required at least three red sequence galaxies for a field to be considered a cluster candidate, although 24 of the 34 $i - [3.6]$ red sequence clusters have at least five red sequence galaxies).  

In this analysis, we use all 39 red sequence cluster candidates identified in \citet{Golden-Marx2019}.  Because we are analyzing the bent radio AGNs relative to the cluster as a whole, we focus on the overdensity measurements using the red sequence cluster center.  We do this despite some fields showing a slight decrease in the overdensity of red sequence galaxies relative to the background (e.g., when centered on the radio source, four red sequence galaxies are detected within 1$\arcmin$, but when centered on the distribution of red sequence galaxies, only three of these galaxies are within 1$\arcmin$).  This is why two of our fields have a red sequence overdensity measurement below 2$\sigma$, even though we include them as cluster candidates.  By measuring the overdensity relative to the red sequence center, we can relate the geometry of each bent radio source to the location of the strongest density peak of red sequence galaxies.  

\subsubsection{Cluster Candidate Subsamples}\label{sect:clustersubsample}
\begin{figure*}
\begin{center}
\subfigure{\includegraphics[scale=0.7,trim={1.82in 3.0in 2.13in 3.505in},clip=true]{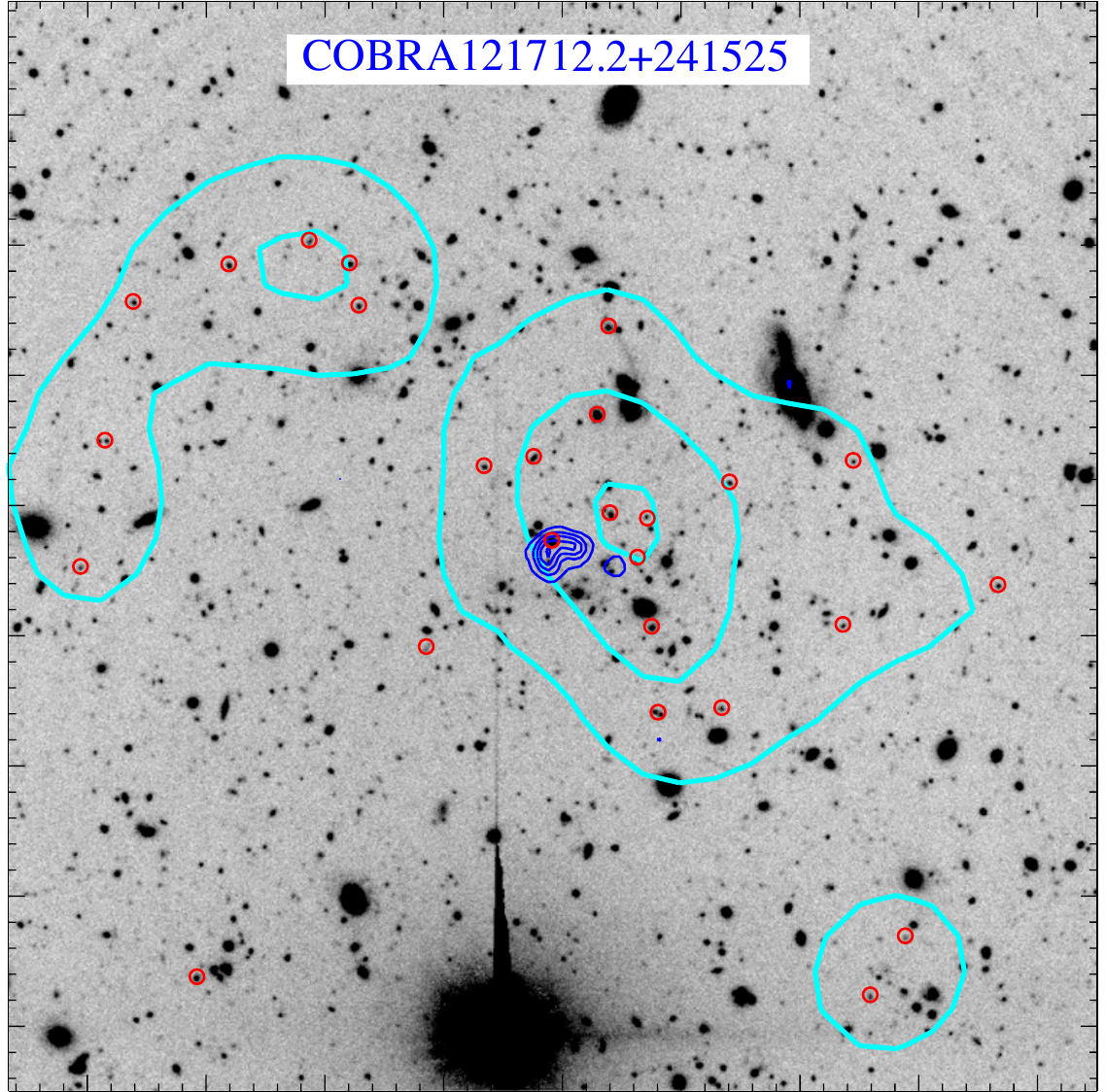}}
\subfigure{\includegraphics[scale=0.64,trim={1.6in 2.8in 1.9in 3.25in},clip=true]{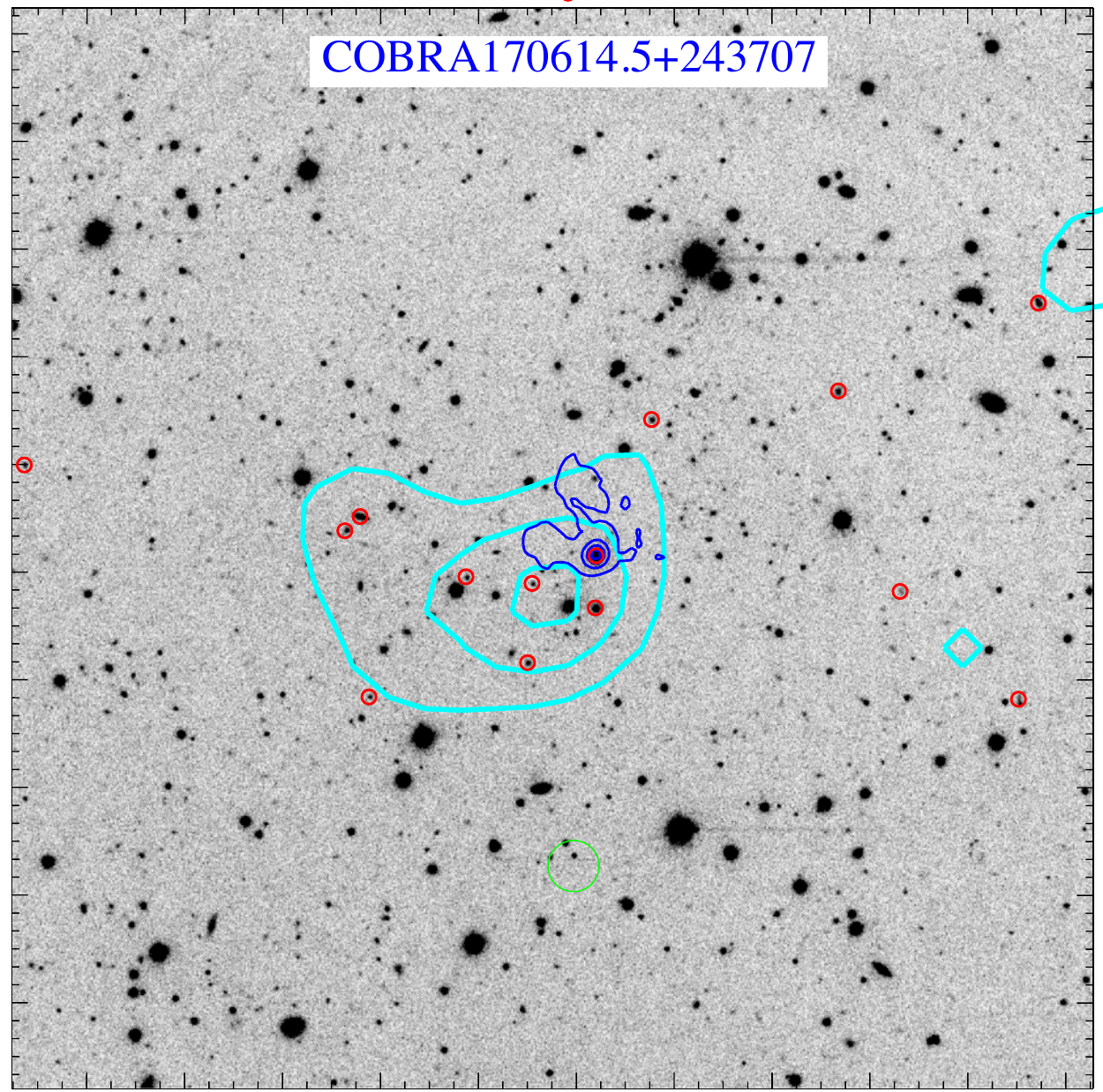}}
\subfigure{\includegraphics[scale=0.7,trim={1.82in 3.0in 2.13in 3.505in},clip=true]{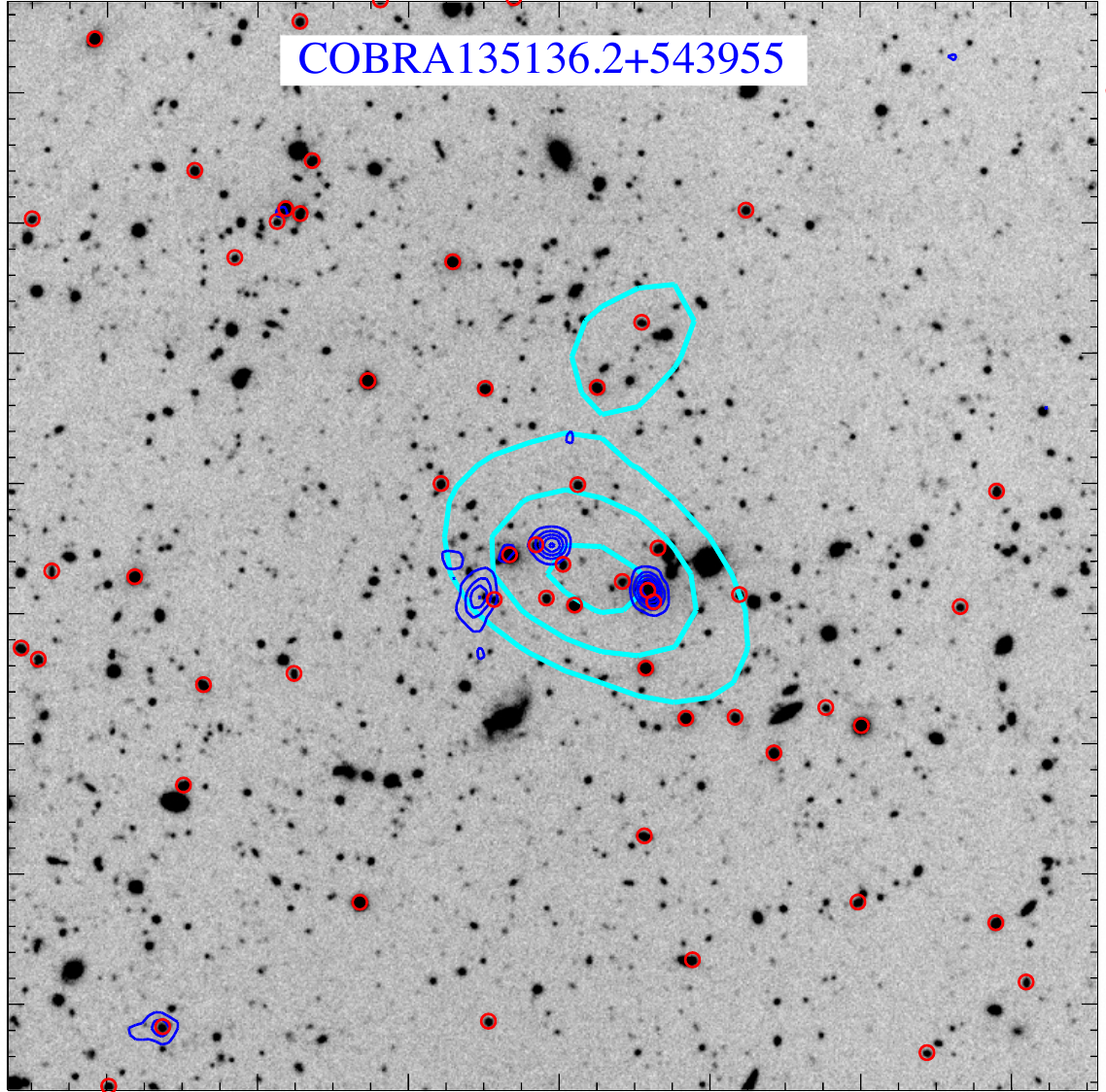}}
\subfigure{\includegraphics[scale=0.7,trim={1.82in 3.0in 2.13in 3.505in},clip=true]{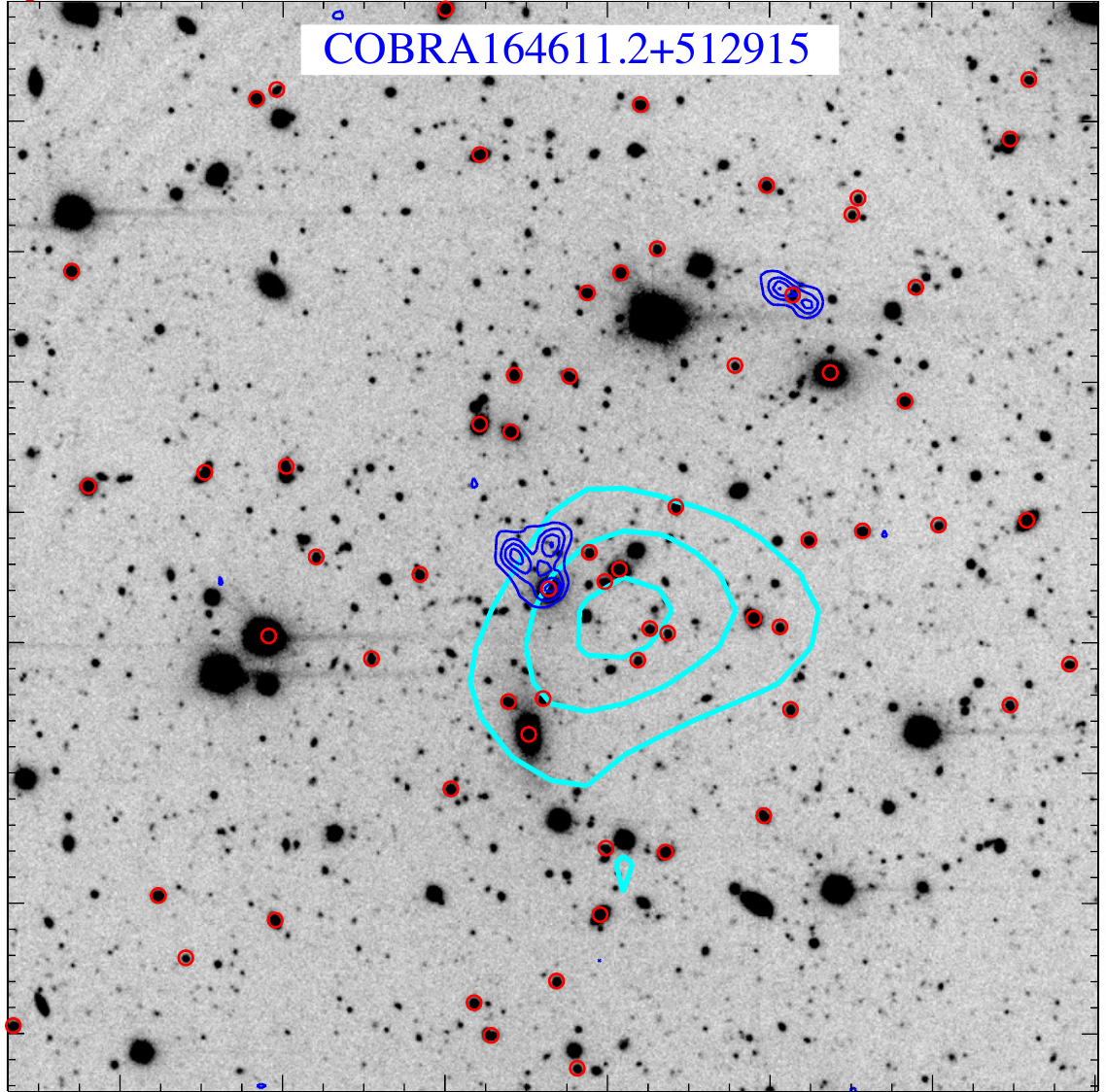}}

\caption{The strongest cluster candidates from the two m*+1 subsamples.  Each panel shows a 5$\arcmin\times$5$\arcmin$ cutout of an LDT LMI $i$-band image centered on the radio host.  The top row shows two examples drawn from the m*+1 $i - [3.6]$ subsample (COBRA0121712.2+241525 at $z$ $\approx$ 0.90 and COBRA170614.5+243707 at $z$ $\approx$ 0.71) and the bottom row shows two examples drawn from the m*+1 $r - i$ subsample (COBRA135136.2+543955 at $z$ = 0.55 and COBRA164611.2+512915 at $z$ = 0.351). The blue contours reflect the 20\,cm VLA FIRST imaging.  The red circles indicate red sequence galaxies.  The cyan contours indicate the surface density of red sequence galaxies as described in \citet{Golden-Marx2019}.  Galaxy overdensity measurements and radio source information for these and all COBRA cluster candidates analyzed here are given in Tables\,\ref{tb:GALAXY} and \ref{tb:RADIO}.}
\label{Fig:m*+1-examples}
\end{center}
\end{figure*}

\begin{figure*}
\begin{center}
\subfigure{\includegraphics[scale=0.7,trim={1.84in 3.0in 2.14in 3.505in},clip=true]{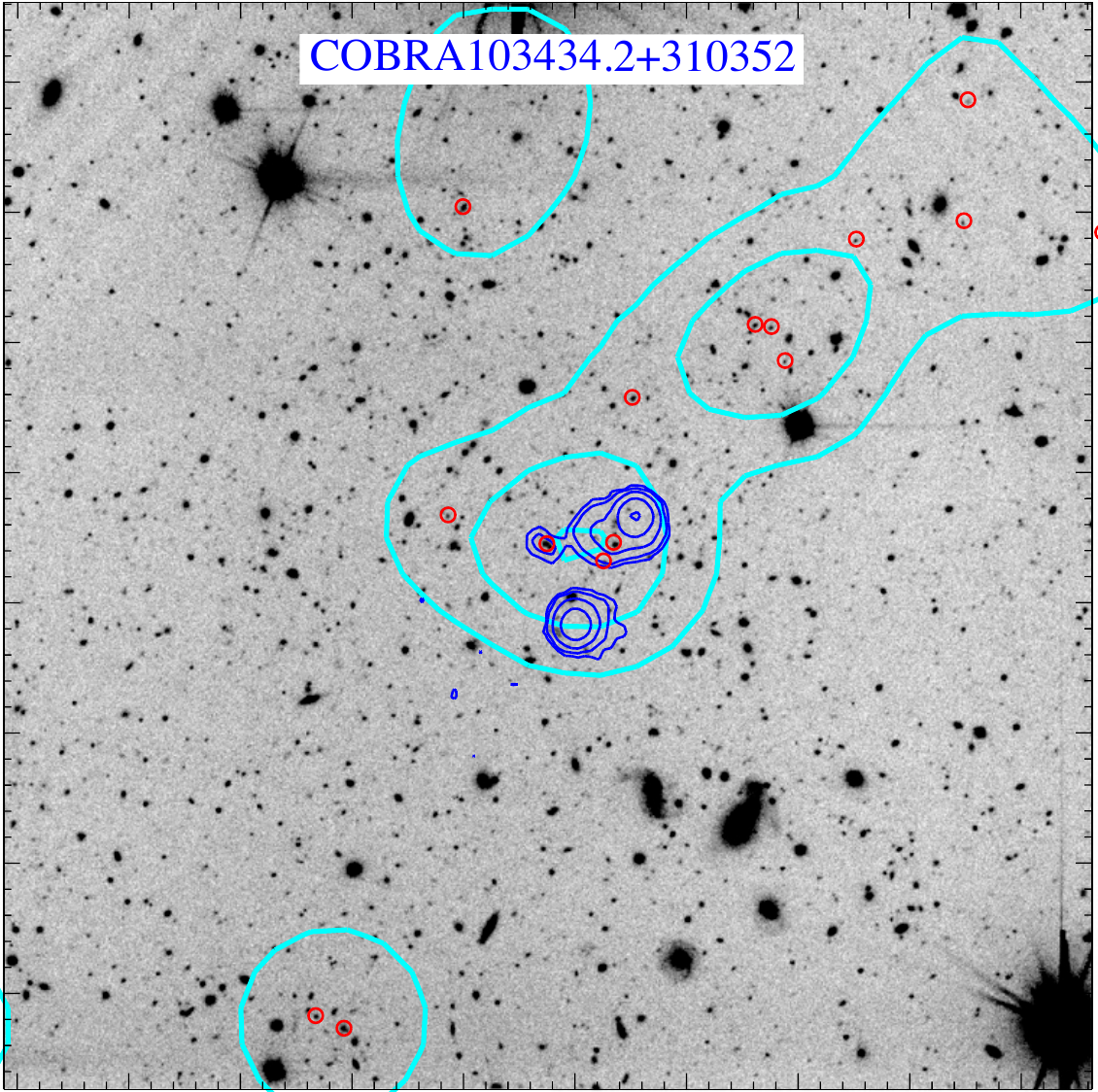}}
\subfigure{\includegraphics[scale=0.7,trim={1.84in 3.0in 2.14in 3.505in},clip=true]{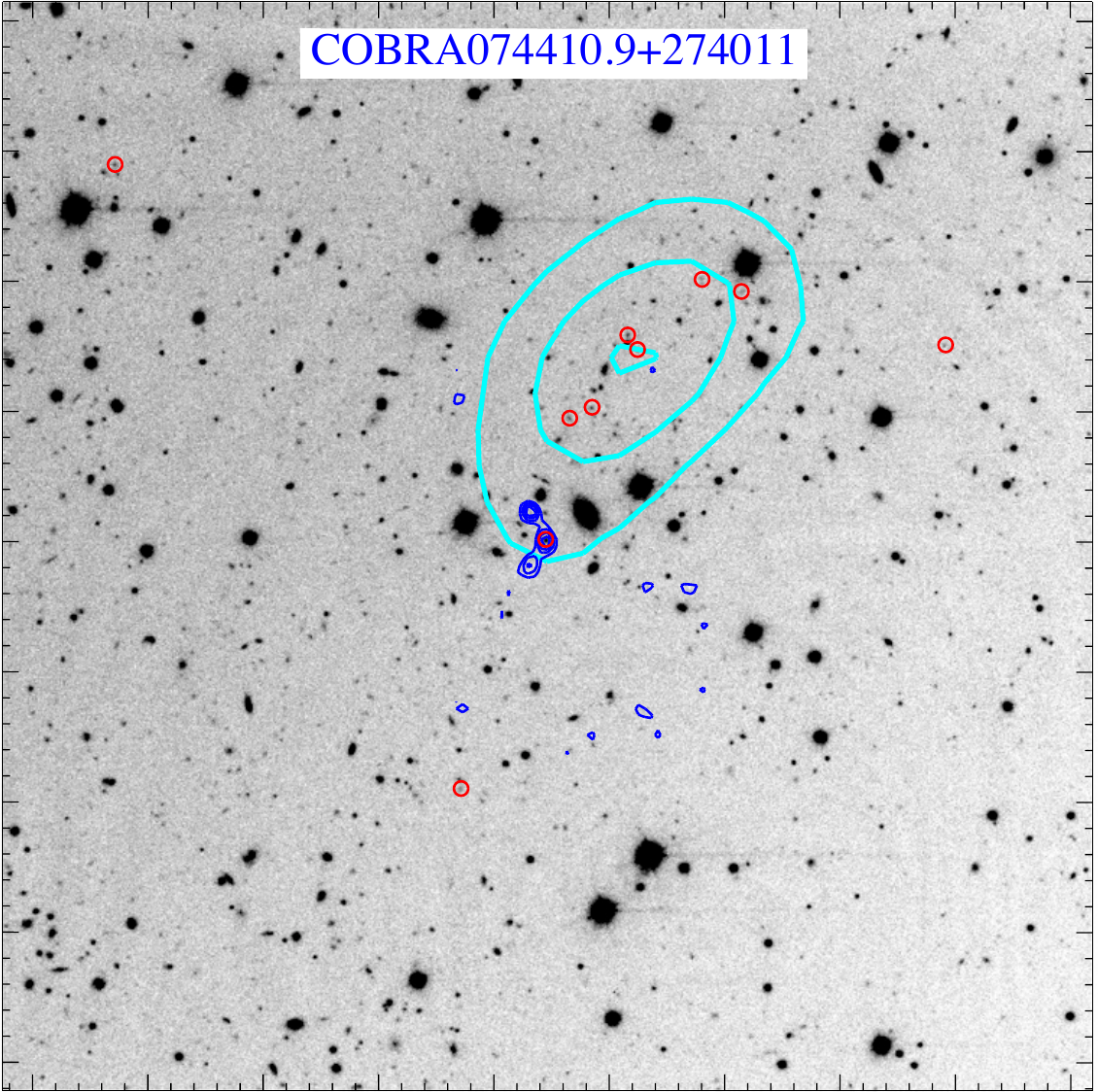}}
\subfigure{\includegraphics[scale=0.7,trim={1.84in 3.0in 2.14in 3.505in},clip=true]{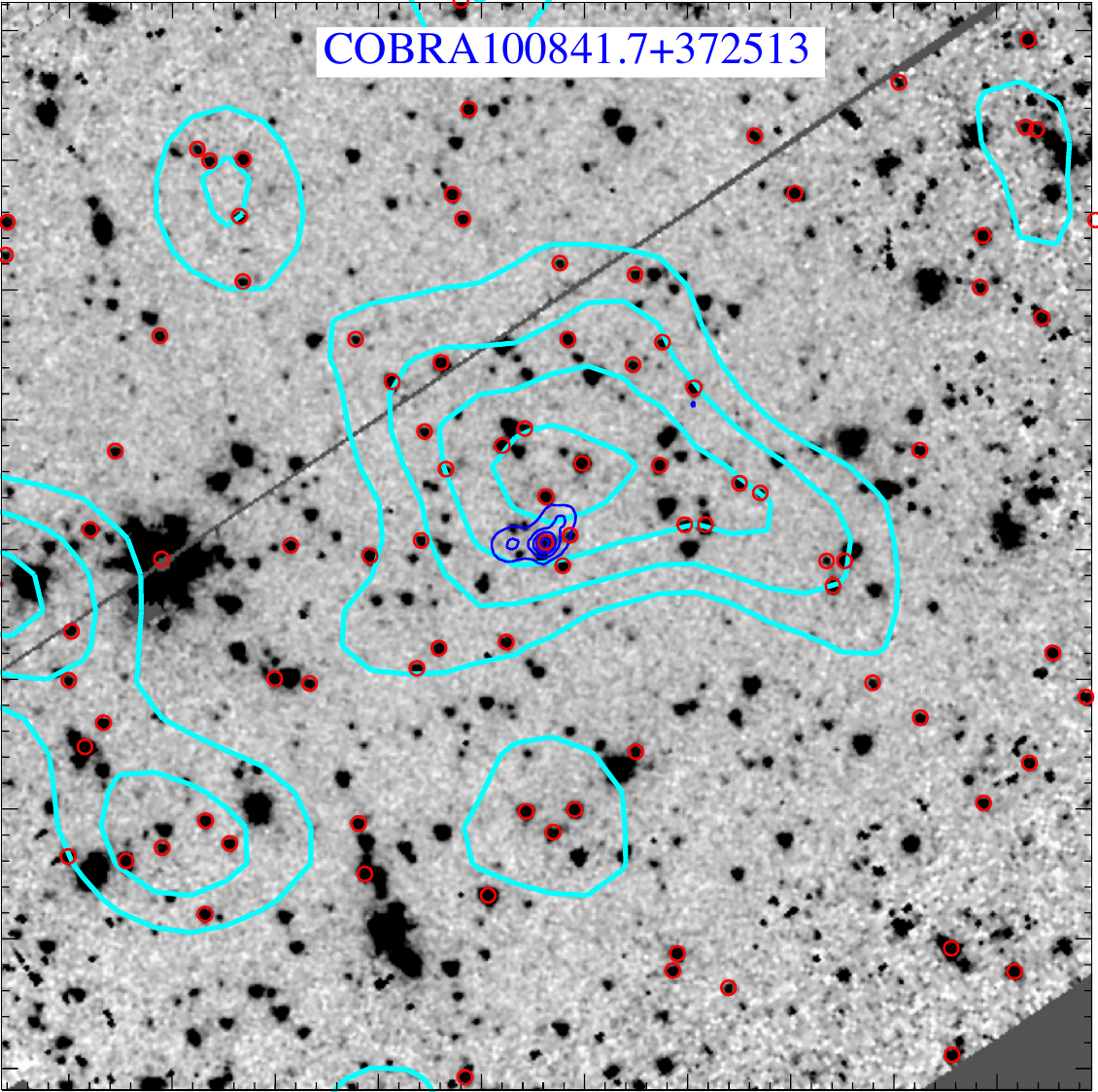}}
\subfigure{\includegraphics[scale=0.7,trim={1.84in 3.0in 2.14in 3.505in},clip=true]{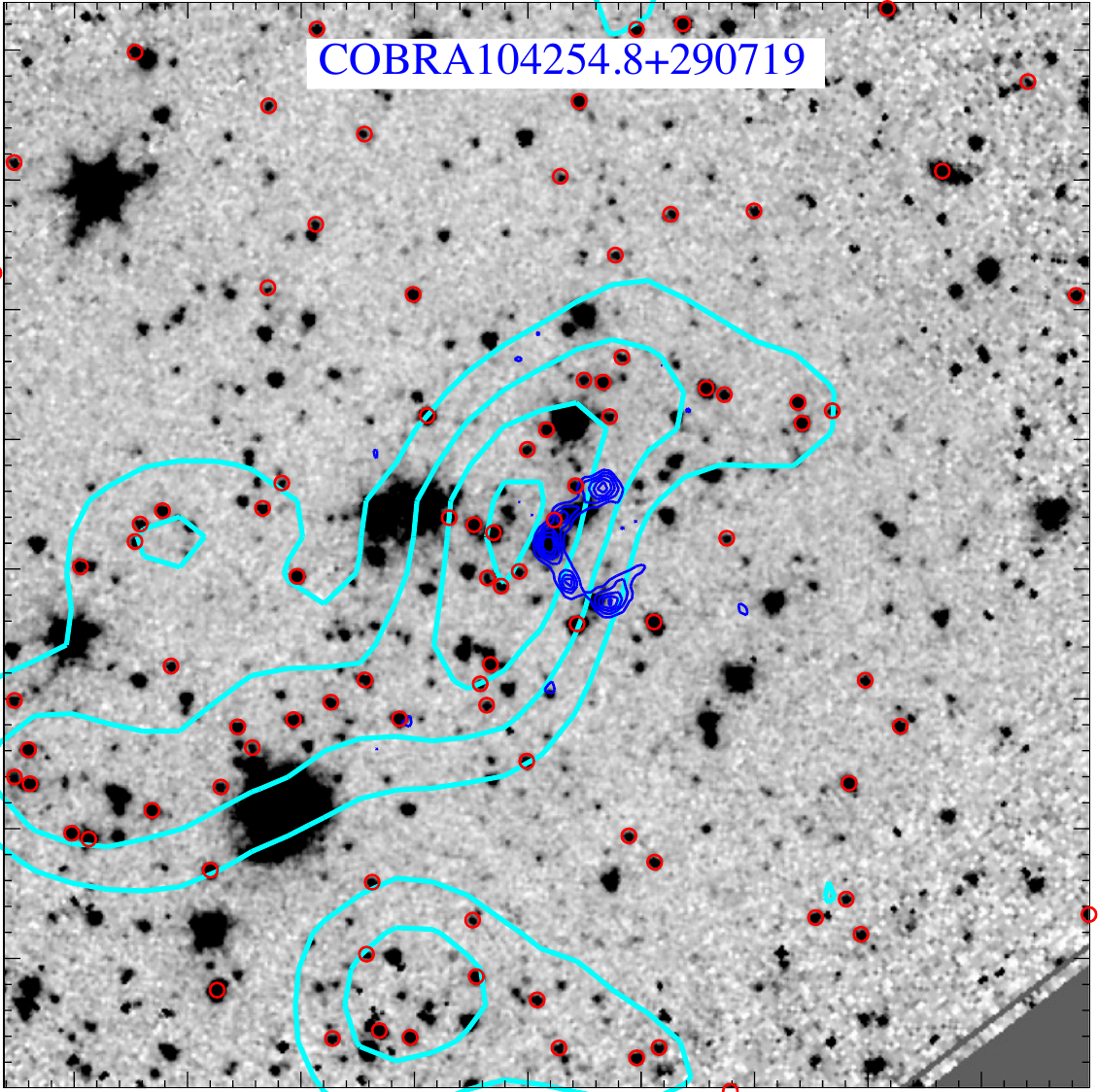}}
\caption{The strongest cluster candidates from the two magnitude-limited subsamples.  The top row shows 5\arcmin$\times$5\arcmin\ cutouts of LDT LMI $i$-band images centered on the radio host, while the bottom row shows 5\arcmin$\times$5\arcmin\ cutouts $Spitzer$ IRAC 3.6\,$\mu$m images centered on the radio host.  The top row shows two examples of the magnitude-limited $i - [3.6]$ subsample (COBRA103434.2+310352 at $z$ $\approx$ 1.20 and COBRA074410.9+274011 at $z$ $\approx$ 1.30) and the bottom row shows two examples of the magnitude-limited $[3.6] - [4.5]$ subsample (COBRA100841.7+372513 and COBRA104254.8+290719 both at $z$ $\approx$ 1.35).  The blue contours reflect the 20\,cm VLA FIRST imaging.  The red circles indicate the red sequence galaxies (or galaxies above the $[3.6] - [4.5]$ color cut for that sample).  The cyan contours indicate the surface density of red sequence galaxies as described in \citet{Golden-Marx2019}.  In COBRA104254.8+290719, the host galaxy does not follow the $[3.6] - [4.5]$ color cut indicative of $z$ $>$ 1.2 galaxies.  However, when we examine the field in $i - [3.6]$, the redshift estimate of the host agrees with the redshift estimate from the surrounding galaxies.  Galaxy overdensity measurements and radio source information for these and all COBRA cluster candidates analyzed here are given in Tables\,\ref{tb:GALAXY} and \ref{tb:RADIO}.}
\label{Fig:maglim-examples}
\end{center}
\end{figure*}

To determine which radio sources lie in cluster candidates, we use three colors ($i - [3.6]$, $[3.6] - [4.5]$, and $r - i$) and two magnitude limits (either a uniform depth of m*+1 depending on the redshift of the cluster or a fixed detected magnitude limit).  The magnitude limit of our 3.6\,$\mu$m observations is 21.4\,mag.  Based on the magnitude of a modeled m* galaxy from EzGal, this limits the m*+1 subsample to cluster candidates at $z$ $<$ 1.1 in the $i - [3.6]$ analysis and $z$ $<$ 1.0 in the $r - i$ analysis.  Using these colors and magnitude limits, we divide our sample into four non-overlapping subsamples: the m*+1 $i - [3.6]$ subsample, the magnitude limited $i - [3.6]$  subsample, the magnitude-limited $[3.6] - [4.5]$ subsample, and the m*+1 $r - i$ subsample (see Figures\,\ref{Fig:m*+1-examples} and \ref{Fig:maglim-examples} for examples of these cluster candidates and the red sequence galaxies selected and Table\,\ref{tb:GALAXY} for the list of cluster candidates in each subsample).
 
Of these subsamples, the m*+1 $i - [3.6]$ subsample is the most statistically robust.  This subsample includes 18 fields that span 0.4 $<$ $z$ $<$ 1.1 (see Table\,\ref{tb:GALAXY} for the list of cluster candidates).  The strength of the $i - [3.6]$ color comes from the linear relationship between color and redshift out to $z$ $\approx$ 2.0, making it less prone to false identification (see Figure 4 in \citealp{Golden-Marx2019}).  Additionally, the 4000\,\AA\, break, one of the strongest and most identifiable features in an early-type galaxy's spectrum, falls in this color range at 0.9 $<$ $z$ $<$ 8.0, making each color easily identifiable with a given redshift.  The strength of the m*+1 completeness limit is the uniform magnitude limit for all of the clusters in this subsample, which makes them easily comparable to one another.  Additionally, of our LDT observations, we have the most complete sample for $i$-band (as opposed to $r$-band) and, due to our observing scheme, these fields generally have seeing $<$ 1$\farcs$2 (although outliers exist, see \citealp{Golden-Marx2019}).  These qualities combined make the m*+1 $i - [3.6]$ subsample the most robust and the primary subsample used in this analysis.  

The second subsample we analyze is the magnitude-limited $i - [3.6]$ subsample.  This subsample includes 9 cluster candidates that span 1.1 $<$ $z$ $<$ 1.4.  This subsample shares many of the same strengths as the m*+1 $i - [3.6]$ subsample.  However, due to the fixed magnitude limit, we probe different depths for each field depending on the redshift of the cluster.

The third subsample we analyze is the magnitude-limited $[3.6] - [4.5]$ subsample.  We use it to identify some of the highest redshift cluster candidates in our sample (1.2 $<$ $z$ $<$ 2.2) and it contains 7 clusters.  Although some fields in this subsample have $i - [3.6]$ cluster detections, we chose to use this color to analyze these fields because the $[3.6] - [4.5]$ color cut is extremely effective and our $i$-band images are relatively shallow at high redshift. The strength of the magnitude-limited $[3.6] - [4.5]$ subsample comes from the shape of the color-redshift relation for this color (see Figure 4 in \citealp{Golden-Marx2019}).  At $z$ $>$ 1.5, this trend is flat and at $z$ $>$ 1.2, a single color cut ($[3.6] - [4.5]$ $>$ $-$0.15) prevents the detection of most foreground galaxies and allows for the detection of most high-$z$ star-forming and quiescent galaxies \citep[e.g.,][]{Papovich2008,Cooke2015}.  \citet{Golden-Marx2019} further reinforced the efficacy of the $[3.6] - [4.5]$ color-cut in detecting high-$z$ galaxies using the extensive multi-band photometric and spectroscopic data from the ORELSE survey \citep{Lubin2009}, finding the fewest low-$z$ interlopers in this color-cut as compared to their red sequence completeness fractions.  However, despite the accuracy in removing foreground galaxies, the $[3.6] - [4.5]$ color cut is limited in isolating galaxies of a specific redshift given the flat nature of the color-redshift distribution, allowing for possible contamination within the sample.  Furthermore, the $[3.6] - [4.5]$ color cut is most effective in the redshift regime beyond our m*+1 magnitude limit, which again means we probe different depths for each field.

The fourth subsample is the m*+1 $r - i$ subsample.  This subsample contains 5 cluster candidates and we use it to probe some of the lowest redshift sources in our sample (0.35 $<$ $z$ $<$ 0.8).  The cluster candidates at similar redshifts within the m*+1 $i - [3.6]$ subsample are not included in this subsample because they either lack $r$-band imaging or have $r$-band observations with poor seeing ($>$ 1$\farcs$2).  Although the m*+1 $r - i$ subsample probes a  uniform depth, the $r - i$ color - redshift relation is only linear out to $z$ $\approx$ 0.6.  However, the 4000\,\AA\, break falls within this color range at 0.6 $<$ $z$ $<$ 0.9, making the color cut effective in this range as well.

\subsection{Determining the Cluster Center}\label{sect:clustercenter}
Because bent, double-lobed radio sources are not always located at the centers of clusters \citep[e.g.,][]{Sakelliou2000, Garon2019}, \citet{Golden-Marx2019} developed a Python pipeline to measure the local surface density of red sequence galaxies in COBRA fields.  \citet{Golden-Marx2019} did this by placing a uniformly spaced 10$\arcsec$ grid over the combined F.O.V. of the $i$-band and 3.6\,$\mu$m images and measuring the number of red sequence galaxies within 10$\arcsec$ of each grid point (they also did this in a uniform manner for the other colors).  \citet{Golden-Marx2019} used Gaussian Kernel smoothing to identify the peak of the red sequence overdensity and determine a unit-weighted cluster center.  In doing this, \citet{Golden-Marx2019} determined where the dense core of red sequence galaxies in an evolved, relaxed cluster is regardless of the location of the bent radio source (see Table\,\ref{tb:GALAXY} for the coordinates of the red sequence cluster centers for each cluster candidate and Figures\,\ref{Fig:m*+1-examples} and \ref{Fig:maglim-examples} for examples of the red sequence surface density contours).  While most clusters, especially the strongest candidates, show a singular strong overdensity, some fields show multiple smaller overdensity peaks.  Because \citet{Sakelliou2000} found that most bent AGNs in their low-$z$ sample of clusters are at clustocentric radii $<$ 300\,kpc (which is $\approx$ 0$\farcm$6 at $z$ $=$ 1.0), \citet{Golden-Marx2019} selected the nearest overdensity to the AGNs in these cases.  

\subsection{Radio Source Parameters}\label{sect:RSparameters}
\subsubsection{Radio Source Opening Angle $\&$ Physical Size}\label{sect:RSopeningSize}
The success of the COBRA survey in identifying galaxy clusters is predicated on bent, double-lobed radio sources being found nearly exclusively in dense cluster and group environments \citep[e.g.,][]{Blanton2000,Blanton2001,Wing2011,Paterno-Mahler2017,Garon2019,Golden-Marx2019}.  Because our radio observations are from the VLA FIRST survey, we measure the size and opening angle of our bent radio sources using the publicly available FIRST Catalog Database updated as of December 17, 2014.  The updated FIRST values result in slightly different parameters than those reported in \citet{Paterno-Mahler2017}, specifically with regards to the total integrated flux of each radio source.  As a result, we remeasure all radio source properties previously reported in \citet{Paterno-Mahler2017}.  Although VLA FIRST was designed to reach a uniform depth over all observed regions \citep{Becker1995}, we probe a wide range of redshifts.  Thus, we are not sensitive to lower power radio sources at higher redshifts.

To verify which VLA FIRST detected radio components are associated with each COBRA bent radio source, we individually search each field for the radio detections within 120$\arcsec$ of the central radio source coordinates reported in \citet{Paterno-Mahler2017}.  We visually inspect each component to confirm which detections are associated with each bent source.  For 36 of the 39 red sequence cluster candidates reported in \citet{Golden-Marx2019}, we identified an obvious bent radio source consisting of at least two radio components (see Table\,\ref{tb:GALAXY} for these fields).  For the remaining three clusters, our further analysis of the radio source left us uncertain whether a bent radio source was truly present in these cases or whether the initial detection was consistent with a chance alignment of unassociated radio sources.  Due to this ambiguity, we remove these three fields from our analysis (these fields are marked in Table\,\ref{tb:GALAXY}).

\begin{figure}
\centering
\epsscale{1}
\includegraphics[scale=0.45,trim={2.5in 0.15in 2.5in 0.15in},clip=True]{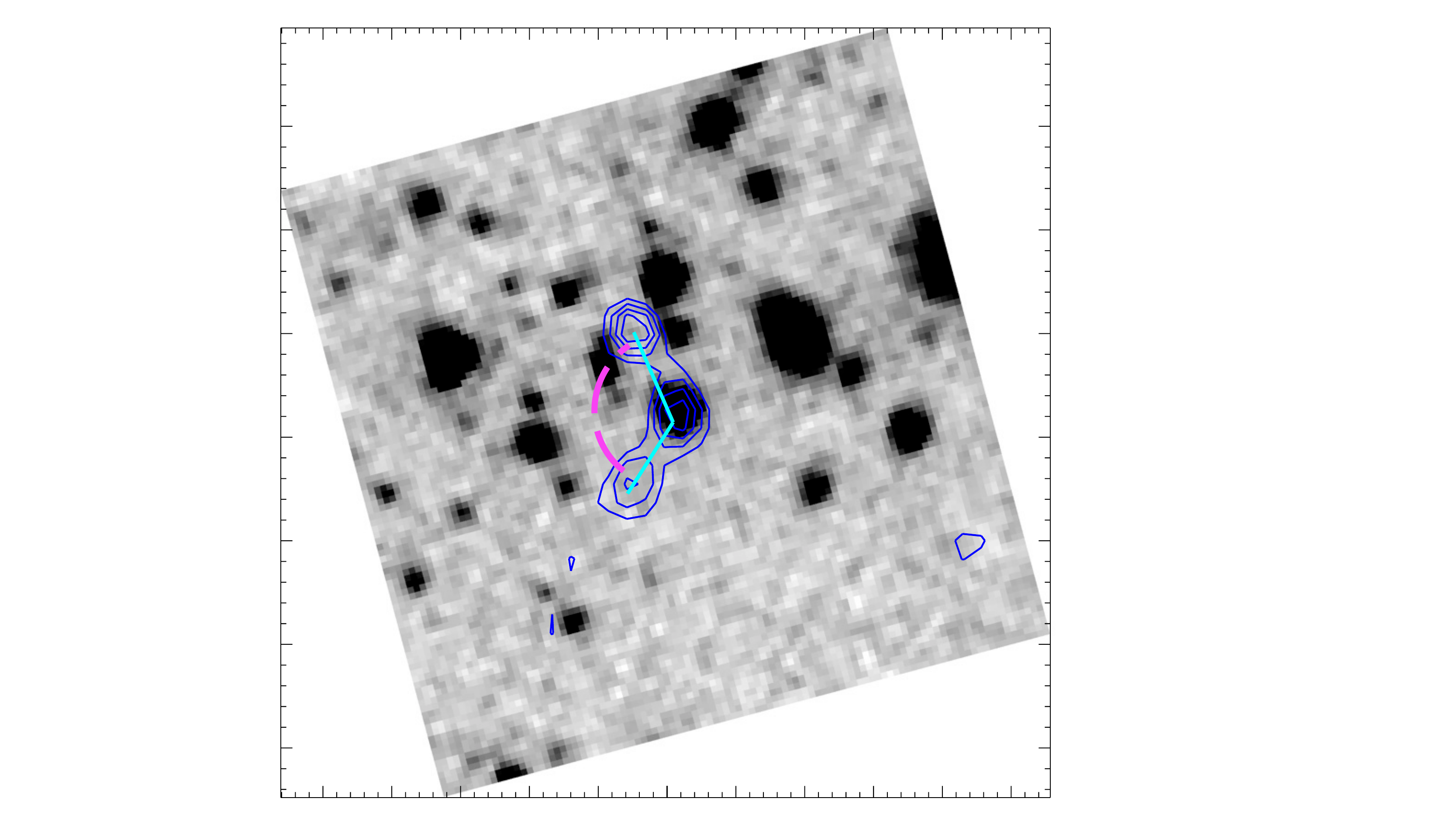}

\caption{Illustration of the opening angle measurement technique, in this particular case for COBRA074410.9+274011.  The greyscale image is an $\approx$ 0$\farcm$5 $\times$ 0$\farcm$5 cutout of the 3.6\,$\mu$m IRAC mosaic.  The VLA FIRST radio contours are overlaid in blue.  The cyan lines that trace both rays of the opening angle are drawn from the center of each lobe component to the radio core.  The magenta arc indicates where the opening angle was measured.  \label{Fig:OpeningAngleDiagnostic}}
\end{figure}

To measure the opening angle of each bent, double-lobed radio source, we use the VLA FIRST radio components identified for all 36 sources.  As mentioned previously, these bent sources consist of two or three VLA FIRST identified radio detections (some bent sources lack a radio detected core; see Table\,\ref{tb:RADIO}).  We note that although these are individual sources, the contours shown in Figures\,\ref{Fig:m*+1-examples}, \ref{Fig:maglim-examples}, and \ref{Fig:OpeningAngleDiagnostic} are displayed to create a single connected radio source when possible.  They are not reported in the VLA FIRST catalog in this manner.  Thus, to measure the opening angle, we use the position of each radio component's central RA and DEC as reported in the VLA FIRST catalog.  By visual inspection, we identify which component is the central radio core associated with the host galaxy and which components are radio lobes.  For sources without a detected radio core, we use the location of the optical/IR host galaxy as the core.  

To measure the opening angle of each bent source, we use the law of cosines, where the Opening Angle is defined as:  
\begin{equation}\label{equation:Arccos}
    \rm{Opening\,\,Angle} = \rm{arccos}\left(\frac{(H)^2 - (L_1)^2 - (L_2)^2}{-2(L_1)(L_2)}\right).
\end{equation}
In Equation\,\ref{equation:Arccos}, L$_{1}$ and L$_{2}$ represent the angular distance between the center of the radio core and the center of each radio lobe based on the reported RA and DEC from the VLA FIRST catalog.  Geometrically, L$_{1}$ and L$_{2}$ represent the sides of the triangle used to measure the opening angle.  H represents the line connecting the centers of both radio lobes.  Figure\,\ref{Fig:OpeningAngleDiagnostic} shows a schematic of how the opening angle is measured.  

From these measurements of the opening angle, we confirm that the majority of our sources are WAT-like sources with only a few sources being narrow-angle tail (NAT)-like sources; 27 of our bent radio sources have opening angles greater than 90$^{\circ}$, the minimum opening angle of typical WAT sources \citep[e.g.,][]{Owen1976,Valentijn1979,ODea1985} and only 2 of our sources are NAT-like, with opening angles less than 45$^{\circ}$ \citep[e.g.,][]{Owen1976,Valentijn1979,ODea1985}.  This difference is vital to our analysis as WATs are typically found near cluster centers and hosted by more massive elliptical and cD galaxies, although counter examples exist \citep[e.g.,][]{Sakelliou2000}, while NATs are typically associated with radio AGNs at larger peculiar velocities farther from cluster centers \citep{Owen1976}.  That our sample is primarily WATs indicates that these sources should be strong tracers of cluster cores and generally near cluster centers (as seen in Figure\,\ref{Fig:OffsetHistogram}; see Sections\,\ref{sect:Richness} and \ref{sect:RichnessBending} for a description of how richness and opening angle relate). 

Beyond the opening angle, we are also interested in the projected physical extent of each radio source.  To measure the physical size of the radio source, we first measure the length of each arm of the ``C" shape, by measuring the distance between the RA and DEC of each component reported in VLA FIRST.  However, using only the distance between the central position of each VLA FIRST source will result in an underestimation of the projected length of the radio source.  To correct for this, we follow \citet{Wing2011} and include the measured size of each radio component reported in the VLA FIRST database.  Specifically, the VLA FIRST database reports the deconvolved major and minor axis of an approximate elliptical fit for each detected radio component.  Thus, to better measure the full extent of each radio source, we add the length of the semi-major axis of each lobe component of the radio source to the distance between the centers of the radio components measured earlier.  We then convert our angular length measurement into a physical measurement by assuming the radio source at is at the host cluster redshift using the Astropy Cosmology tool \citep{astropy:2013,astropy:2018}.%by assuming H$_{0}$=70\,km\,s$^{-1}$\,Mpc$^{-1}$, $\Omega_{\Lambda}$ = 0.70, and $\Omega_{M}$ = 0.30.} 

Although we aimed to measure the projected physical size uniformly for each radio source, for COBRA074410.9+274011 (see Figure\,\ref{Fig:OpeningAngleDiagnostic}), the second lobe has no deconvolved semi-major or semi-minor axis because the angular size is smaller than the deconvolution.  For this source, we use the reported convolved semi-major axis from the VLA FIRST survey catalog as our estimate of the additional length of the radio source, although this yields an upper limit.  For the few sources where there are more than three components (i.e., multiple components in a single lobe), we determine the size by measuring the distance between the core and the radio component at the greatest distance from the core (again adding an additional factor of the beam-deconvolved semi-major axis for that component).   

We estimate the error in the opening angle and angular size of the radio source (see Table\,\ref{tb:RADIO}) by propagating the reported positional uncertainty from VLA FIRST through the law of cosines and inverse cosine equations used to determine the length of the radio lobes and the size of the opening angle.  For the size of the radio sources, we convert the angular size of each radio source to physical size and account for an additional factor of uncertainty associated with the redshift.    

\begin{figure}
\centering
\epsscale{1}
\includegraphics[scale=0.48,trim={0.3in 0.1in 0.0in 0.2in},clip=True]{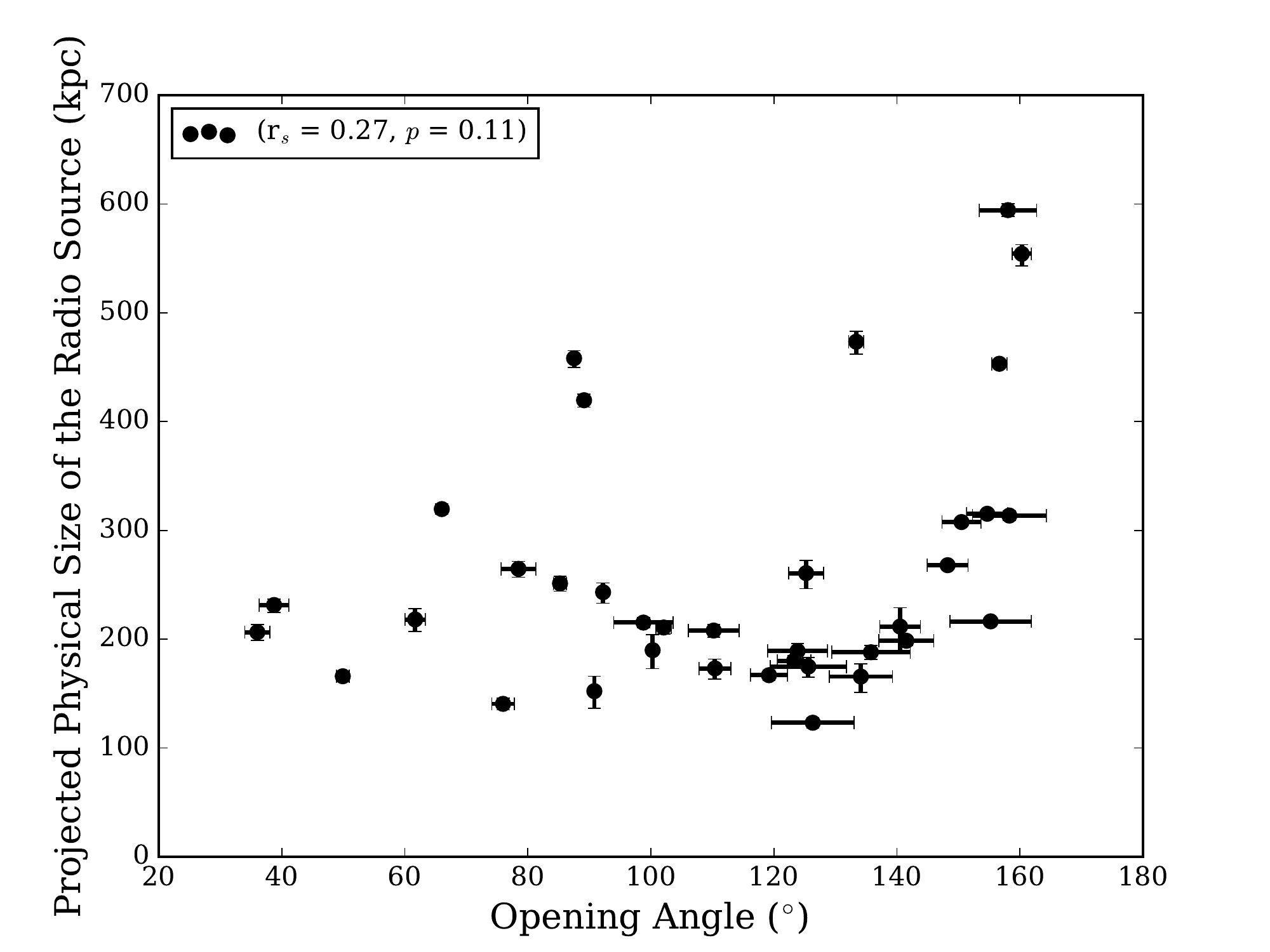}

\caption{The projected physical size of each cluster candidate's bent, double-lobed radio source as a function of the opening angle, measured as described in Section\,\ref{sect:RSopeningSize}.  Although no strong correlation exists between opening angle and physical size, we do find that the radio sources with the largest physical sizes have the largest opening angles.  \label{Fig:OpeningAnglePhysSize}}
\end{figure}

For the 36 bent AGNs, we compare the physical size of the radio source and the opening angle, to one another in Figure\,\ref{Fig:OpeningAnglePhysSize}.  We adopt a Spearman Test (see Appendix \ref{Appendix}) to determine if a correlation exists between the opening angle and the projected physical size.  It returns r$_{s}$ = 0.27 and $p$ = 0.11, which corresponds to a weak correlation with insufficient evidence to reject the null hypothesis, thus implying no strong correlation between these measures.  When we evaluate the mean and normalized mean absolute deviation for bent AGNs with opening angles below and above 80$^{\circ}$, we find that these two samples are statistically similar.   Interestingly, as shown in Figure\,\ref{Fig:OpeningAnglePhysSize} the only difference appears to be that the largest bent AGNs all have opening angles above 80$^{\circ}$, with the largest being the least bent.  It is possible that the difference in the spread of the physical size of the radio sources is the result of small number statistics.  We discuss if these radio properties are linked to any cluster properties in Sections\,\ref{sect:AGNcluster} and \ref{sect:discussion}.

\subsubsection{Radio Source Power}\label{sect:RSluminosity}
To measure the radio power of each AGN, we first sum the integrated 20\,cm flux density for each component reported in the VLA FIRST Catalog Database to determine a total radio flux for each bent source.  We convert this flux into power using Equation\,\ref{Eq:1}, 
\begin{equation}\label{Eq:1}
    P_{1.44} = 4\pi\,D^{2}_{L}S_{1.44}\,(1+z)^{\alpha - 1},
\end{equation}
where the (1+z)$^{\alpha - 1}$ includes both the k-correction to 1.4\,GHz and distance dimming.  In Equation\,\ref{Eq:1}, D$_{L}$ is the luminosity distance at the redshift of the AGN, which is calculated in the same manner as for the size of the radio source, S$_{1.44}$ is the summed integrated flux density of each component of our bent radio sources at the reference frequency of the VLA FIRST survey (1.44\,GHz), and $\alpha$ is the spectral index.  Although there is some observed scatter in the measured values of $\alpha$, with typical values ranging from 0.7 to 0.8 \citep[e.g.,][]{Kellermann1988,Condon1992,Peterson1997,Lin2007,Miley2008}, we chose to use a spectral index of 0.8 for all of our bent radio sources, as this value is typical of such extragalactic radio sources \citep{Sarazin1988} and used in similar studies of AGN in high-$z$ clusters \citep[e.g.,][]{Paterno-Mahler2017,Moravec2019,Moravec2020}.  We find a range of radio powers at 1.44\,GHz, from 6.4$\times$10$^{24}$ - 3.5$\times$10$^{27}$\,W\,Hz$^{-1}$.  As expected, the least powerful sources are our lowest redshift radio sources (COBRA15313.0$-$001018 at $z$ $\approx$ 0.44 and COBRA164611.2+512915 at $z$ $\approx$ 0.351), while the four most powerful radio sources, all above 10$^{27}$\,W\,Hz$^{-1}$, are at $z$ $\geq$ 1.2.  Three of these four sources are SDSS-identified quasars, with COBRA103434.2+310352 being the only very powerful source not identified as an SDSS quasar.  The bulk of our radio source powers are $\approx$ 10$^{25}$W\,Hz$^{-1}$, making our sample similar to that of \citet{Moravec2019}.  Also, as discussed in \citet{Golden-Marx2019}, the majority of our sample are fainter than the minimum radio power of the CARLA sample (P$_{500\,MHz}$ $\approx$ 10$^{27.5}$\,W\,Hz$^{-1}$ \citep[e.g.,][]{Wylezalek2013,Cooke2015}.  

The typical P$_{1.44}$ divide between FRI and FRII radio sources is $\approx$ 10$^{25}$\,W\,Hz$^{-1}$ \citep[e.g.,][]{Fanaroff1974,Ledlow1996}, and all but our two faintest and lowest redshift radio sources are above this threshold.  Additionally, 16 of the 36 bent radio sources in our sample fall within the range of the values reported for WATs (10$^{24.75}$\,W\,Hz$^{-1}$ $<$ P$_{1.44}$ $<$ 10$^{25.75}$\,W\,Hz$^{-1}$) \citep[e.g.,][]{Blanton2000,Blanton2001,Wing2011}.  Given the large range of measured radio power, this implies that bent morphology is not strongly tied to radio luminosity, though we do not report any sources that are more than two orders of magnitude above the FRI/FRII divide.%  From these radio powers, we should expect to trace a variety of strong and poor cluster environments.          

\begin{figure}
\centering
\epsscale{1}
\includegraphics[scale=0.48,trim={0.25in 0.1in 0.0in 0.4in},clip=True]{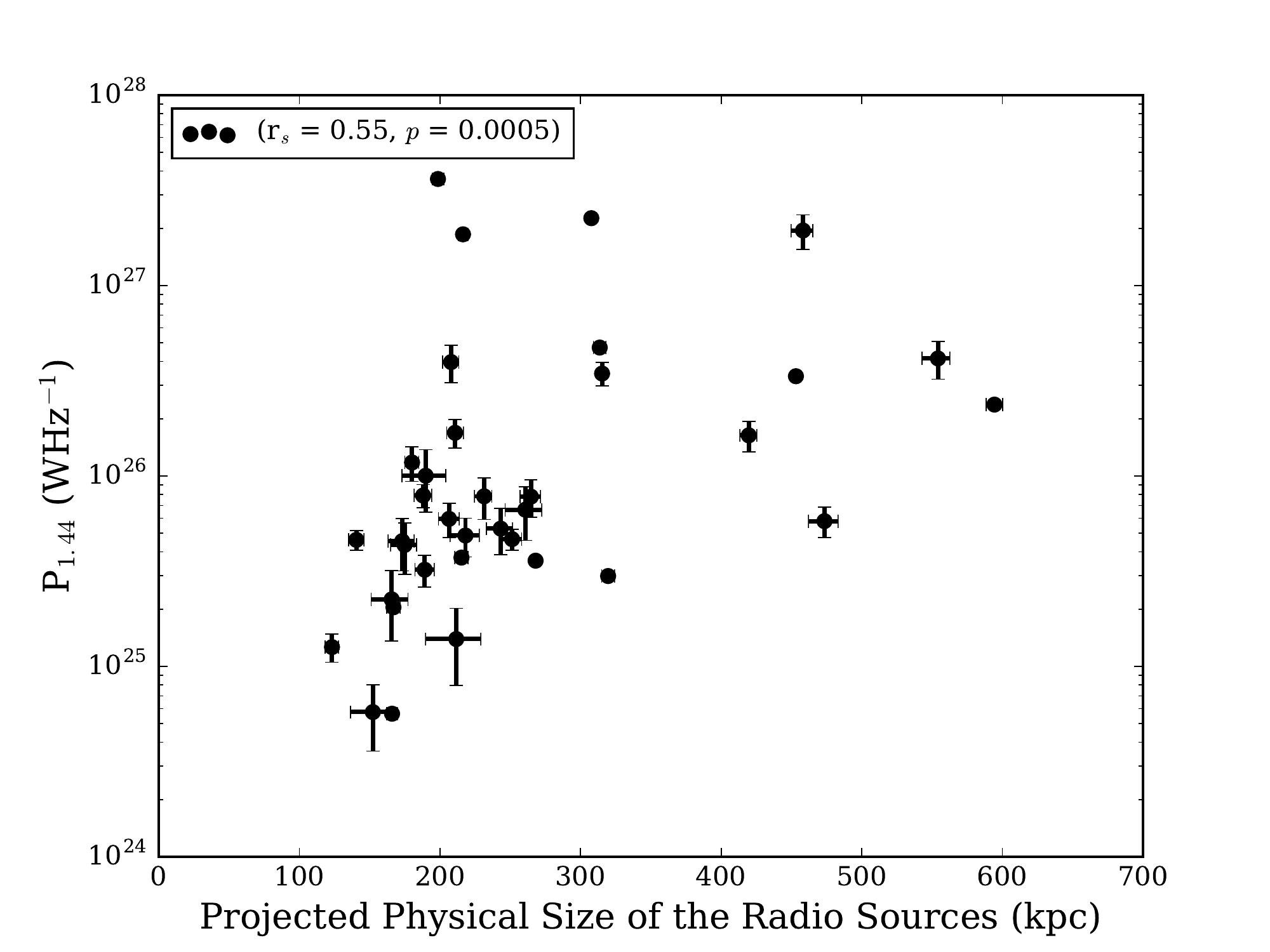}

\caption{The radio power (P$_{1.44}$) of each radio source as a function of the projected physical size of each radio source.  We measure the radio power in Section\,\ref{sect:RSluminosity}.  We determine the physical size of each source using the redshift estimates reported in \citet{Golden-Marx2019} and the radio source positions and component sizes from VLA FIRST.  We use the radio fluxes reported in the FIRST Catalog Database to measure our radio power.  We find strong evidence for a positive correlation between radio power and physical size.  \label{Fig:RadioLuminosityPhysicalSize}}
\end{figure}

As the radio sources in our sample are not singular point sources, we compare the physical extent of each radio source with the power of the source.  As expected, in Figure\,\ref{Fig:RadioLuminosityPhysicalSize}, we find that generally the most energetic radio sources are also those with the largest physical extent, in agreement with \citet{Moravec2020}.  We further verify the strength of this correlation using the Spearman test.  We obtain r$_{s}$ = 0.55 and $p$ = 0.0005, which strongly suggests a positive correlation and a rejection of the null hypothesis.  However, despite the uniformity of the VLA FIRST survey \citep{Becker1995}, these detections and any analysis are subject to Malmquist bias due to the wide redshift range.  In order to address the validity of this correlation, we note that the 1\,mJy flux limit for VLA FIRST \citep{Becker1995} corresponds to $\approx$ 2.8$\times$10$^{25}$W\,Hz$^{-1}$ at $z$ = 2.2, the redshift of our highest redshift cluster candidate.  Thus, we are likely underestimating the true number of faint, high-$z$ bent radio sources.  In terms of the physical size of these radio sources, however, we are not limited by the angular resolution of VLA FIRST.  The 5$\arcsec$ beam size corresponds to $\approx$ 41\,kpc at $z$ = 2.2, much smaller than our smallest radio source.  Thus, the lack of small ($<$ 200\,kpc), bright ($>$ 10$^{26}$W\,Hz$^{-1}$) radio sources is likely real, while the lack of large ($>$ 200\,kpc), faint ($<$ 10$^{25}$W\,Hz$^{-1}$) radio sources may not be.    

\subsection{Radio Source Offsets from the Cluster Center}\label{sect:RSoffset}
\begin{figure}
\centering
\epsscale{1}
\includegraphics[scale=0.5,trim={0.45in 0.1in 0.0in 0.4in},clip=True]{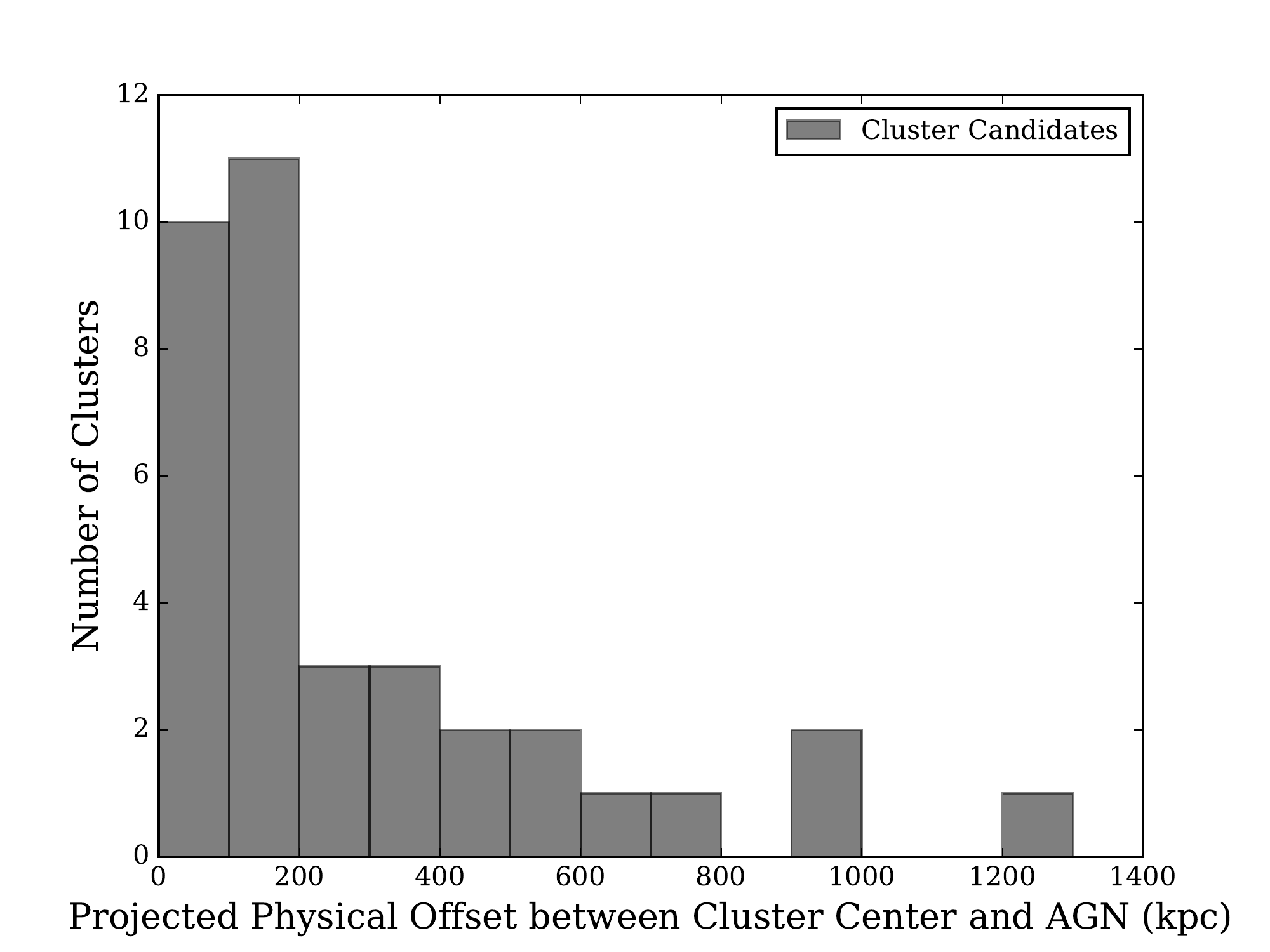}

\caption{Histogram showing the distribution of projected physical offsets between the bent, double-lobed radio sources and the red sequence cluster center.  These offsets use the measurements of the cluster center presented in \citet{Golden-Marx2019}.  We find that most radio sources are less than 200\,kpc from the red sequence cluster center, although we see sources offset by as much as $\approx$ 1200\,kpc.  \label{Fig:OffsetHistogram}}
\end{figure}

A cluster center based on the surface density of red sequence galaxies was determined by \citet{Golden-Marx2019}.  Although \citet{Golden-Marx2019} showed that the majority of their bent radio sources are offset from the cluster center, we refine that measurement by focusing only on the cluster candidates.  As mentioned in Section\,\ref{sect:clustersubsample}, we determine the cluster center using a unit-weighted red sequence surface density measurement, where we treat all red sequence galaxies equally, using either the $i - [3.6]$, $r - i$, or $[3.6] - [4.5]$ analysis.  As seen in Figure\,\ref{Fig:OffsetHistogram}, and in agreement with \citet{Sakelliou2000}, the majority of sources are offset from the cluster center by $<$ 200\,kpc.  This is consistent with the vast majority of our sources being near the cluster center because they are either brightest cluster galaxies (BCGs), other massive central galaxies, infalling galaxies near the cluster center, or outgoing galaxies that have fallen through the cluster center.  To address whether the host galaxy is a BCG, we explore the 3.6\,$\mu$m absolute magnitude of our host galaxies in relation to the other surrounding red sequence galaxies and the orientation of the bent radio lobes relative to our cluster center (see Sections \ref{sect:BCG} and \ref{sect:direction}).  We treat the 3.6\,$\mu$m absolute magnitude as a proxy for stellar mass as IR magnitudes at such wavelengths show no bias to stellar mass estimates \citep{Lemaux2012} and typical, non-quasar, radio detected AGN should have IR bands that are generally uncontaminated by the AGN).

\begin{figure}
\centering
\epsscale{1}
\includegraphics[scale=0.5,trim={0.45in 0.1in 0.0in 0.4in},clip=True]{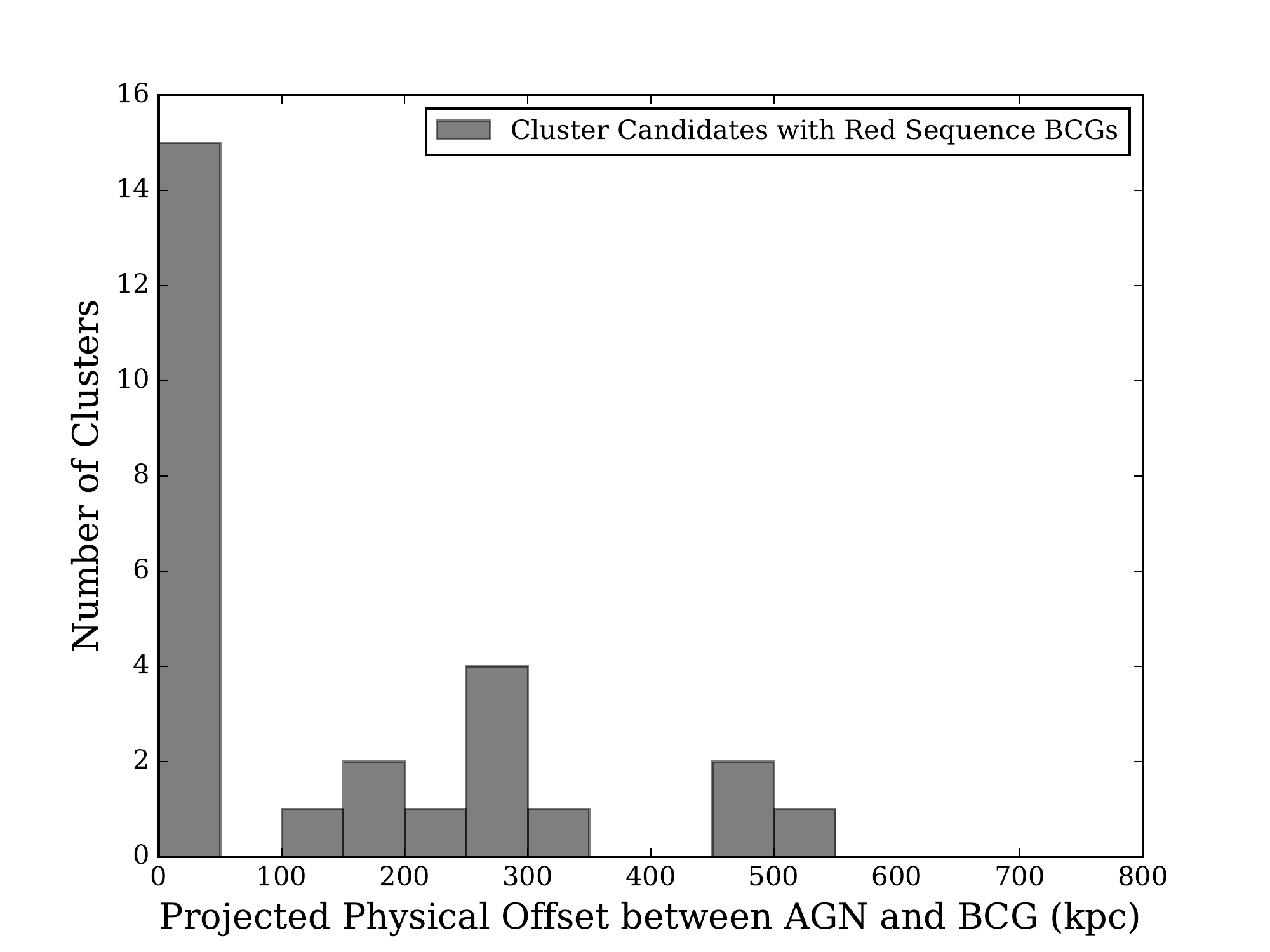}
\includegraphics[scale=0.5,trim={0.45in 0.1in 0.0in 0.4in},clip=True]{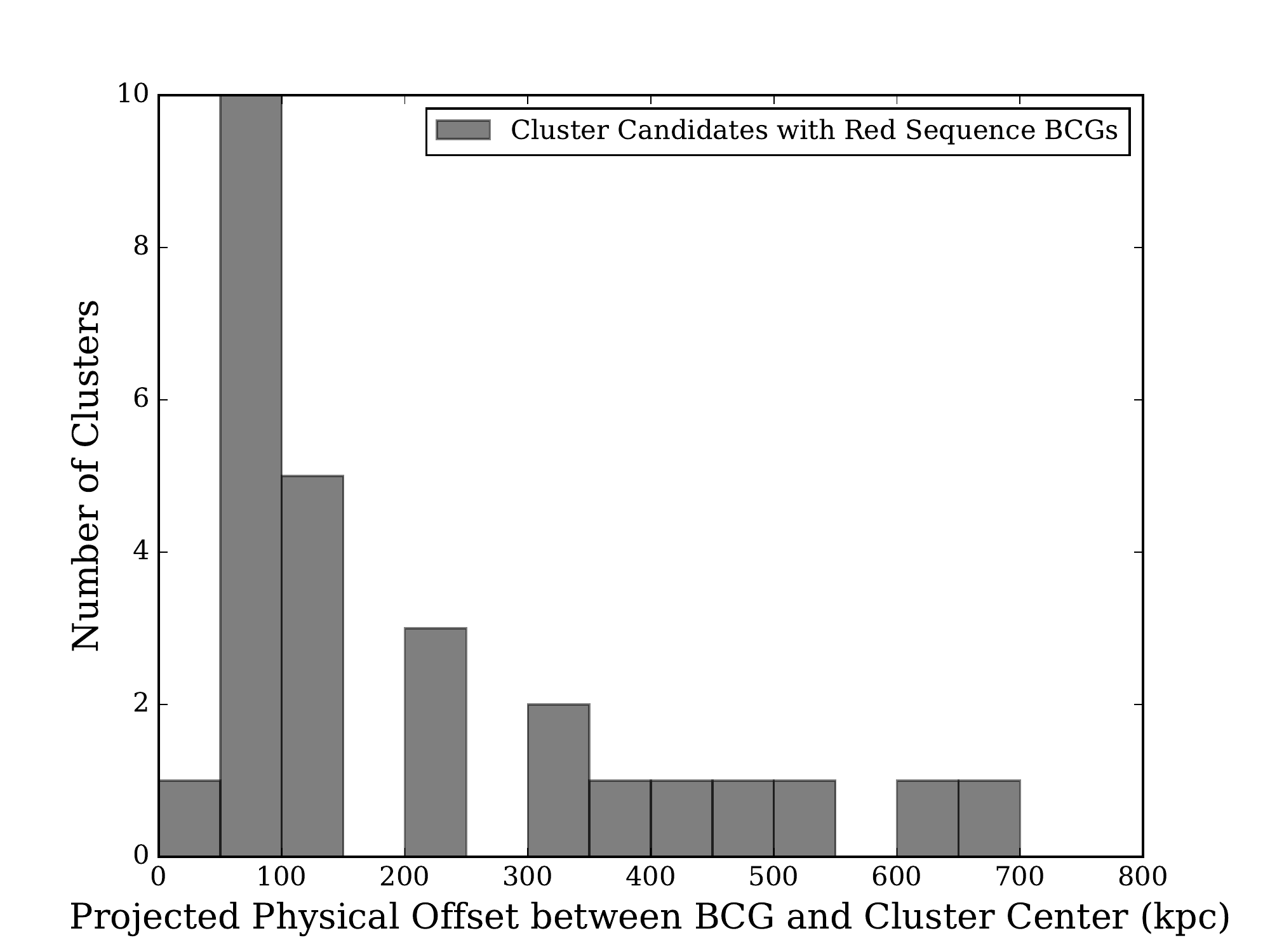}

\caption{Histograms showing the distribution of the offset between the red sequence BCG and the AGN or the new cluster center.  The top panel shows the distribution of offsets between the AGN and the BCG, while the bottom panel shows the distribution of offsets between the BCG and the red sequence cluster center.  In each case, we remove those sources without a red sequence selected host galaxy (primarily the quasars).  In the top histogram, we find that 55$\%$ of our radio AGN are BCGs.  In the bottom histogram, we find general agreement between the location of the BCG and the red sequence cluster center, implying that these different estimates of the cluster center yield similar results. \label{Fig:BCG-AGN}}
\end{figure}

However, the red sequence surface density measurement used to determine the cluster center does not weight the brightness/mass of a red sequence galaxy, nor does it account for quasar host galaxies, which are typically bluer than the red sequence.  If we instead treat the brightest red sequence galaxy as the BCG  and thus as the cluster center, we find that 15 of the 27 cluster candidates where the host is a red sequence galaxy have a bent AGN residing within the BCG (see Section\,\ref{sect:BCG} for a complete discussion of how the BCGs are selected using our red sequence analysis and the 3.6$\mu$m observations).  To determine the difference between the two measurements of the cluster center, we compare the offset of the AGN to the BCG and to the new cluster center.  As seen in the top panel of Figure\,\ref{Fig:BCG-AGN}, most sources remain within 300\,kpc, with the maximum offset between the AGN and BCG being $\approx$ 530\,kpc.  As compared to the values from \citet{Sakelliou2000}, this is well within the expected range of bent radio source offsets.  We also compare the location of the BCG to our red sequence cluster centers.  We find strong agreement between the two locations, with 16 of the 27 sources being offset by less than 150\,kpc.  For these sources, our red sequence center should not impact our analysis, although the sources with larger offsets are obviously more impacted, especially in regards to the infall angle (see Section\,\ref{sect:direction}).  Additionally, while misidentification of cluster members is more likely for fainter galaxies, it is possible that some of our red sequence BCGs, especially those not hosting a bent AGN, are not true cluster members.  That we see similar values between the location of the BCG and the red sequence cluster center reinforces the use of the red sequence surface density, which should be less impacted by the misidentification of a single galaxy.  

\section{Comparing AGN properties to Cluster Population Properties}\label{sect:AGNcluster}
In this section, we compare the properties of COBRA bent, double-lobed radio sources to properties of their host clusters.  Our goal is to elucidate the details of the evolutionary history of COBRA clusters.  We begin by determining the relationship between the opening angle of the bent radio source and cluster richness (Section\,\ref{sect:Richness}).  We then determine the BCG fraction (Section\,\ref{sect:BCG}) and examine the relationship between the AGN's offset from the cluster center and the opening angle of the bent radio source (Section\,\ref{sect:BendingOffset}). Furthermore, we measure the orientation of each bent radio AGN relative to the cluster center (Section\,\ref{sect:direction}).  Lastly, we examine any correlation between the power of each radio source and the richness of the surrounding cluster (Section\,\ref{sect:PowerRichnes}).

\subsection{Cluster Richness}\label{sect:Richness}
As bent radio AGNs are found in a wide variety of clusters and groups, we aim to determine how the cluster environment impacts the bent radio AGN morphology, specifically in regards to the opening angle.  As both \citet{Hardcastle2005} and \citet{Morsony2013} showed, the morphology of bent, double-lobed radio sources depends on the density of the ICM, as well as a number of other factors relating to both the cluster and the AGN.  For our analysis, we quantify the cluster environment using the red sequence overdensities and combined overdensities as proxies for richness as reported in \citet{Golden-Marx2019}.  We use the significance and not the overall number of galaxies because we have clusters at redshifts between 0.35 $<$ $z$ $<$ 2.2 and we expect that the red sequence population will grow over time.  Moreover, there are differences in the background contamination levels of the different color cuts and completeness limits of our four subsamples (m*+1 $i - [3.6]$, m*+1 $r - i$, magnitude-limited $i - [3.6]$, and magnitude-limited $[3.6] - [4.5]$; see Section\,\ref{sect:clustersubsample}).  We use the significance measured when centered on the cluster centers identified in \citet{Golden-Marx2019} to compare each field.  

\begin{figure}
\centering
\epsscale{1}
\includegraphics[scale=0.5,trim={0.45in 0.1in 0.0in 0.4in},clip=True]{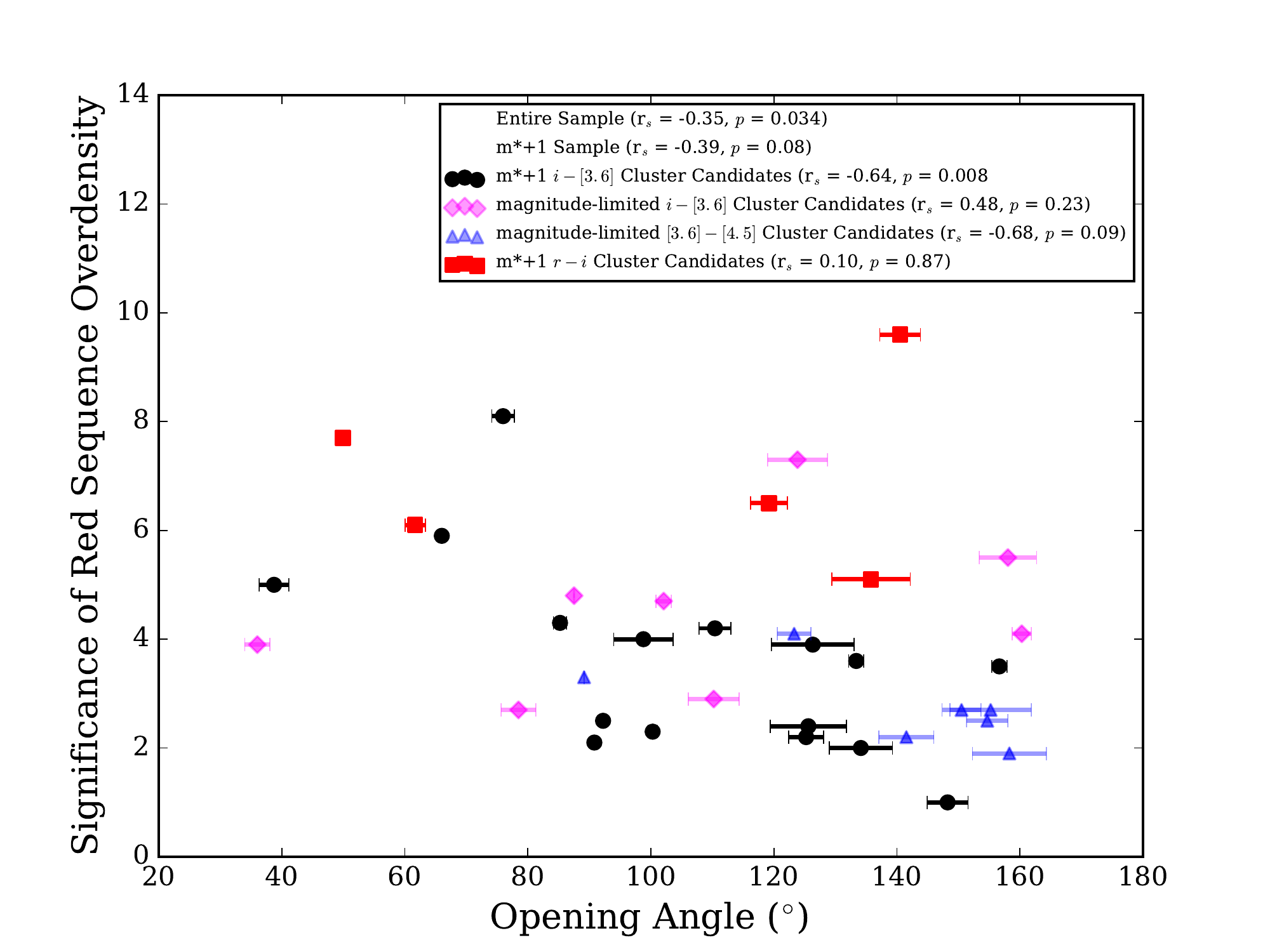}
\includegraphics[scale=0.5,trim={0.45in 0.1in 0.0in 0.4in},clip=True]{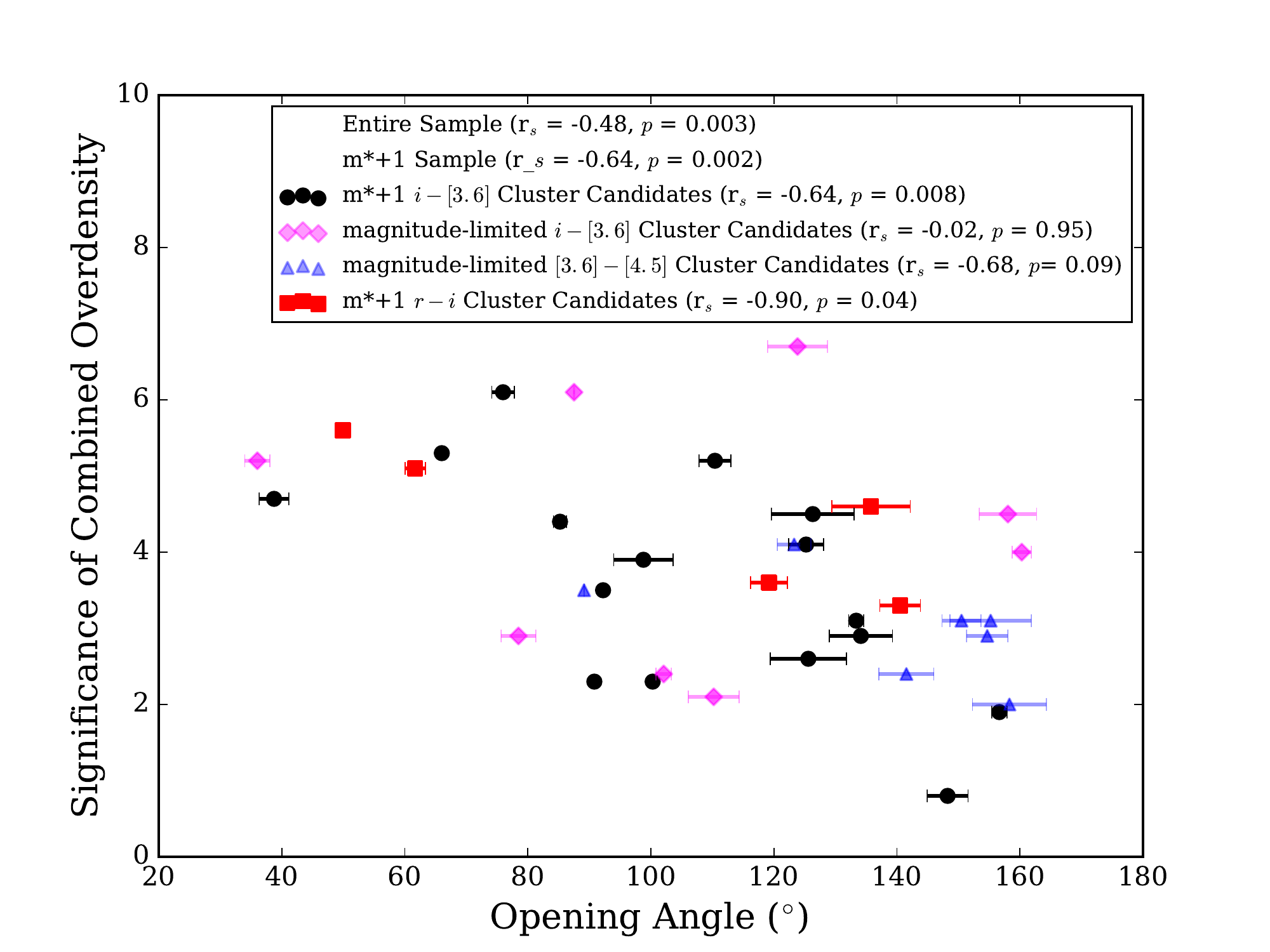}

\caption{The significance of the red sequence overdensity (top) and combined overdensity (bottom) as a function of the opening angle of the bent radio source.  The m*+1 $i - [3.6]$ sample is indicated by black circles, the magnitude-limited $i - [3.6]$ sample is indicated by pink diamonds, the magnitude-limited $[3.6] - [4.5]$ sample is indicated by blue triangles, and the m*+1 $r - i$ sample is indicated by red squares.  We bold the two m*+1 samples because they are statistically similar samples based on the magnitude of a modeled m* galaxy with EzGal.  All significances reported are from \citet{Golden-Marx2019}.  We find a weak to moderate correlation between the narrowness of the opening angle and both overdensity measurements, which is reinforced when examining the m*+1 $i - [3.6]$ subsample. \label{Fig:SignificanceOpeningAngle}}
\end{figure}

Using the sample of 36 red sequence selected COBRA clusters, we find a weak linear anti-correlation between the significance of the red sequence overdensity and the narrowness of the opening angle of the bent radio source hosted by the given cluster; narrower bent sources generally reside in richer cluster candidates (top panel of Figure\,\ref{Fig:SignificanceOpeningAngle}).  We confirm this anti-correlation using the Spearman test (r$_{s}$ = $-$0.35) and find suggestive evidence for rejecting the null hypothesis ($p$ = 0.034).  If we focus on the two m*+1 subsamples, both of which are at $z$ $<$ 1.1 and use a uniform magnitude limit, we measure a slightly stronger anti-correlation (r$_{s}$ = $-$0.39) and slightly weaker evidence for rejecting the null hypothesis ($p$ = 0.08).  Ultimately, both the entire sample and the m*+1 subsample show similar degrees of certainty as to the likelihood of this trend.

Since the overdensity significance is measured differently in each subsample with respect to completeness and galaxy type, we examine the individual subsets.  Because the m*+1 $i - [3.6]$ sample is the primary focus of our analysis and the most statistically robust sample, that we detect evidence of a moderate anti-correlation (r$_{s}$ = $-$0.64) and convincing evidence against a null hypothesis ($p$ = 0.008) strengthens the likelihood that this correlation is real.  The clusters in this sample are statistically similar as they are limited to galaxies brighter than m*+1, and thus the same relative absolute magnitude, as opposed to the magnitude-limited sample.  If we assume the significance of the overdensity of red sequence members is indicative of the overall cluster mass \citep[e.g.,][]{Rykoff2014,Gonzalez2019}, then our result implies more massive clusters have narrower bent AGNs.  As more massive clusters will, all else being equal, have a denser ICM and ICM density directly impacts the opening angle, it follows that massive clusters should host narrower bent sources.  

We further verify the anti-correlation between our measure of cluster richness and the opening angle of the bent AGN by comparing the significance of the combined overdensity to the opening angle (bottom panel of Figure\,\ref{Fig:SignificanceOpeningAngle}).  As the combined overdensity weights the likelihood each red sequence galaxy is at the redshift we estimate via EzGal, as well as the likelihood that the galaxies that are redder and bluer than our red sequence color range are at our target redshift using detailed photometric redshift estimates from the ORELSE survey (see \citealp{Golden-Marx2019} for a complete description of the combined overdensity and \citealp{Lubin2009} for the ORELSE survey), seeing a stronger relation confirms that the opening angle is tied to cluster richness (r$_{s}$ = $-$0.48 and $p$ = 0.003 for the entire sample, while r$_{s}$ = $-$0.64 and $p$ = 0.002 for the m*+1 subsample).  

These correlations between cluster richness and opening angle point towards the cluster environment being a strong indicator of the opening angle.  We also find no evidence of a relation between the absolute magnitude of the host galaxy and the opening angle, which further points to the cluster environment being the primary driver of the bent appearance of each radio source. 

However, the opening angle of a bent radio source depends on the ICM density and velocity of the host galaxy relative to the ICM \citep[e.g.,][]{Hardcastle2005,Morsony2013}.  Because massive clusters have galaxies with a greater velocity dispersion at large offsets from the cluster center, it is possible that the anti-correlation between cluster richness and opening angle may be linked to the velocity of the host galaxy, especially for any radio sources that are farther from the cluster center.  To determine if the distance from the cluster center is the driving force behind the relationship we measure between opening angle and overdensity significance for both red sequence and combined overdensity, we restrict ourselves to clusters where the bent AGN is within 300\,kpc of the cluster center.  We find a similar relationship between red sequence overdensity and opening angle for the clusters where the AGN is closer to the cluster center.  Using the Spearman test, we measure a similar degree of certainty for the entire sample and this subsample (r$_{s}$ = $-$0.39 for both the complete sample and the subset of bent AGNs near the cluster center, although the $p$ value increases from 0.034 for the full sample to 0.12 for the subsample).  However, within the m*+1 $i - [3.6]$ sample, we see stronger evidence of this anti-correlation for the clusters with closer AGN (r$_{s}$ = -0.80 and $p$ = 0.003), making this anti-correlation very likely.  Similarly, when we compare the combined overdensity to the opening angle for the subsample of clusters where the AGN is near the cluster center, we find agreement with our previous result (r$_{s}$ = -0.55 and $p$ = 0.004 for the entire subsample and r$_{s}$ = -0.70 and $p$ = 0.016 for the m*+1 $i - [3.6]$ subsample).  That these relationships mirror one another for the complete sample and the subsample of clusters where the AGN is closer to the cluster center indicates that AGN offset is not a latent parameter in the correlation between opening angle and cluster richness and that the bent nature may be more linked to the overall cluster richness, or ICM density if these two parameters scale.   

Additionally, there is a greater possibility that these sources at larger offsets are not associated with the clusters with which we identify them.  \citet{Golden-Marx2019} note that using the color cuts, especially the $i - [3.6]$ color cut, and red sequence range ($\pm$ 0.15\,mag), there is a very low level of background contamination of similarly colored galaxies.  \citet{Golden-Marx2019} found that $\approx$ 5 - 10 $\%$ of their 1$\arcmin$ background regions would be detected as cluster candidates, depending on the color (see Section 6.3 in \citealp{Golden-Marx2019} for the complete description of background contamination).  These random regions had a separation of $\approx$ 15$\arcmin$ on average, making random associations with bent sources unlikely.  However, as noted by the analysis using the ORELSE data for the combined overdensity, some fraction of redder and bluer galaxies at a given redshift are cluster members, meaning it is possible that we could be misidentifying the cluster center or the cluster entirely.

\begin{figure}
\centering
\epsscale{1}
\includegraphics[scale=0.5,trim={0.45in 0.1in 0.0in 0.4in},clip=True]{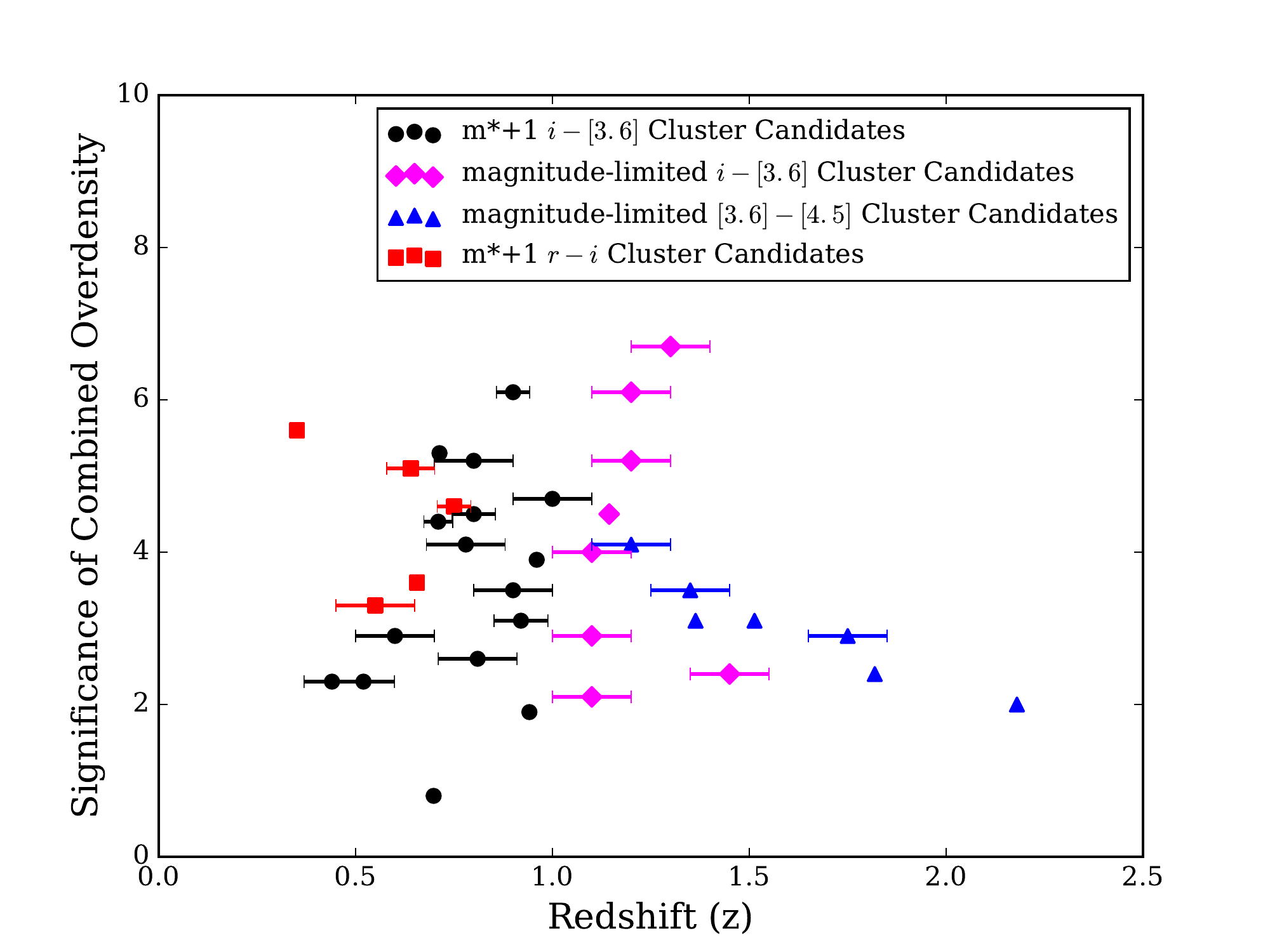}

\caption{The significance of the combined overdensity as a function of redshift.  The same legend is used as in Figure\,\ref{Fig:SignificanceOpeningAngle}.  As before, all significances reported are from \citet{Golden-Marx2019}.  Excluding the magnitude-limited $[3.6] - [4.5]$ cluster subset, there is no correlation between the strength of our overdensity measurements and redshift.  \label{Fig:SignificanceRedshift}}
\end{figure}

To further strengthen the result that narrower bent sources are in richer clusters, we plot the significance of cluster candidates as a function of redshift (Figure\,\ref{Fig:SignificanceRedshift}) to show that this correlation does not come from redshift differences in our sample since the significance of the combined overdensity is not a function of redshift.  Specifically, the two m*+1 samples show no correlations with redshift, meaning that we are not biasing these measurements towards the lowest redshift sources in our sample.  This highlights that the detected correlation between richness and opening angle is not an artifact of our cluster detection, but representative of a real effect and further strengthens that the opening angle of the bent source is directly correlated with the surrounding cluster environment.

\subsubsection{Measurements of the BCG Fraction and a Comparison between the Host Galaxies and Cluster}\label{sect:BCG}
To further understand the relationship between bent, double-lobed radio sources and the cluster environment, we aim to determine if the host galaxies of these radio sources are BCGs.  Because we are examining clusters at a range of redshifts (0.35 $<$ $z$ $<$ 2.2), and BCGs continue to grow hierarchically via mergers with the surrounding galaxies throughout this epoch \citep[e.g.,][]{DeLucia2007,Lidman2012,Lin2013,Ascaso2014,Burke2015,Zhang2016}, we do not expect to find a dominant BCG in each cluster.  Rather, we expect some clusters to have multiple massive galaxies that may evolve into the singular dominant low-$z$ BCG through galaxy-galaxy mergers \citep[e.g.,][]{Lidman2012,Ascaso2014}.  

To determine what fraction of our host galaxies are BCGs, we compare the host galaxy for each COBRA bent radio source, identified in \citet{Paterno-Mahler2017} using the $Spitzer$ 3.6\,$\mu$m observations, to the surrounding red sequence galaxies.  Because the red sequence is our best tool for identifying potential cluster members, we focus only on the red sequence galaxies within either the 1$\arcmin$ ($\approx$ 480\,kpc at $z$ = 1) region centered on the red sequence cluster center or the 1$\arcmin$ region centered on the bent radio AGN.  We examine both regions because in a few cases the center of the red sequence surface density is offset from a BCG that appears when centered on the AGN, and this BCG may be more representative of the cluster than a fainter bright galaxy.  Based on the host galaxies, our red sequence color cut, and our redshift estimates, we find that 27 of the host galaxies lie along the red sequence.  The remaining 9 host galaxies are either SDSS-identified quasars, have colors that differ from their redshift estimate because either the redshift estimate is from SDSS or from the color of the surrounding galaxies, or lack an identified host galaxy.  Although the SDSS-identified quasar may have host galaxies that are the BCG, the photometry is dominated by the emission of the AGN \citep{Stern2012}, making these galaxies bluer than our expected red sequence members and thus the host magnitude a poor proxy for stellar mass.    

Of this subsample of 27 red sequence galaxy cluster candidates, 15 host galaxies are BCGs ($\approx$ 55.5$\%$).  This is in line with many of our host galaxies being located near the cluster center.  Since all host galaxies are among the three brightest galaxies we detect, we follow \citet{Ascaso2014}, \citet{Shen2017}, and \citet{Shen2019} in defining all of our host galaxies as BCG candidates. 

\begin{figure}
\centering
\epsscale{1}
\includegraphics[scale=0.5,trim={0.1in 0.1in 0.0in 0.2in},clip=True]{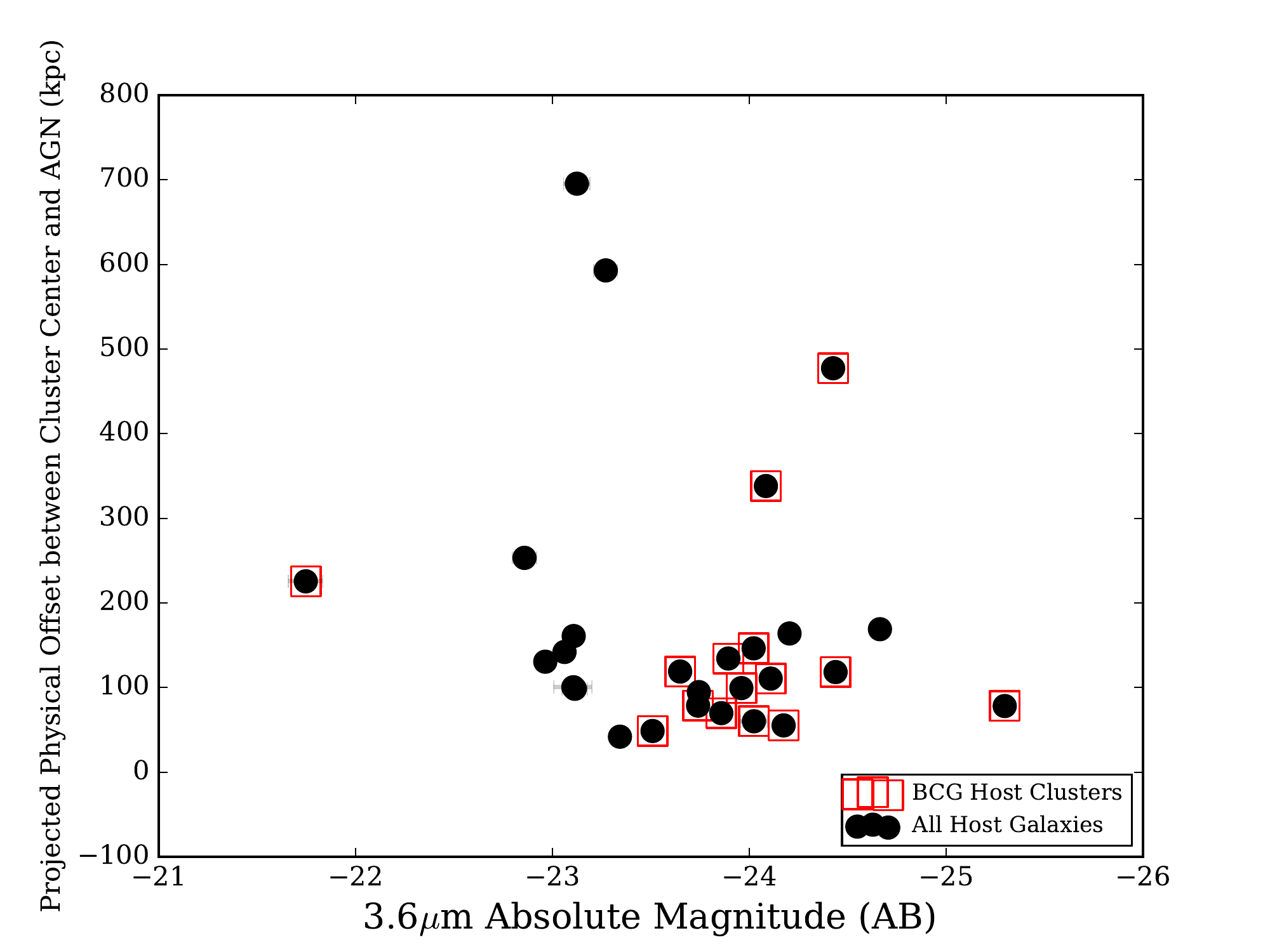}
\caption{The projected physical offset between the host galaxy of the bent radio AGNs and the cluster centers (kpc) as a function of the host galaxy's 3.6\,$\mu$m absolute magnitude.  Host galaxies that are BCGs are identified by a red box.  We see a dichotomy, where most of the galaxies brighter than $-$23.5\,mag are BCGs.  We note that two fainter host galaxies at large offsets that are not BCGs and thus may be examples of fainter infalling/outgoing galaxies where the faster host galaxy velocity is responsible for bending the radio lobes.\label{Fig:absmagredshiftoffset}}
\end{figure}

To further estimate if these host galaxies are BCGs, we measure their 3.6\,$\mu$m absolute magnitude as a function of the offset between the cluster center and the AGN.  To have a uniform proxy for stellar mass for the entire sample, we k-correct each host galaxy's 3.6\,$\mu$m apparent magnitude via EzGal using identical parameters to those reported in \citet{Golden-Marx2019}.  We note that the host galaxy in COBRA164951.6+310818 is an outlier (see Figure\,\ref{Fig:absmagredshiftoffset}).  This galaxy is the faintest host galaxy by far (M$_{3.6}$ = $-$21.75\,mag) and is also a BCG.  We verified that this is indeed the only apparent host galaxy for this bent radio source.  Because this source is located in one of the weakest galaxy cluster candidates, and given its lower redshift nature ($z$ $\approx$ 0.52), is likely a poor galaxy group.

In showing the projected physical offset between the AGN and the cluster center as a function of the host galaxy's 3.6\,$\mu$m absolute magnitude in Figure\,\ref{Fig:absmagredshiftoffset}, we reiterate that most of our host galaxies are near the cluster center.  However, we also show that although almost all of our host galaxies are luminous, all but one of the galaxies fainter than $-$23.5 mags are not BCGs, which creates a clear dichotomy between the absolute magnitudes of typical BCG host galaxies and the other bright galaxies that host bent radio AGN.  As expected for non-BCGs, we see that some of these galaxies are the most offset from the cluster center ($>$ 400\,kpc).  In Figure\,\ref{Fig:absmagredshiftoffset}, we see that two of the host galaxies at the largest offsets have M$_{3.6}$ $\approx$ $-$23.00\,mag, placing them among the fainter host galaxies (though still brighter than an M* galaxy) and in the population of non-BCG galaxies.  Although we are limited by a small sample size, this offset and fainter host magnitude could indicate that these host galaxies may be infalling/outgoing cluster galaxies as opposed to dominant cluster BCGs.  Given the possibility of infalling/outgoing galaxies near the cluster center this could also be true for the other fainter host galaxies that are not BCGs.  However, there are also two BCGs at offsets between 300\,kpc and 500\,kpc.  These sources might represent offset radio BCGs like those found in \citet{Moravec2020} or be part of merging galaxy clusters, where our cluster center estimate may not accurately map the entire distribution.   

\begin{figure}
\centering
\epsscale{1}
\includegraphics[scale=0.5,trim={0.45in 0.1in 0.0in 0.4in},clip=True]{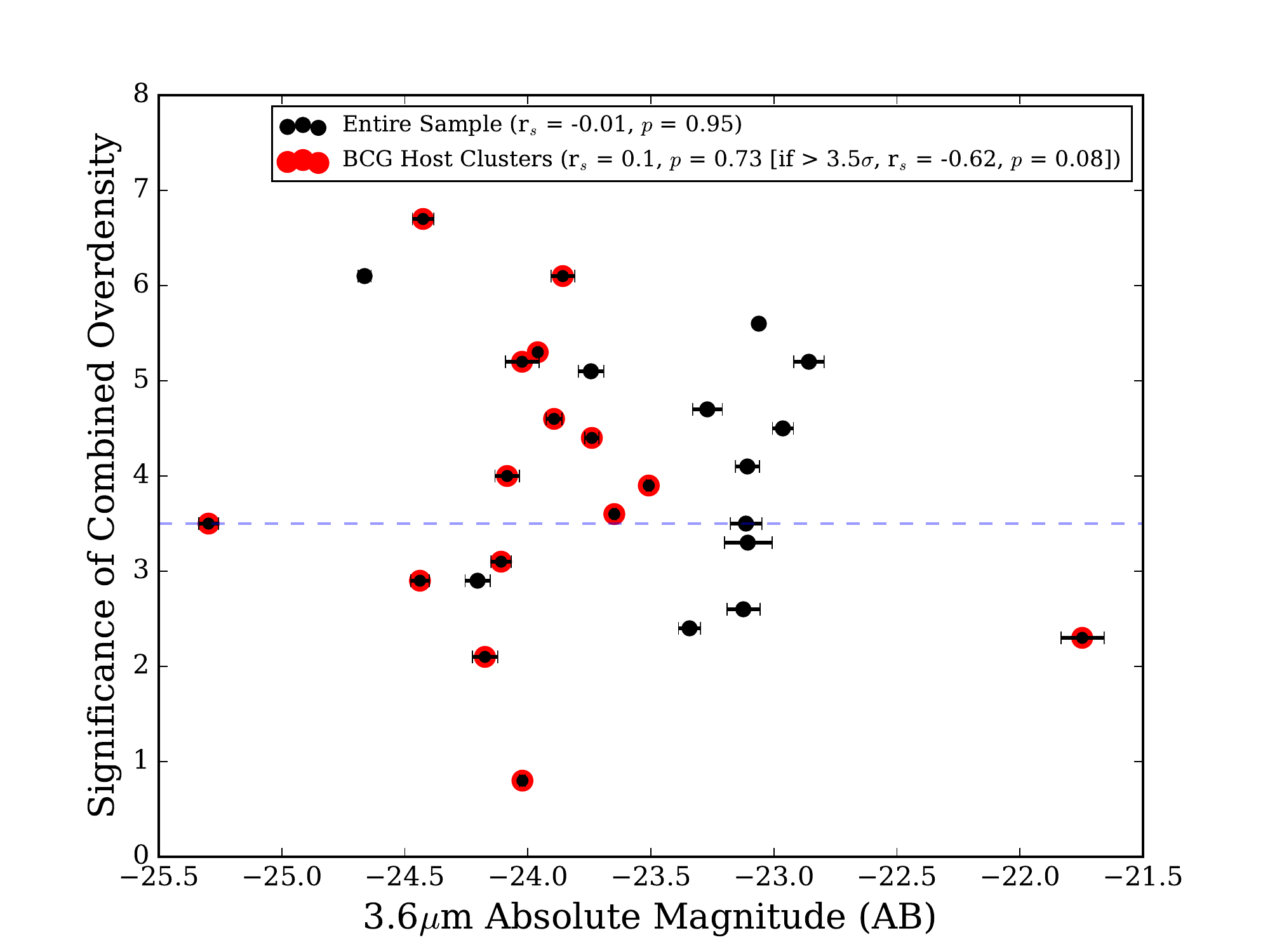}
\caption{The significance of the combined overdensity of each cluster candidate as a function of the host galaxy's 3.6\,$\mu$m absolute magnitude (AB).  All of the overdensities are measurements presented in \citet{Golden-Marx2019}.  We see no correlations between the 3.6\,$\mu$m absolute magnitude of the host galaxy and either measurement of richness for the entire sample.  When we divide out the bent AGNs hosted by BCGs, we measure a strong anti-correlation between the combined overdensity significance and host absolute magnitude for the richest clusters ($>$ 3.5$\sigma$ overdensity).   \label{Fig:absmagcombinedover}}
\end{figure}

Although our small sample size limits what evolutionary effects can be traced by comparing the absolute magnitude of the BCG to the overall cluster richness, we note that in the richest COBRA clusters ($>$ 3.5$\sigma$) where the host galaxy is a BCG, we see a strong correlation (see Figure\,\ref{Fig:absmagcombinedover}; r$_{s}$ = $-$0.62 and $p$ = $-$0.08).  As developed clusters show a well-studied relationship between the luminosity of the dominant BCG and cluster mass, this might hint that our richest clusters are also some of the more evolved systems in our sample \citep[e.g.,][]{Lin2004,Wechsler2018}.

\subsection{The Relationship Between the Opening Angle and the Cluster Offset}\label{sect:BendingOffset}
\begin{figure}
\centering
\epsscale{1}
\includegraphics[scale=0.48,trim={0.35in 0.1in 0.0in 0.4in},clip=True]{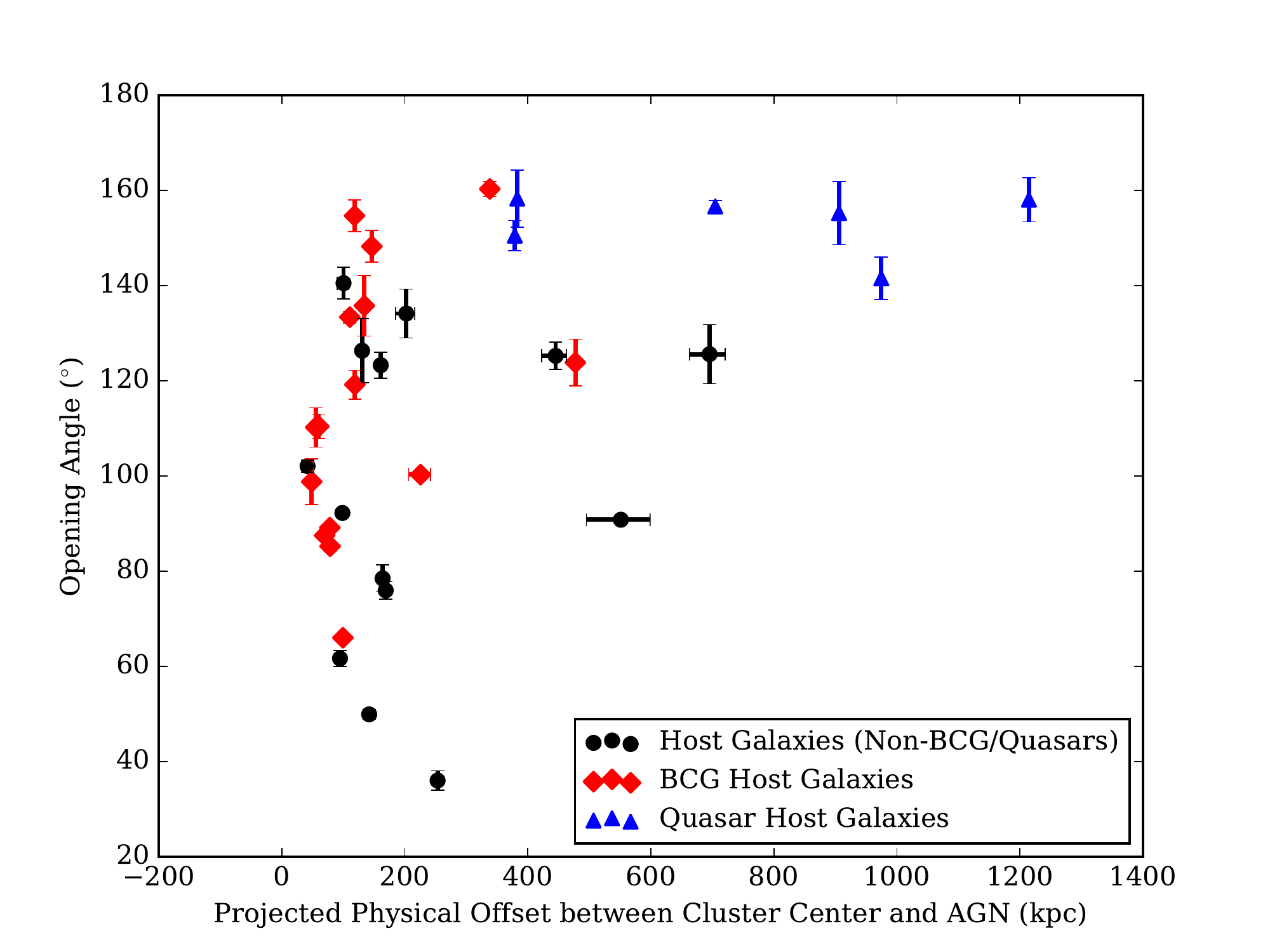}

\caption{The opening angle of each bent AGN as a function of the projected physical offset between the cluster center and the AGN host galaxy.  We separate the host galaxies into three different types, quasars (blue triangles), BCGs (red diamonds; see Section\,\ref{sect:BCG} for a description on how we determine if a host galaxy is a BCG), and non-BCG/quasar host galaxies (black circles).  Both the BCG and non-BCG/quasar host galaxy populations follow similar distributions, indicating that all COBRA radio host galaxies may be proto-BCG candidates. \citet{Golden-Marx2019}.\label{Fig:OffsetOpeningAngle}}
\end{figure}

As shown in Section\,\ref{sect:RSoffset}, bent, double-lobed radio sources are found both near to and offset from the cluster center.  Additionally, Figure\,\ref{Fig:absmagredshiftoffset} highlights that some of our faint host galaxies are at the largest offsets from the cluster center.  Here, we aim to determine if the distributions of offsets from the red sequence cluster center or the opening angle of the bent AGNs differ among our BCG and non-BCG galaxy populations.  As we see in Figure\,\ref{Fig:OffsetHistogram}, the majority of the 36 bent radio sources in our sample have offsets from the cluster center less than 200\,kpc, similar to that of \citet{Sakelliou2000}.  Figure\,\ref{Fig:OffsetOpeningAngle} shows that there is a smaller range of physical offsets for narrower opening angles.  This may be an artifact of the relatively few narrow radio sources in our sample.  However, we see a similar trend between the BCG host galaxies and the non-BCG/quasar host galaxies.  We confirm this using a Kolmogorov-Smirnov (KS) test for each of the two parameters and find that for both the projected physical offset between the AGN and cluster center and the opening angle of the bent radio source, there is no evidence that these samples are drawn from different distributions (if we assume $p$ = 0.1 as the minimum value indicating two quantities are drawn from the same population, we find $p$ = 0.14 for the offsets and $p$ = 0.30 for the opening angles).  This reinforces that our host galaxies are likely from the same population of cluster galaxies, with all of our host galaxies being high-$z$ BCGs or BCG candidates/proto-BCGs.  

Interestingly, all of the quasars have relatively large opening angles and four of the six quasars are offset at distances greater than 600\,kpc.  Similarly, \citet{Mo2018} found that the density of optical, mid-IR, and Type II AGNs peaks near the cluster center, but that the population of Type I AGNs peaks away from the cluster center and approaches field values near the center.  As stated previously, the host galaxies of most bent AGNs can be modeled as typical early-type galaxies, where the AGN's nucleus is obscured (Type II AGNs).  Thus, our results, which show most of the bent AGNs are near the cluster center, agree with \citet{Mo2018}.  Since most quasars are Type I AGNs, the COBRA quasars could be representative of such a population of radio sources typically offset from the cluster center. 

Although we determine the BCG differently from \citet{Garon2019}, who also study a sample of bent sources made up primarily of WATs, they also find that non-BCG galaxies in their 0.02 $<$ $z$ $<$ 0.8 cluster sample tend to host narrower bent sources closer to the cluster center.  Interestingly, we see that for our entire higher redshift sample, narrower sources are generally closer, but closer sources are not necessarily narrow.  We only observe two bent AGNs at $\theta$ $<$ 90$^{\circ}$ and offset more than 200\,kpc.  Given the higher redshift nature of our sample, it is possible that these narrower sources are similar to some of the sources with smaller opening angles measured in the \citet{Garon2019} sample (our lowest redshift source, COBRA164611.2+512915, follows this trend by being a non-BCG host galaxy that is very narrow).   Within our sample, the lack of lower-mass infalling host galaxies with narrow opening angles might be due to either an evolutionary effect or the detection limits of the VLA FIRST survey.   

\subsection{The Direction of Radio Sources Relative to the Cluster Center}\label{sect:direction} 
\begin{figure}
\centering
\epsscale{1}
\includegraphics[scale=0.3,trim={3.5in 0.1in 0.5in 0.in},clip=True]{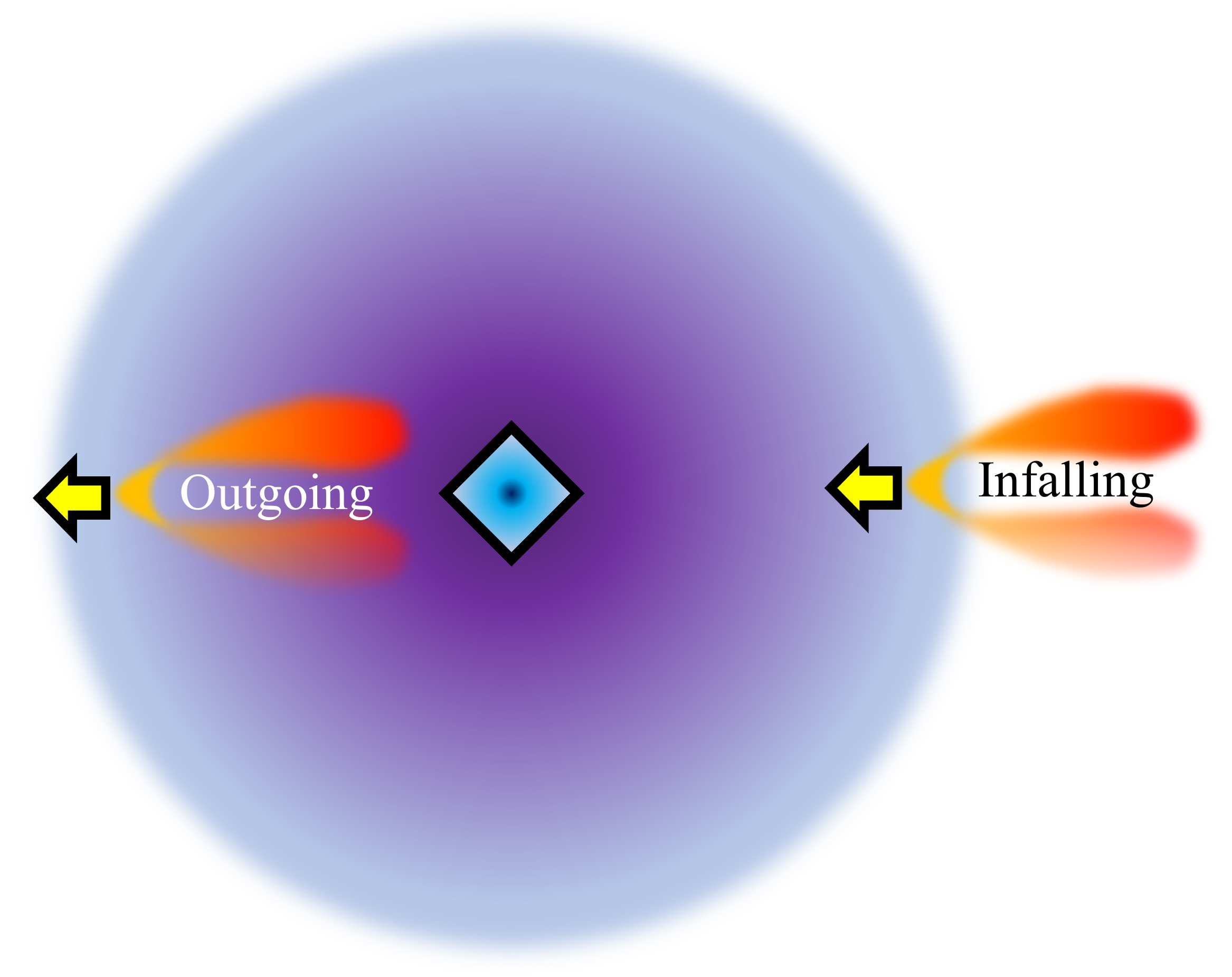}

\caption{Illustration of the difference in orientation between a directly infalling and outgoing radio source.  Here, the cluster is depicted via the large purple circle and the cluster center is shown by the blue triangle.  The infalling and outgoing radio sources, both shown in red/orange, are labeled and have yellow velocity vectors to describe the direction of motion relative to the cluster center.  \label{Fig:RadioSourceDiagram}}
\end{figure}

In order to determine what physical mechanism might be responsible for bending the radio lobes in these COBRA clusters (e.g., an infalling/outgoing host galaxy or a central host galaxy involved in either a major or minor merger; \citealp[e.g.,][]{Sakelliou2000,Douglass2011,Paterno-Mahler2013}), we examine the directional component of the radio source relative to the cluster center.  Although we previously constrain which radio sources are BCGs (see Section\,\ref{sect:BCG}), because we do not treat these BCGs as the cluster center, we treat all bent radio sources as potentially infalling/outgoing galaxies in this analysis.

If the host galaxy of the bent AGN is infalling radially toward the cluster center, the orientation of the opening angle of the bent radio source should point toward the cluster center (see Figure\,\ref{Fig:RadioSourceDiagram}).  If the radio source is outgoing, it will open toward the cluster center.  Additionally, dynamical friction acting on the bent sources as they pass through the cluster center should result in outgoing galaxies being closer to the cluster center than the most offset infalling galaxies. 

\begin{figure}
\centering
\epsscale{1}
\includegraphics[scale=0.48,trim={0.3in 0.1in 0.0in 0.4in},clip=True]{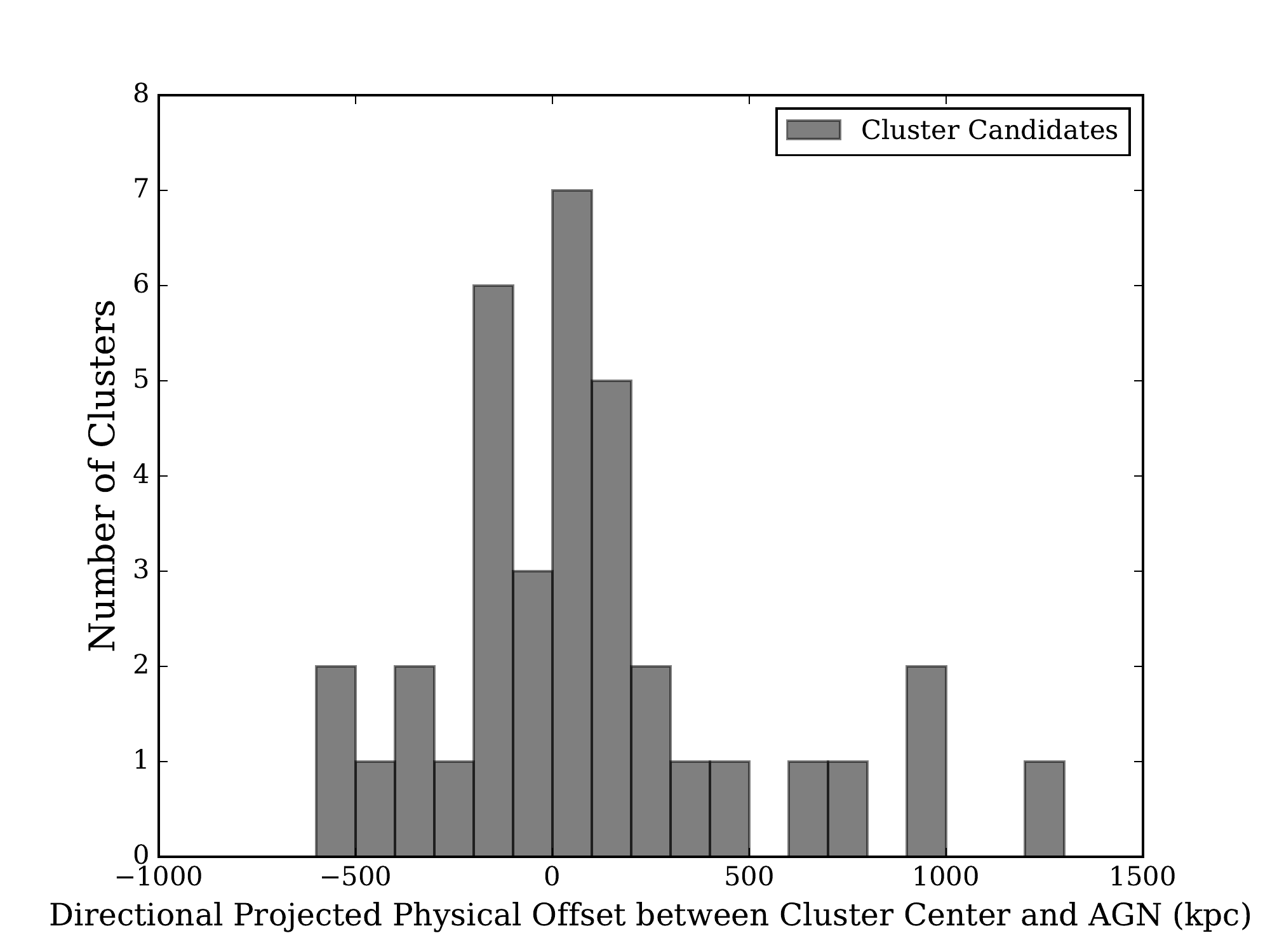}

\caption{Histogram showing the distribution of the directional offset of each bent radio source from the cluster center.  To differentiate radio sources that are infalling and outgoing, we display outgoing radio sources as negative distances and infalling radio sources as positive distances.  The magnitude of these distances are identical to those shown in Figure\,\ref{Fig:OffsetHistogram}, only adjusted for direction.  We find that most radio sources are $\pm$ 200\,kpc from the cluster center, but see that the spread of possible separations is greater for infalling radio sources, as expected due to dynamical friction.  \label{Fig:DirectionOffsetHistogram}}
\end{figure}

After identifying the infalling and outgoing bent sources, we adjust the offsets shown in Figure\,\ref{Fig:OffsetHistogram} to account for this dichotomy. We report outgoing galaxies as being at negative distances and infalling galaxies as being at positive distances.  As seen in Figures\,\ref{Fig:OffsetHistogram} and \ref{Fig:DirectionOffsetHistogram}, the majority of our radio sources are within $\pm$ 200\,kpc of the cluster center.  For sources within $\pm$ 100\,kpc, it is possible that some of the discrepancy between infalling and outgoing sources might result from error in our cluster center measurement since we do not weight the cluster center by galaxy mass or luminosity.  

The major takeaway from Figure\,\ref{Fig:DirectionOffsetHistogram} comes from the spread in the offsets.  Although we see a similar number of infalling and outgoing radio sources (21 to 15), infalling radio sources extend up to $\approx$ 1200\,kpc from the cluster center, while the most offset outgoing source is $\approx$ 600\,kpc from the center, which is expected due to dynamical friction.  However, all three infalling sources at distances greater than 1000\,kpc are quasars, where the possibility of miscentering is higher given the quasars non-red color and our use of the red sequence to identify the cluster center.  Thus, the difference in the offset populations may be an artifact of our analysis, not a physical property.  However, it is alternatively possible that these quasars represent AGNs that are triggered by infall into the cluster as shown by their large offsets.  

For all measurements of the infall angle, projection effects are highly problematic.  We measure the opening angle assuming the radio source is moving on a flat 2D projection of space, when the AGNs could be moving into or out of the plane of the sky at any possible angle.  When the third dimension is unfolded, it is possible to have infalling radio sources that appear to be outgoing from our viewing angle (and vice versa).  Thus, although we determine which side the bent radio source is pointing relative to the cluster center, these measurements have a large associated uncertainty.  Furthermore, it is possible that some of the host galaxies classified as infalling and outgoing galaxies are actually central BCGs where the bent lobes form via the sloshing or large-scale merger motions of the ICM rather than infalling/outgoing cluster galaxies.  As not all bent sources are opening directly away from or towards the cluster center, this is particularly important.    

\begin{figure}
\centering
\epsscale{1}
\includegraphics[scale=0.45,trim={3.in 0.15in 0.15in 0.15in},clip=True]{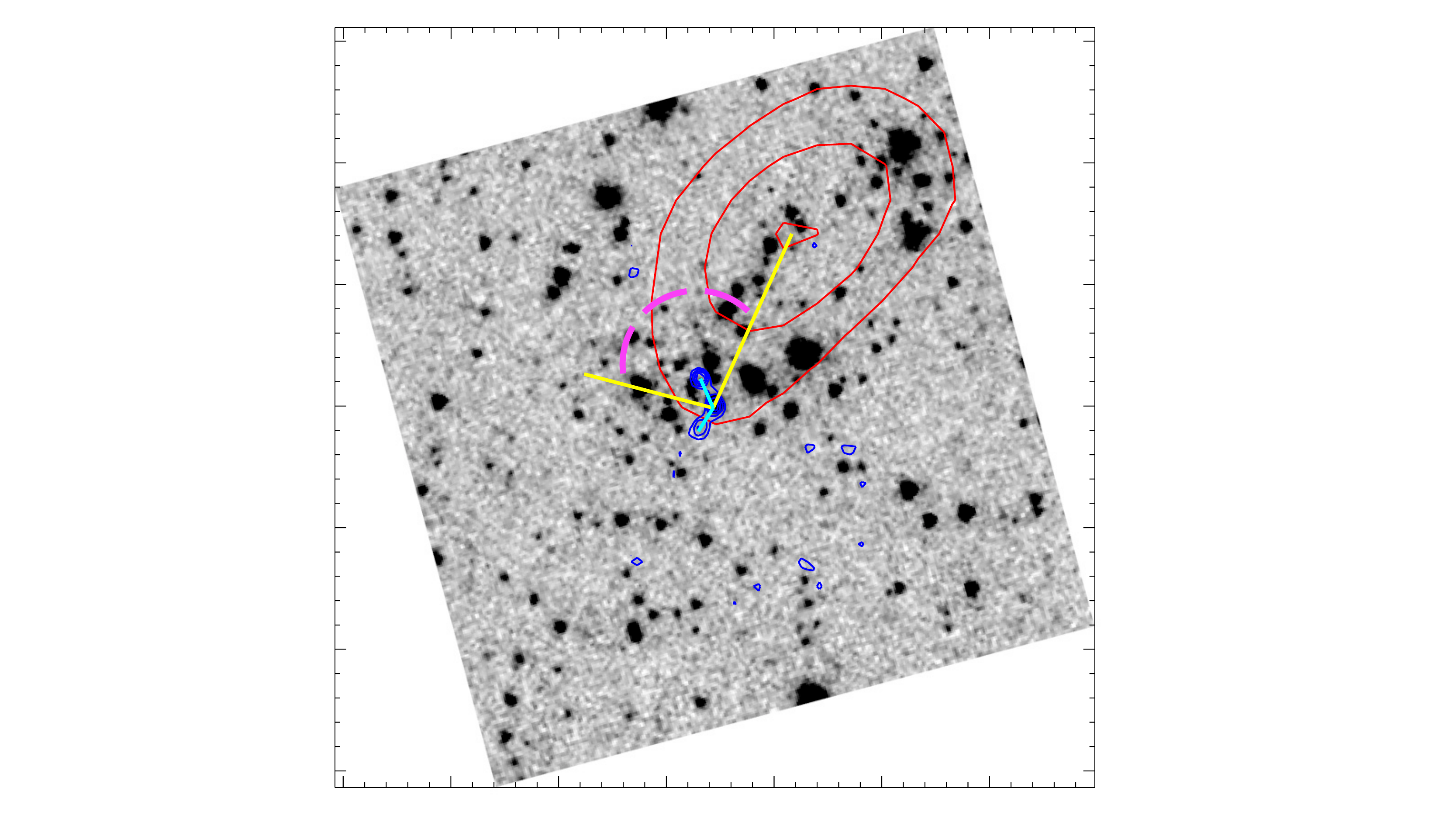}

\caption{Illustration of the infall angle measurement technique, for the case of COBRA074410.9+274011.  The grey-scale image is an $\approx$ 2$\arcmin$ $\times$ 2$\arcmin$ cutout of the 3.6\,$\mu$m IRAC mosaic.  The VLA FIRST radio contours are overlaid in blue, while the surface density of red sequence galaxies is indicated by the red contours.  The cyan lines trace the opening angle of the VLA FIRST radio components.  The yellow lines trace the line connecting the cluster center to the center of the radio source and the center of the radio source about the bisector.  The magenta arc indicates where the infall angle is measured.  Using our initial methodology, this radio source is found to be infalling at a distance of $\approx$ 477.3\,kpc.  However, using our second methodology, we reclassify this source as being an intermediate source given its infall angle of 110.5$^{\circ}$.  \label{Fig:InfallDiagnostic}}
\end{figure}

\begin{figure}
\centering
\epsscale{1}
\includegraphics[scale=0.48,trim={0.4in 0.1in 0.0in 0.4in},clip=True]{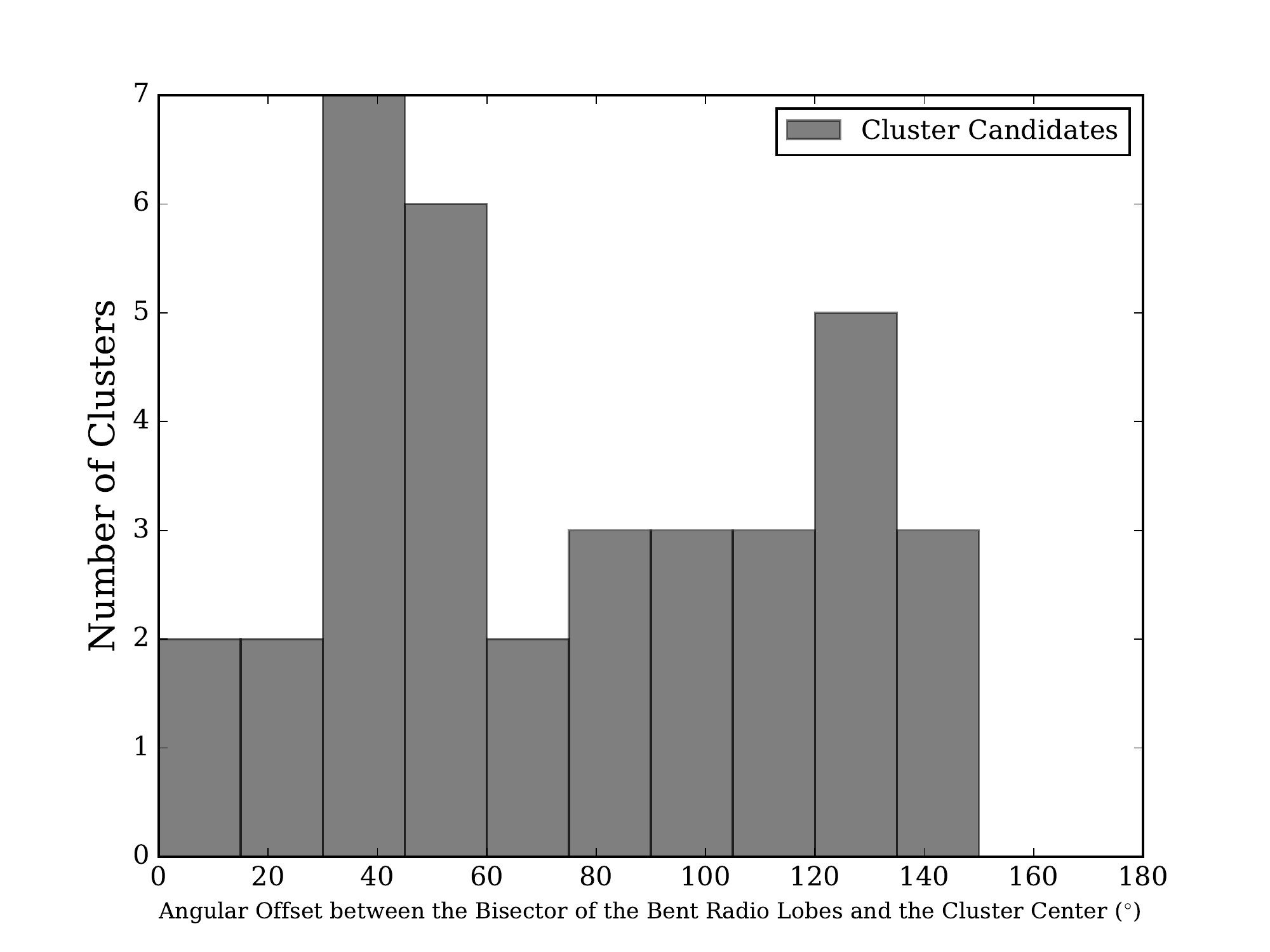}

\caption{Histogram showing the distribution of infall angles of bent radio source relative to the cluster center.  We define the infall angle as the projected angle between the bisector of the bent radio source's opening angle and the line connecting the bent radio source to the cluster center. Angles of 0$^{\circ}$ - 45$^{\circ}$ corresponds to a galaxy that is outgoing, while an angle of 135$^{\circ}$ - 180$^{\circ}$ corresponds to a galaxy that is infalling.  We find few radio sources are directly infalling or outgoing.  Rather, most of our sources are at intermediate angles, though we see a large population of sources that are outgoing.  \label{Fig:AngularOffsetHistogram}}
\end{figure}

Although bent sources can be infalling or outgoing, they do not always follow directly radially paths.  As our previous measurement assumed this, we explore whether our bent sources are infalling or outgoing by measuring the infall angle relative to the cluster center.  We do this by measuring the angle between the line bisecting our bent radio sources and the line connecting our host galaxies to our red sequence cluster center (this is done in a similar manner to \citealp{Sakelliou2000}; see Figure\,\ref{Fig:InfallDiagnostic} for an example).  Unlike the previous measurement, we do not use a strict dichotomy, but instead, are less restrictive in our designation.  By construction, an angle of 0$^{\circ}$ corresponds to an outgoing galaxy that has fallen through the cluster center along a radial path while an angle of 180$^{\circ}$ corresponds to a perfectly radially infalling bent radio source.  We define bent sources that are outgoing as having angles less than 45$^{\circ}$ and bent sources that are infalling as having angles greater than 135$^{\circ}$.  We find peaks in our sample around 35$^{\circ}$ and 130$^{\circ}$.  Although few of our radio sources follow direct radial paths relative to our red sequence cluster centers, they peak near the infalling and outgoing regime (see Figure\,\ref{Fig:AngularOffsetHistogram}).  While we have a number of radio sources that are infalling/outgoing, we see a large population of radio sources on less radial paths.  It is possible that these sources represent bent sources following circular paths or that these bent AGNs are actually central BCGs, where the bent nature is due to interactions in the ICM (e.g., sloshing spiral clusters as in \citealp{Paterno-Mahler2013} or large-scale cluster mergers as in \citealp{Douglass2011}).  

As with the previous measurements, projection effects impact our measurement of the direction of a galaxy's motion and will impact our measurement of infall angle.  In this case, projection effects may account for the large number of sources at intermediate angles in our measurement and mean that our errors are likely underestimated. 

\subsection{Radio Power and Cluster Richness \label{sect:PowerRichnes}}
\begin{figure}
\centering
\epsscale{1}
\includegraphics[scale=0.5,trim={0.25in 0.1in 0.0in 0.4in},clip=True]{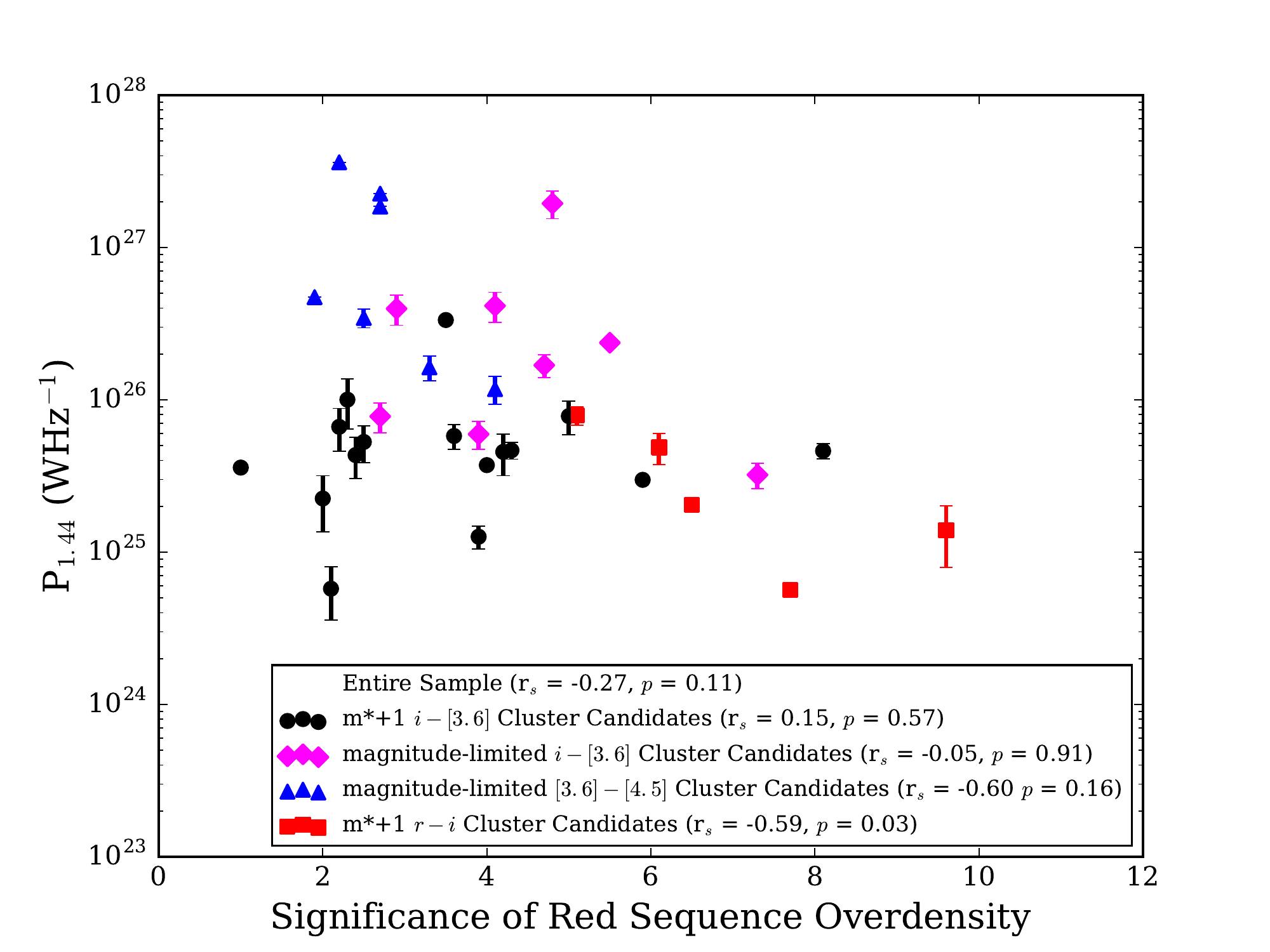}
\includegraphics[scale=0.5,trim={0.25in 0.1in 0.0in 0.4in},clip=True]{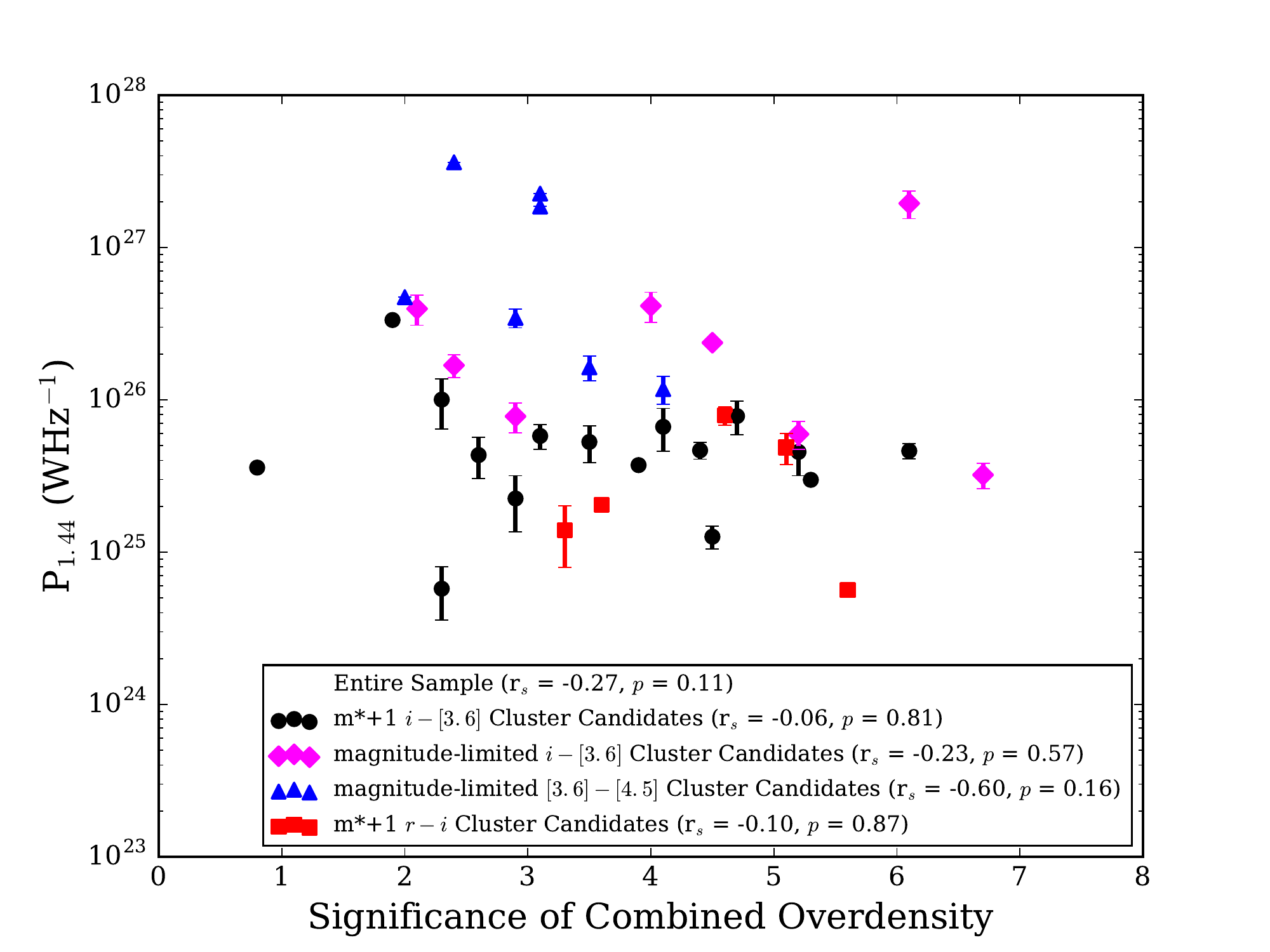}

\caption{The radio power of bent AGNs as a function of the significance of the red sequence overdensity (top) and combined overdensity (bottom).  The same legend as in Figure\,\ref{Fig:SignificanceOpeningAngle} is used.  All significances reported are from \citet{Golden-Marx2019}.  For the entire sample, we see a weak anti-correlation between significance of the overdensity measurement and the radio power.  However, the $p$ value for this trend makes the likelihood that this is a real trend unlikely. Furthermore, the lack of an apparent trend within the m*+1 $i - [3.6]$ subsample might instead point to weak to no evidence of a correlation between these quantities. \label{Fig:SignificanceRadioLuminosity}}
\end{figure}

Our earlier analysis of the anti-correlation between cluster richness and the opening angle of the radio source leads us to ask whether a similar relationship exists between the power of the radio source and the surrounding cluster environment, quantified via richness.  Figure\,\ref{Fig:SignificanceRadioLuminosity} presents our measurement of radio power versus cluster richness, which shows a slight anti-correlation; richer clusters host less powerful radio sources.  However, the Spearman test shows that while there is a weak anti-correlation (r$_{s}$ = $-$0.27), there is little to no evidence to reject the null hypothesis ($p$ = 0.11).  Using the Spearman test on the m*+1 $i - [3.6]$ subsample, we again find a very weak anti-correlation, and no evidence to reject the null hypothesis.  

The lack of any strong correlations is similar to other high-$z$ cluster surveys.  Using the complete CARLA survey (1.3 $<$ $z$ $<$ 3.2), \citet{Wylezalek2013} found no correlation between the radio source power and the density of IRAC detected $[3.6] - [4.5]$ $>$ $-$0.1 sources.  Similarly, \citet{Moravec2020} found only a very weak positive correlation between these two quantities in their sample of MaDCoWS clusters at $z$ $\approx$ 1.0.  Thus, our results generally agree with both \citet{Wylezalek2013} and \citet{Moravec2020} that radio power is not a tracer of cluster richness.

Although we do not see any trends, we do note that at low redshift (0.08 $<$ $z$ $<$ 0.40), \citet{Croston2019} found a strong positive relationship between the radio power of Low Frequency Array (LOFAR) detected radio sources and cluster richness (much stronger than the one presented in \citealp{Moravec2020}), while \citet{Wing2011} saw no correlation in their sample of clusters with bent AGNs at similar redshifts.  Because our clusters are selected via a similar methodology to \citet{Wing2011} and \citet{Croston2019} saw this trend only among their richest clusters, this might indicate that such a relationship only develops within the richest and most massive clusters at low-redshift, which we do not expect our sample to include. 

\section{Discussion}\label{sect:discussion}

As shown in Section\,\ref{sect:AGNcluster}, by correlating the properties of bent, double-lobed radio sources with their host clusters, we gain insight into what bends double-lobed radio sources.  As highlighted in our introduction, although COBRA is unique in its use of bent radio AGNs as cluster finders at high redshift, there are a number of large radio AGN targeted cluster surveys, some of which also include bent radio sources.  To contextualize our high-$z$ bent, double-lobed radio sources with respect to other samples of high-$z$ clusters with radio galaxies and other samples of bent sources, we compare our results to other surveys and explore additional potential correlating properties of radio galaxies and their host clusters.  We first compare our anti-correlation of cluster richness and opening angle to other cluster surveys in Section\,\ref{sect:RichnessBending}.  We then compare the infall angle of COBRA clusters to similar low-$z$ samples of bent sources in Section\,\ref{sect:discussionoffset}.  Lastly, we see how these offsets scale with the radio parameters in Section\,\ref{sect:OffsetRSsizeBending}.   

\subsection{Cluster Richness and Bending Angle}\label{sect:RichnessBending}
As shown in Figure\,\ref{Fig:SignificanceOpeningAngle}, we see an anti-correlation between the strength of the overdensity and the size of the opening angle, with narrower bent sources being found in richer environments.  As mentioned previously, this could indicate that these richer clusters have a more dense ICM, if all else is equal.  Currently, six COBRA clusters, five of which are in this sample, have X-ray observations, either with XMM-Newton, $Chandra$, or Swift (Blanton et al. in prep, Paterno-Mahler et al. in prep) and we are applying for additional Chandra and XMM-Newton observations to characterize the ICM.  With these future observations, we aim to determine how well the optical/IR red sequence overdensities trace correspondingly strong ICMs as well as the dynamical state of that ICM.  

To further validate that the opening angle of a bent AGN is narrower in richer clusters, we measured the opening angles of bent sources at similar redshifts that are not in cluster candidates.  Because there is no large sample of field galaxies hosting bent AGNs at high redshift, we compare the distribution of opening angles for the 36 bent AGNs in our cluster sample to the sample of 38 bent AGNs with optical observations that are not red sequence cluster candidates in \citet{Golden-Marx2019}.  Although this is not a true representation of field bent AGNs because our optical follow-up observations were chosen based on richer 3.6\,$\mu$m overdensity measurements from \citet{Paterno-Mahler2017}, the cluster environment of these sources is measured uniformly to the sample presented here. 

Although we again find a range of opening angles, the distributions between these two samples do not mirror one another.  Within cluster candidates, we find 25.0$\%$ of bent radio sources have opening angles $<$ 90$^{\circ}$, while 7.9$\%$ of bent radio sources in poorer environments have opening angles $<$ 90$^{\circ}$.  We find 41.7$\%$ of bent radio sources in cluster environments have opening angles at 90$^{\circ}$ $<$ $\theta$ $<$ 135$^{\circ}$, while 34.2$\%$ of bent sources in poor environments have opening angles in that range. Among the least bent sources, only 30.5$\%$ of bent sources in clusters have opening angles $>$ 135$^{\circ}$, while 57.9$\%$ of bent radio sources in poor environments are in this range.  The bent AGNs not in red sequence clusters tend to, on average, be less bent, which agrees with \citet{Wing2011}, who found that bent sources of any opening angle are more commonly in low-$z$ cluster than their straight counterparts.  However, despite this apparent agreement, \citet{Wing2011} did not find any correlation between the opening angle of the bent radio source and cluster richness.  This difference may result from our different richness measurements, as \citet{Wing2011} did a single band overdensity based on the number of galaxies brighter than M$_{r}$ = $-$19.0\,mag and did not account for color or redshift. 

Similar to our own measurement, \citet{Garon2019} correlated the opening angle with cluster mass (estimated via the SZ and X-ray calibrated richness measurement from \citealp{Wen2015}) for their sample of lower redshift bent radio AGNs from the Radio Galaxy Zoo.  \citet{Garon2019} found that the local surface density of bent AGNs has a higher peak value surrounding narrower sources than wider bent sources.  This is in agreement with our results, and is further points to narrower bent sources being found in richer environments, possibly due to a denser ICM creating a greater ram pressure.

\subsection{Cluster Offsets and Infall Angle}\label{sect:discussionoffset}
Not all COBRA bent, double-lobed radio sources are at the center of our clusters (see Figure\,\ref{Fig:OffsetHistogram}).  \citet{Sakelliou2000}, with their small sample of low-$z$ bent sources, found similar results in terms of the distributions of the offsets of the bent radio sources from the cluster center using X-ray observations to locate the cluster centroid.  \citet{Sakelliou2000} similarly measured the angle of the bent radio source relative to the cluster center, although they define the angles slightly differently than our measurement in Section\,\ref{sect:direction}.  \citet{Sakelliou2000} found that 11 of their 17 sources are infalling, 4 are outgoing, and 2 are at intermediate angles.  By contrast, we find that 3 of our 36 sources are infalling, 11 are outgoing, and 22 are at intermediate angles.  Although the distributions of infalling and outgoing galaxies differ between our two samples, both show that the farthest outgoing radio sources are closer to the cluster center than the farthest infalling sources.  As discussed in Section\,\ref{sect:direction}, this may result from dynamical friction slowing down infalling galaxies. 

To determine if the differences in distribution are the result of inaccuracies in our cluster center, we also measure the infall angle using the BCG as the cluster center.  Using the purely directional measure of determining if sources are infalling or outgoing relative to the BCG, we find that 6 of the 12 non-BCG host galaxies are infalling and 6 are outgoing relative to that BCG (all are at distances less than 530\,kpc).  If we instead measure the infall angle relative to the BCG, we find that 4 are outgoing (at angles $<$ 45$^{\circ}$), none are infalling, and 8 at intermediate angles.  While it is intriguing that none of our sources are directly infalling, that a third of the sources appear outgoing supports the notion that we are detecting a real population of outgoing cluster galaxies.

One cluster with an outgoing radio source is COBRA135136.2+543955, which was previously identified in \citet{Wen2012}.  The AGN is offset less than 200\,kpc from the cluster center, which does not correspond to the BCG.  If the BCG is the actual cluster center, then COBRA135136.2+543955 still appears to have an outgoing radio source, though at a smaller angle.  Interestingly, COBRA135136.2+543955 is one of the few clusters in our sample with two distinct radio sources at the same redshift, the second of which is associated with the BCG (COBRA151458.0$-$011749 is another example of this), both of which are detected as galaxies with SDSS BOSS spectra.  The identification of systems with multiple radio loud AGN may be indicative of a unique merger history for these clusters \citep[e.g.,][]{Moravec2020b}.

Since our measured distribution of infall angles differs from \citet{Sakelliou2000}, and does not appear to be the result of our estimate of the cluster centers, we examined whether this is the result of weaker cluster detections skewing our results.  Among the weakest cluster candidates, it is possible that our correlation with the radio source is the result of a false detection.  As measured in \citet{Golden-Marx2019}, 2$\sigma$ and greater overdensities appear in $\approx$ 5 - 10\,$\%$ of the random background fields.  If some of the low richness cluster sources are bent sources that randomly align with other galaxies to create the appearance of a cluster, these bent sources could instead be in poorer groups, filament structures, or fossil groups.  These measurements would inaccurately characterize the infalling and outgoing radio source population, which could slightly skew our measurements for all populations. 

To answer this question, we measure the fraction of infalling, outgoing, and intermediate radio sources in clusters with either a red sequence or combined overdensity above 3$\sigma$. Using the red sequence overdensity, we include 22 cluster candidates in our sample and find 1 is infalling, 7 are outgoing, and 14 are at intermediate angles.  These values correspond to very similar fractions of sources as the total sample of all cluster candidates, with all values falling within Poisson error values (4.5$\%$ vs 8.3$\%$ for infalling sources, 31.8$\%$ vs 33.3$\%$ for outgoing sources, and 63.6$\%$ vs 58.3$\%$ for intermediate sources).  Similarly, using the combined overdensity, we include 24 cluster candidates and find 1 is infalling, 8 are outgoing, and 15 are at intermediate angles.  Again, our fractions mirror those of the total sample (4.2$\%$ vs 8.3$\%$ for infalling sources, 33.3$\%$ vs 33.3$\%$ for outgoing sources, and 62.5$\%$ vs 58.3$\%$ for intermediate sources).  That we see a similar distribution for this richer cut implies that our measurements are not biased by our weakest, less well constrained, cluster candidates.

\citet{Sakelliou2000} hypothesized that the lack of a large populations of outgoing radio sources may result from the dense ICM at the center of a cluster that would be shock heated by cluster-cluster mergers, which may also bend the radio lobes.  However, the clusters in the \citet{Sakelliou2000} sample are all at $z$ $<$ 0.24, meaning that the cluster candidates in this COBRA sample are likely, on average, younger systems in terms of their dynamical history and the development of their ICM.  Thus, it is possible that the ICM in our clusters is less dense, which could result in the different populations of infalling, outgoing, and intermediate radio sources in these two samples. If the ICM density is proportional to our richness measurements, we expect that some of our poorest clusters will have outgoing galaxies.  Among the poorest clusters (overdensity less than 3$\sigma$) we find 4 outgoing fields (3 have weaker combined overdensities).  Additionally, we find that 6 of our 11 outgoing radio sources are at $z$ $>$ 1.0.  Because we expect to find less dense ICMs at high redshift, these characteristics support that a less dense ICM may be responsible for some outgoing radio sources.  When combined with our earlier results that the richer sources host a similar number of outgoing radio sources, this may imply that on a whole, our sample contains clusters hosting a less dense ICM.       

The \citet{Sakelliou2000} sample also differs greatly from our sample in terms of the number of intermediate sources.  \citet{Sakelliou2000} find 2 of their 17 sources are at intermediate angles, while we report 22 of our 36 (11.7$\%$ vs 61.1$\%$).  Some of this is likely due to host galaxies being BCGs where the infall angle is inaccurate since the BCG is a better estimate of the cluster center or bent radio sources caused by gas sloshing in minor cluster mergers \citep[e,.g.,][]{Paterno-Mahler2013}.  However, these radio sources could also follow circular orbits about the cluster center or radial orbits where they do not pass through the cluster center.  In these scenarios, the pericenter of the orbits could be relatively far from the cluster center, even for those that are outgoing, preventing the bent source from passing through the most dense regions of the ICM and allowing it to survive regardless of the central ICM density. 

Furthermore, as shown in \citet{Edwards2010}, bent sources can trace cluster filaments as well as clusters.  If any of our poorer clusters (3 red sequence galaxies) are actually elongated cluster filaments, this could cause the bent source to not point toward the center of our red sequence galaxies, but instead another cluster system entirely (one that could be outside our shared LDT - $Spitzer$ IRAC F.O.V.).  Similarly, \citet{Novikov1999} showed that bent radio sources are preferentially oriented along supercluster axes.  Thus, this discrepancy could be explained if we are not identifying the entire cluster structure. 
 
\subsection{Cluster Offset, Radio Source Size, and Bending Angle}\label{sect:OffsetRSsizeBending}
\citet{Moravec2019} presented a relationship between the projected physical offset of an AGN from the cluster center and the size of the radio source and further strengthened this trend using the expanded MaDCoWS sample in \citet{Moravec2020}.  As discussed in \citet{Moravec2019}, the size of the radio source should depend on the power of the jet, the density of the surrounding medium (both our sample and the \citealp{Moravec2019} sample highlight that radio loud AGN are not always at the center of the cluster, allowing the ICM density to be a function of the offset from the cluster center), and the age of the radio source.  Like the \citet{Moravec2019} sample, our sample has no direct measurement of the ICM density and the age of the radio sources.  As such, we follow \citet{Moravec2019} and normalize the radio powers to a fiducial power, creating a normalized radio source size using Equation\,\ref{Eq:2}, 
\begin{equation}
    Le_{norm} = Le\left(\frac{P_{0}}{P_{1.44}}\right)^{\frac{1}{5}}, \label{Eq:2}
\end{equation}
and compare this to the offset between the radio source and the cluster center from \citet{Golden-Marx2019}. We re-write the relationship in Equation\,\ref{Eq:3},
\begin{equation} \label{Eq:3}
    Le_{norm} = 10^{-0.8\pm0.6}D^{1.4\pm0.4},   
\end{equation}
where D is the distance between the cluster center and the radio source, Le$_{norm}$ is the normalized size of the radio source, P$_{0}$ is the normalizing radio power at 1.44\,GHz (2$\times$10$^{26}$\,W\,Hz$^{-1}$), and P$_{1.44}$ is the power of the radio source in 1.44\,GHz.  

\begin{figure}
\centering
\epsscale{1}
\includegraphics[scale=0.48,trim={0.2in 0.1in 0.0in 0.4in},clip=True]{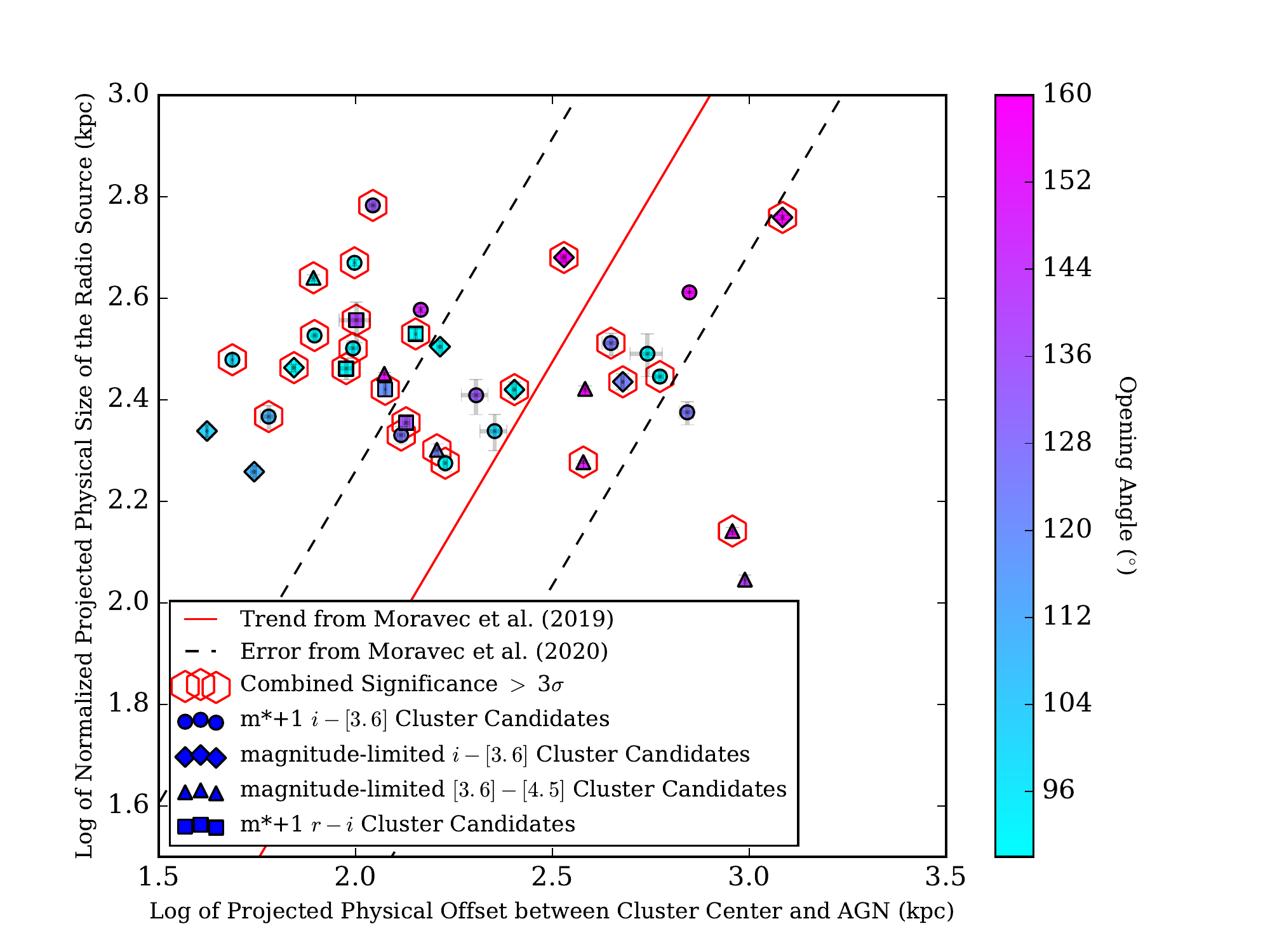}

\caption{The log of the normalized physical size of each radio source as a function of the log of the physical offset between the cluster centers and bent AGNs.  We color each point depending on the opening angle, noting that all sources with opening angles below 90$^{\circ}$ are colored cyan.  The points surrounded by red hexagons show clusters with combined overdensity significances above 3$\sigma$. The trend from \citet{Moravec2019} is overlaid in red and the 1$\sigma$ error calculated using the full sample from \citet{Moravec2020} is shown in the dashed black line.  We find little agreement with the trend reported in \citet{Moravec2019} and see instead that the most narrow bent radio sources tend to be offset from this trend, though they follow a similar slope.   \label{Fig:OffsetSizeNormal}}
\end{figure}

The MaDCoWS subsample from \citet{Moravec2019} followed the trend with only minimal scatter at smaller cluster offsets, while the larger sample in \citet{Moravec2020} showed more scatter (the 1$\sigma$ scatter on the trend in Figure\,\ref{Fig:OffsetSizeNormal} is calculated using the full sample from \citealp{Moravec2020}). 
However, as Figure\,\ref{Fig:OffsetSizeNormal} shows, we see no agreement with \cite{Moravec2019}, with most of our sources falling outside the 1$\sigma$ scatter. 

Because the sample in \citet{Moravec2020} only includes three bent AGNs, we include the opening angle within Figure\,\ref{Fig:OffsetSizeNormal} to determine if the opening angle may be a latent parameter responsible for this discrepancy.  At small offsets ($<$ 100\,kpc), \citet{Moravec2020} also see a separate population of sources well above the trend from \citet{Moravec2019}.  The disagreement between the trend and our measurements in this regime could be the result of the most central radio sources being fundamentally different, as shown in the relationship between cluster offset and opening angle in \citet{Garon2019}. When combined with Figure\,\ref{Fig:OffsetOpeningAngle}, this may suggest that centrally located radio sources, which include the full range of radio source luminosities and opening angles for our sample, are more impacted by the dense central ICM, rather than the uniform ICM assumed in this relation. 

Interestingly, the bulk of our sources follow a slope similar to the one reported in \citet{Moravec2019}.  These sources are also primarily the narrowest bent radio AGNs, making it possible that the offset from the trend in \citet{Moravec2019} is due to an unaccounted factor which results from the bending.  Since \citet{Moravec2020} only has three bent sources in their sample and most of our narrow sources are in our richest clusters (with a combined overdensity above 3$\sigma$), where the offset is better constrained, it is even more likely that the bending yields in a fundamental difference.    

Similar to the least offset sources, most of our more offset sources (beyond 300\,kpc) follow a similar slope to the trend in \citet{Moravec2019}, although fewer are in rich clusters.  These points are primarily wider bent sources and mostly fall within the 1$\sigma$ error estimate from \citet{Moravec2020}.  The slightly better agreement with \citet{Moravec2019} may further indicate that the opening angle is a latent parameter.  However, the divergence from this trend at the farthest offsets could be due to AGNs having a finite size.  \citet{Moravec2019} and \citet{Moravec2020} raise the possibility that the radio source size is limited by the density of the ICM acting to prevent the AGN's outflow.  The lack of this trend at greater distances may indicate that the density of the ICM only inhibits an AGN's size up to a certain point, where the intrinsic properties of the AGN become dominant and a shock front forms creating a lobe.  Such an environmental impact could also be similarly linked to the large population of wide opening angles at large offsets (Figure\,\ref{Fig:OffsetOpeningAngle}) and in less rich cluster environments (Figure\,\ref{Fig:SignificanceOpeningAngle}). 

Alternatively, the MaDCoWs sample and detection methods may differ enough to create the offsets between COBRA sources and the trend from \citet{Moravec2019}.  While we use a uniformly weighted red sequence surface density to determine the cluster center, MaDCoWs uses a combination of a number and flux weighted measurement \citep{Gonzalez2019}.  Additionally, MaDCoWs selects some of the most massive clusters at 0.7 $<$ $z$ $<$ 1.5, while COBRA probes a wider range of cluster masses, especially at the low-mass end.  If cluster mass is a latent parameter in the trend from \citet{Moravec2019}, it is possible that this may resolve the discrepancy.  \citet{Shen2020} also examined the relationship between size and distance from the cluster center for the extended radio sources in their sample of clusters from the high-$z$ ORELSE survey.  After breaking their sources into high and low density bins, \citet{Shen2020} found that like \citet{Moravec2019}, more compact radio sources are generally closer to the cluster center, with most of their more massive structures falling within 1$\sigma$ of the trend from \citet{Moravec2019}.  However, they were also unable to recreate the trend from \citet{Moravec2019} using their own data.  That \citet{Shen2020} saw more agreement among their massive systems could further illuminate the discrepancy in our data since we expect bent sources to trace both rich and poor clusters.  Furthermore, as discussed in \citet{Shen2020}, it is also possible that the disagreement between our values and the trend from \citet{Moravec2019} stems from the VLA FIRST observations having a 5$\arcsec$ resolution, while the JVLA observations from \citet{Moravec2019} have 1$\arcsec$ - 2$\arcsec$ resolution.  This difference would primarily inflate the size of our smallest sources, which is one regime where we find little agreement with the trend from \citet{Moravec2019}.

\section{Conclusion}\label{sect:Conclusions}
 
This is the third in our series of high-$z$ COBRA papers and the first to evaluate the radio properties of each cluster's radio host galaxy beyond reporting the radio luminosity.  We investigate the radio properties of 39 red sequence cluster candidates identified in \citet{Golden-Marx2019} and combine our optical/IR imaging with the VLA FIRST radio observations to re-evaluate the radio properties of the bent, double-lobed radio sources in each cluster.  Below is a summary of our findings.  
\begin{itemize}
    \item $Radio$ $Morphology$ $and$ $Power$: From our sample of 39 red sequence cluster candidates, we confirm bent AGNs in 36 cluster candidates.  By measuring the radio power and physical size of each source, we find that the brightest radio sources are generally the largest radio sources, although no correlations between the opening angle of the bent AGN and the size of the AGN are detected.
    \item  $Cluster$ $Richness$ $and$ $Opening$ $Angle$: Using the sample of 36 bent radio sources, we examine the richness of the surrounding cluster relative to the opening angle.  We find that richer clusters (measured using the significance of our red sequence and combined overdensities) host narrower bent radio AGN.  If all things are equal, this implies that richer clusters in our sample have denser ICMs, which are responsible for bending the radio lobes. 
    \item $BCG$ $Fraction$: From our sample of 36 cluster candidates, we use the red sequence measurements from \citet{Golden-Marx2019} to measure the fraction of host galaxies that are BCGs.  For this analysis, we remove the six SDSS-identified quasars, two radio sources where the model redshifts of the host galaxy do not match the SDSS photometry or the color of the surrounding galaxies, and one radio source where no host is found.  Of the remaining 27 radio sources, 15 are BCGs.  The remaining 12 host galaxies are among the three brightest galaxies we identify in each cluster, making them all galaxies that may evolve into BCGs.
    \item $Correlating$ $the$ $Host$ $Galaxy$ $with$ $the$ $Radio$ $Source$ $Properties$: Of the 27 red sequence identified host galaxies, the distribution of opening angles as a function of offsets from the cluster center follow a similar distribution between BCGs and non-BCGs, implying that all of our host galaxies are drawn from a single population (as verified by a KS test).
    \item $Infalling$ $and$ $Outgoing$ $Radio$ $Sources$: For our sample of 36 bent, double-lobed radio sources, we classify each as infalling, outgoing, or intermediate radio sources regardless of the host galaxy being a BCG.  Our first measurement is solely based on whether the radio source opens toward or away from the cluster center.  We find 21 sources appear infalling while 15 appear outgoing.  To account for clusters being three dimensional, we refine this measurement by measuring the infall angle relative to the cluster center.  With this measurement, we find 3 sources are infalling, 11 are outgoing, while 22 are at intermediate angles.  The large number of outgoing radio sources, especially when compared to the low-$z$ sample from \citet{Sakelliou2000}, might imply that either the central ICM in these high-$z$ clusters is at a low enough density to allow radio sources to pass through without the radio sources being disrupted or that their non-directly outgoing paths imply they pass their pericenter at large offsets from that center.  The lack of a strong dichotomy could result from bent AGNs following more circular paths or belonging to more complex merging clusters than at low redshift.  Alternatively, much of this measurement is subject to projection effects, creating large amounts of potential uncertainty, which could account for the large number of intermediate sources.  
\end{itemize}

To further explore the relationships between bent AGNs and their host clusters, we will continue to explore the cluster environments of these radio sources, particularly using X-ray observations to trace the ICM.  To resolve some of the ambiguity of our determination of infalling/outgoing/intermediate radio sources, we plan to do complimentary follow-up NIR or IR observations of the extended cluster regions to do a background-subtracted companion analysis in color-space of the regions surrounding the bent AGNs.  We aim to determine if the bent AGNs lie among spherically symmetric large scale galaxy structures or are biased toward a given direction.  Additionally, spectroscopic follow-up will aid in our identification of infalling and outgoing host galaxies and all cluster members as well as the orbital dynamics of the host galaxy. 

We plan to complement this study by combining it with the similarly selected sample of clusters hosting lower redshift bent, double-lobed radio sources from \citet{Wing2011} to trace how the cluster properties evolve as a whole.  Specifically, by doing a red sequence analysis on the clusters identified in \citet{Wing2011}, we can create a uniform sample of low- and high-$z$ clusters hosting bent radio sources to determine how the radio properties, cluster properties, and host galaxy properties evolve and see if the anti-correlation between richness and opening angle extends to low-$z$ clusters.  Building on this, we plan to use new radio surveys to continue to identify bent radio sources in even higher redshift clusters and protoclusters.  Additionally, we plan to further refine our red sequence analysis by modeling the red sequence slope to account for variety in red sequence populations.  
 
\appendix
\section{Spearman Test}\label{Appendix}
Within this paper, we used the Spearman Test as a way to determine the strength of a given correlation (estimated by r$_{s}$, which spans values between $-$1 and +1) and the likelihood that a given correlation rejects the null hypothesis (estimated by $p$, which spans values between 0 and 1; 1 - $p$ can be thought of as the confidence level of the detection) and is thus a statistically real correlation.  We define the strength of a trend using the following metric based on the discussion given in \citet{Fowler2009}: if $|$r$_{s}|$ = 0.00 to 0.19, this a very weak to no correlation; if $|$r$_{s}|$ = 0.20 to 0.39, this as a weak correlation; if $|$r$_{s}|$ = 0.40 to 0.69, this as a moderate correlation; if $|$r$_{s}|$ = 0.70 to 0.89, this as a strong correlation; and if $|$r$_{s}|$ = 0.90 to 1.00, this as a very strong correlation.  Similarly, we defined the $p$ value using the following metric: if $p$ $>$ 0.1, this is very weak to no evidence to reject the null hypothesis; if $p$ = 0.05 to 0.1, this is weak evidence to reject the null hypothesis; if $p$ = 0.01 to 0.05, this is strong evidence to reject the null hypothesis; and if $p$ $<$ 0.01, this is very strong evidence to reject the null hypothesis.  Generally, we define a likely detection as having a $p$ value less than 0.1 (or greater than 90$\%$ certainty), although we do report all $p$ values, even those which do not reject the null hypothesis.  It is possible to have a strong detection, shown by a higher value of $|$r$_{s}|$, but have weak or no evidence to reject the null hypothesis, $p$ $>$ 0.1, or vice versa.  In such a case, despite the strength of the trend, it is statistically unlikely to be a real trend.  Thus, we use the combination of the two parameters to report the strength and likelihood of a given correlation.  When we perform a Kolmogorev-Smirnov test, we use this same criteria for those $p$ values.
 
\acknowledgments
EGM would like to thank the referee for their very helpful comments on this paper.  EGM would also like to thank Jesse Golden-Marx for reading drafts of this paper and useful discussions.  EGM would also like to thank Zheng Cai for useful discussions and for reading drafts of this paper. EGM would also like to thank the LDT telescope operators for their help with taking observations.  Additionally, EGM would like to thank the organizers of the Early Stages of Galaxy Cluster Formation 2017 Conference and the Tracing Cosmic Evolution with Clusters of Galaxies 2019 Conference for fostering stimulating discussions that led to ideas addressed in this paper.    

This work has been supported by the National Science Foundation, grant AST-1309032.

EM acknowledges the support of the EU-ARC.CZ Large Research Infrastructure grant project LM2018106 of the Ministry of Education, Youth and Sports of the Czech Republic.

These results made use of the Lowell Discovery Telescope at Lowell Observatory. Lowell is a private, non-profit institution dedicated to astrophysical research and public appreciation of astronomy and operates the LDT in partnership with Boston University, the University of Maryland, the University of Toledo, Northern Arizona University, and Yale University. LMI construction was supported by a grant AST-1005313 from the National Science Foundation.

This work is based in part on observations made with the  $\sl{Spitzer}$ Space Telescope, which is operated by the Jet Propulsion Laboratory, California Institute of Technology under a contract with NASA. Support for this work was provided by NASA through an award issued by JPL/Caltech (NASA award RSA No. 1440385).

Funding for SDSS-III has been provided by the Alfred P. Sloan Foundation, the Participating Institutions, the National Science Foundation, and the U.S. Department of Energy Office of Science. The SDSS-III web site is http://www.sdss3.org/.

SDSS-III is managed by the Astrophysical Research Consortium for the Participating Institutions of the SDSS-III Collaboration including the University of Arizona, the Brazilian Participation Group, Brookhaven National Laboratory, Carnegie Mellon University, University of Florida, the French Participation Group, the German Participation Group, Harvard University, the Instituto de Astrofisica de Canarias, the Michigan State/Notre Dame/JINA Participation Group, Johns Hopkins University, Lawrence Berkeley National Laboratory, Max Planck Institute for Astrophysics, Max Planck Institute for Extraterrestrial Physics, New Mexico State University, New York University, Ohio State University, Pennsylvania State University, University of Portsmouth, Princeton University, the Spanish Participation Group, University of Tokyo, University of Utah, Vanderbilt University, University of Virginia, University of Washington, and Yale University.

IRAF is distributed by the National Optical Astronomy Observatory, which is operated by the Association of Universities for Research in Astronomy (AURA) under a cooperative agreement with the National Science Foundation.

This research made use of Astropy,\footnote{http://www.astropy.org} a community-developed core Python package for Astronomy \citep{astropy:2013, astropy:2018}.

\facilities{LDT, $Spitzer$, Sloan}

\clearpage
\newpage
\mbox{~}
\clearpage
\newpage
\begin{deluxetable*}{llllllrrrc}
    \tablecolumns{10}
    \tabletypesize{\footnotesize}
    \tablecaption{COBRA Red Sequence Clusters and Radio Source Properties \label{tb:GALAXY}}
    \tablewidth{0pt}
    \tabletypesize{\footnotesize}
    
    \setlength{\tabcolsep}{0.04in}
    \tablehead{
    \colhead{Field}&
    \colhead{Redshift}&
    \multicolumn{2}{c}{Host Coordinates}&
    \multicolumn{2}{c}{RS Cluster Center}&
    \colhead{AGN Offset}&
    \colhead{Infall Angle}&
    \multicolumn{2}{c}{Overdensity Significance}
    \cr
    \colhead{}&
    \colhead{z}&
    \colhead{RA\tablenotemark{a}}&
    \colhead{DEC}&
    \colhead{RA}&
    \colhead{DEC}&
    \colhead{(kpc)}&
    \colhead{($^{\circ}$)}&
    \colhead{RS}&
    \colhead{Combined}
    \cr
    }
    
\startdata
\cutinhead{m*+1 $i - [3.6]$ cluster candidates}
COBRA005837.2+011326& 0.71& 00 58 37.03&+01 13 27.8 & 00 58 36.47 & +01 13 26.4 & 78.6 &133.9& 4.3& 4.4\\
COBRA014741.6$-$004706& 0.60& 01 47 41.60 & $-$00 47 06.0 & 01 47 41.29 & $-$00 46 36.1 & $-$202.4&3.1 & 2.0& 2.9\\
COBRA015313.0$-$001018& 0.44& 01 53 12.96&$-$00 10 19.9 & 01 53 19.46 & $-$00 10 22.1& $-$551.2& 32.5 & 2.1& 2.3\\
COBRA113733.8+300010& 0.96& 11 37 33.80 & +30 00 10.0 & 11 37 33.36 & +30 00 07.8& 48.6 & 55.5 & 4.0& 3.9\\
COBRA121712.2+241525& 0.90& 12 17 12.20 & +24 15 25.0 & 12 17 10.69 & +24 15 31.8& 169.0& 66.2 & 8.1& 6.1\\
COBRA123940.7+280828& 0.92& 12 39 40.70 & +28 08 28.0 & 12 39 39.68 & +28 08 32.2& 110.6& 104.4 & 3.6& 3.1\\
COBRA125047.4+142355& 0.9& 12 50 47.40 & +14 23 55.0 & 12 50 47.39 & +14 24 07.6& 98.5& 104.4 & 2.5& 3.5\\
COBRA134104.4+055851\tablenotemark{b}& 0.90& 13 41 04.40 & +05 58 41.0 & 13 41 03.82 & +05 58 41.4& \nodata& \nodata & 2.2& 2.6\\
COBRA135838.1+384722& 0.81& 13 59 38.10 & +38 47 22.0 & 13 58 30.31 & +38 47 36.6& 695.4& 144.5 & 2.4& 2.6\\
COBRA142238.1+251433& 1.00& 14 22 38.10 & +25 14 33.0 & 14 22 34.16 & +25 13 54.7& $-$592.9& 21.3 & 5.0& 4.7\\
COBRA151458.0$-$011749& 0.80& 15 14 58.00 & $-$01 17 50.0 & 15 14 56.97 & $-$01 17 57.0& $-$130.6& 46.0 & 3.9& 4.5\\
COBRA154638.3+364420& 0.941& 15 46 38.30 & +36 44 20.0 & 15 46 30.89 & +36 44 12.6& 704.2& 82.7 & 3.5& 1.9\\
COBRA162955.5+451607& 0.78& 16 29 55.50 & +45 16 07.0 & 16 29 55.56 & +45 15 07.2& $-$445.2& 56.9 & 2.2& 4.1\\
COBRA164951.6+310818& 0.52 & 16 49 52.36 & +31 08 07.8 & 16 49 53.33 & +31 07 49.4& 225.5& 51.8 & 2.3& 2.3\\
COBRA170105.4+360958& 0.80& 17 01 05.40 & +36 09 58.0 & 17 01 04.98 & +36 10 04.2& $-$60.1& 32.8 & 4.2& 5.2\\
COBRA170614.5+243707& 0.71& 17 06 14.50 & +25 37 07.0 & 17 06 14.54 & +24 36 53.2& 99.2& 87.3 & 5.9& 5.3\\
COBRA171330.9+423502& 0.698& 17 13 30.90 & +42 35 02.0 & 17 13 29.85 & +42 35 18.9& $-$146.3& 60.7 & 1.0& 0.8\\
COBRA221605.1$-$081335\tablenotemark{b}& 0.70& 22 16 05.10 & $-$08 13 35.0 & 22 16 02.88 & $-$08 14 14.6& \nodata& \nodata & 2.4& 1.9\\
\cutinhead{m*+1 $r - i$ cluster candidates}
COBRA012058.9+002140& 0.75&01 20 58.87&+00 21 41.7 & 01 20 57.87 & +00 21 49.8 & 122.5 & 121.0& 5.1& 4.6\\
COBRA075516.6+171457& 0.64&07 55 17.35&+17 14 54.9 & 07 55 17.54 & +17 14 59.9& 95.0& 57.8 & 6.1& 5.1\\
COBRA100745.5+580713& 0.656& 10 07 45.60&+58 07 15.2 & 10 07 45.09 & +58 07 29.8& $-$119.2& 42.9& 6.5& 3.6\\
COBRA135136.2+543955& 0.55& 13 51 36.20 & +54 39 55.0 & 13 51 34.70 & +54 39 46.3& $-$82.9& 31.8 & 9.6& 3.3\\
COBRA164611.2+512915& 0.351& 16 46 11.20 & +51 29 15.0 & 16 46 09.03 & +51 28 54.7& $-$184.9& 20.2 & 7.7& 5.6\\
\cutinhead{magnitude-limited $i - [3.6]$ cluster candidates}
COBRA074025.5+485124 & 1.10& 07 40 25.51&+48 51 25.2 & 07 40 24.24 & +48 51 08.3& $-$163.9& 22.5& 2.7& 2.9\\
COBRA074410.9+274011 & 1.30& 07 44 10.92&+27 40 12.7 & 07 44 09.15 & +27 41 03.0& 477.3& 110.5& 7.3& 6.7\\
COBRA103434.2+310352 & 1.20& 10 34 34.20&+31 03 52.2 & 10 34 33.55 & +31 03 53.0& $-$69.7& 32.7& 4.8& 6.1\\
COBRA130729.2+274659 & 1.144& 13 07 29.20 & +27 46 59.0 & 13 07 37.96 & +27 48 33.2& 1215.6& 103.6& 5.5& 4.5\\
COBRA133507.1+132329\tablenotemark{b} & 1.25& 13 35 07.10 & +13 23 39.0 & 13 35 08.01 & +13 23 20.1& \nodata& \nodata& 4.5 & 2.5\\
COBRA145023.3+340123 & 1.20& 14 50 20.90 & +34 01 33.0 & 14 50 23.34 & +34 01 29.5& 253.3& 85.1& 3.9& 5.2\\
COBRA150238.1+170146 & 1.10& 15 02 38.10 & +17 01 46.0 & 15 02 38.33 & +17 01 51.9& $-$55.3& 0.8& 2.9 & 2.1\\
COBRA152647.5+554859 & 1.10& 15 26 47.50 & +55 48 59.0 & 15 26 52.36 & +55 49 05.0& 338.2& 110.1& 4.1& 4.0\\
COBRA172248.2+542400 & 1.45& 17 22 48.20 & +54 24 00.0 & 17 22 48.62 & +54 23 56.7& 41.9& 125.3& 4.7& 2.4\\
\cutinhead{magnitude-limited $[3.6] - [4.5]$ cluster candidates}
COBRA072805.2+312857 & 1.75& 07 28 05.35&+31 28 59.5 & 07 28 06.26 & +31 28 52.1& 121.9& 120.0& 2.5 & 2.9\\
COBRA100841.7+372513 & 1.35& 10 08 41.71&+37 25 14.2 & 10 08 41.53 & +37 25 32.3& $-$160.8& 36.4& 4.1 & 4.1\\
COBRA103256.8+262335 & 2.180& 10 32 56.83&+26 23 36.2 & 10 32 54.49 & +26 23 33.3& $-$382.6& 50.4& 1.9& 2.0\\
COBRA104254.8+290719 & 1.35& 10 42 54.74&+29 07 19.6 & 10 42 55.45 & +29 07 22.7& 78.4 & 145.1& 3.3& 3.5\\
COBRA121128.5+505253 & 1.364& 12 11 28.50 & +50 52 53.0 & 12 11 25.79 & +50 52 16.0& $-$378.5 & 45.0& 2.7 & 3.1\\
COBRA141155.2+341510 & 1.818& 14 11 55.20 & +34 15 10.0 & 14 11 46.31 & +34 14 35.5& 975.1& 137.1& 2.2& 2.4\\
COBRA222729.1+000522 & 1.513& 22 27 29.10 & +00 05 22.0 & 22 27 36.06 & +00 04 58.4& 905.6& 127.4& 2.7& 3.1\\
\enddata
\tablenotetext{a}{All coordinates are given in J2000.}
\tablenotetext{b}{Cluster candidates without evidence of a bent radio AGN and for which we do not report any measurements relating to the AGN.
}
\end{deluxetable*}

\clearpage
\newpage
\mbox{~}
\clearpage
\newpage
\begin{deluxetable*}{lrcrrcc}
    \tablecolumns{7}
    \tabletypesize{\footnotesize}
    \tablecaption{COBRA Radio Source Properties \label{tb:RADIO}}
    \tablewidth{0pt}
    \tabletypesize{\footnotesize}
    
    \setlength{\tabcolsep}{0.05in}
    \tablehead{
    \colhead{Field}&
    \colhead{Opening Angle}&
    \colhead{Radio Source Length}&
    \colhead{Radio Flux}&
    \colhead{Radio Power (P$_{1.44}$)}&
    \multicolumn{2}{c}{Host Galaxy}
    \cr
    \colhead{}&
    \colhead{($^{\circ}$)}&
    \colhead{(kpc)}&
    \colhead{Density}&
    \colhead{(10$^{25}$\,W\,Hz$^{-1}$)}&
    \colhead{m$_{3.6\mu\,m}$}&
    \colhead{M$_{3.6\mu\,m}$}
    \cr
    \colhead{}&
    \colhead{}&
    \colhead{}&
    \colhead{(mJy)}&
    \colhead{}&
    \colhead{(AB)}&
    \colhead{(AB)}
    }
    
\startdata
\cutinhead{m*+1 $i - [3.6]$ cluster candidates}
COBRA005837.2+011326 & 85.2$\pm$1.1 & 251.3$^{+6.8}_{-7.1}$ & 24.93 & 4.66$^{+0.60}_{-0.59}$ & 18.02 & $-$23.74\\
COBRA014741.6$-$004706 & 134.13$\pm$5.1 & 165.7$^{+6.1}_{-6.4}$ & 18.14 & 2.25$^{+0.93}_{-0.89}$ & 23.10 & $-$18.41 \\
COBRA015313.0$-$001018 & 90.8$\pm$0.1 & 152.3$^{+13.6}_{-14.6}$ & 9.78 & 0.58$^{+0.23}_{-0.22}$ & 18.13 & $-$22.98\\
COBRA113733.8+300010 & 98.8$\pm$4.8 & 215.3$^{+4.7}_{-4.7}$ & 9.49 & 3.72$^{+0.05}_{-0.05}$ & 18.75 & $-$23.51\\
COBRA121712.2+241525 & 76.0$\pm$1.8 & 140.6$^{+5.4}_{-5.5}$ & 13.79 & 4.61$^{+0.54}_{-0.53}$ & 18.49 & $-$24.66\\
COBRA123940.7+280828 & 133.4$\pm$1.2 & 473.4$^{+9.7}_{-11.3}$ & 16.38 & 5.79$^{+1.07}_{-1.05}$ & 18.08 & $-$24.11\\
COBRA125047.4+142355 & 92.2$\pm$0.2 & 243.2$^{+8.5}_{-10.2}$ & 15.8 & 5.29$^{+1.47}_{-1.43}$ & 19.04 & $-$23.11\\
COBRA135838.1+384722 & 125.6$\pm$6.2 & 174.8$^{+8.5}_{-9.9}$ & 16.8 & 4.34$^{+1.34}_{-1.30}$ & 18.86 & $-$23.12\\
COBRA142238.1+251433 & 38.7$\pm$2.4 & 231.4$^{+5.5}_{-6.9}$ & 18.0 & 7.82$^{+1.96}_{-1.91}$ & 19.06 & $-$23.27\\
COBRA151458.0$-$011749 & 126.3$\pm$6.7 & 123.2$^{+4.7}_{-4.7}$ & 5.04 & 1.26$^{+0.22}_{-0.21}$ & 19.00 & $-$22.96\\
COBRA154638.3+364420\tablenotemark{a} & 156.6$\pm$1.2 & 452.8$^{+2.0}_{-2.0}$ & 89.29 & 33.21$^{+0.03}_{-0.03}$ & 16.57 & \nodata\\
COBRA162955.5+451607 & 125.2$\pm$2.8 & 260.7$^{+11.7}_{-14.3}$ & 28.25 & 6.65$^{+2.13}_{-2.06}$ & \nodata & \nodata\\
COBRA164951.6+310818 & 100.3$\pm$0.7 & 189.9$^{+14.3}_{-16.9}$ & 10.03 & 10.03$^{+3.73}_{-3.58}$ & 19.59 & $-$21.75\\
COBRA170105.4+360958 & 110.4$\pm$2.6 & 173.2$^{+8.5}_{-10.0}$ & 18.16 & 4.55$^{+1.42}_{-1.38}$ & 17.94 & $-$24.02\\
COBRA170614.5+243707 & 66.0$\pm$0.2 & 319.0$^{+4.6}_{-4.6}$ & 15.79 & 2.95$^{+0.00}_{-0.00}$ & 17.80 & $-$23.96\\
COBRA171330.9+423502 & 148.2$\pm$3.3 & 268.0$^{+3.0}_{-3.0}$ & 20.01 & 3.5$^{+0.00}_{-0.00}$ & 17.71 & $-$24.02\\
\cutinhead{m*+1 $r - i$ cluster candidates}
COBRA012058.9+002140 & 135.8$\pm$6.4 & 188.0$^{+6.1}_{-6.4}$ & 36.93 & 7.90$^{+1.11}_{-1.09}$ & 17.96 & $-$23.89\\
COBRA075516.6+171457 & 61.7$\pm$1.6 & 218.1$^{+9.8}_{-11.0}$ & 33.56 & 4.87$^{+1.14}_{-1.11}$ & 17.86 & $-$23.74\\
COBRA100745.5+580713 & 119.2$\pm$3.0 & 166.9$^{+4.8}_{-4.8}$ & 13.27 & 2.04$^{+0.00}_{-0.00}$ & 17.99 & $-$23.65\\
COBRA135136.2+543955\tablenotemark{b} & 140.5$\pm$3.3 & 211.6$^{+17.4}_{-21.8}$ & 13.85 & 1.39$^{+0.63}_{-0.60}$ & 18.30 & $-$23.11\\
COBRA164611.2+512915 & 49.9$\pm$1.0 & 165.9$^{+3.8}_{-3.8}$ & 16.32 & 0.56$^{+0.00}_{-0.00}$ & 17.70 & $-$23.06\\
\cutinhead{magnitude-limited $i - [3.6]$ cluster candidates}
COBRA074025.5+485124 & 78.5$\pm$2.8 & 264.7$^{+6.8}_{-7.7}$ & 14.16 & 7.79$^{+1.77}_{-1.73}$ & 18.31 & $-$24.20\\
COBRA074410.9+274011\tablenotemark{c} & 123.8$\pm$4.9 & 189.1$^{+6.8}_{-6.9}$ & 3.87 & 3.21$^{+0.62}_{-0.61}$ & 18.49 & $-$24.43\\
COBRA103434.2+310352 & 87.5$\pm$0.1 & 458.2$^{+7.0}_{-8.4}$ & 285.36 & 194.66$^{+40.53}_{-39.75}$ & 18.85 & $-$23.86\\
COBRA130729.2+274659\tablenotemark{a} & 158.0$\pm$4.7 & 594.4$^{+5.9}_{-5.9}$ & 39.13 & 23.72$^{+0.31}_{-0.31}$ & 17.00 &\nodata\\
COBRA145023.3+340123 & 36.0$\pm$2.0 & 206.5$^{+7.2}_{-7.5}$ & 8.73 & 5.96$^{+1.24}_{-1.22}$ & 19.85 & $-$22.86\\
COBRA150238.1+170146 & 110.2$\pm$4.1 & 208.0$^{+5.4}_{-6.1}$ & 72.11 & 39.66$^{+9.02}_{-8.92}$ & 18.34 & $-$24.17\\
COBRA152647.5+554859 & 160.3$\pm$1.6 & 554.2$^{+8.4}_{-11.3}$ & 75.34 & 41.44$^{+9.42}_{-9.22}$ & 18.43 & $-$24.08\\
COBRA172248.2+542400 & 102.1$\pm$1.3 & 210.7$^{+5.9}_{-6.0}$ & 15.48 & 16.86$^{+2.89}_{-2.85}$ & 19.86 & $-$23.34\\
\cutinhead{magnitude-limited $[3.6] - [4.5]$ cluster candidates}
COBRA072805.2+312857 & 154.7$\pm$3.4 & 315.3$^{+2.1}_{-1.9}$ & 19.94 & 34.52$^{+4.88}_{-4.82}$ & 19.20 & $-$24.44\\
COBRA100841.7+372513 & 123.3$\pm$2.7 & 180.1$^{+4.8}_{-5.1}$ & 17.26 & 11.77$^{+2.45}_{-2.40}$ & 19.60 & $-$23.11\\
COBRA103256.8+262335\tablenotemark{a} & 158.3$\pm$6.0 & 313.6$^{+4.5}_{-4.5}$ & 15.94 & 47.28$^{+0.01}_{-0.01}$ & 19.32 & \nodata\\
COBRA104254.8+290719 & 89.1$\pm$0.0 & 419.6$^{+5.7}_{-6.3}$ & 17.91 & 16.35$^{+3.02}_{-2.97}$ & 17.72 & $-$25.30\\
COBRA121128.5+505253\tablenotemark{a} & 150.5$\pm$3.2 & 307.7$^{+2.3}_{-2.3}$ & 241.55 & 226.21$^{+0.17}_{-0.17}$ & 17.39 & \nodata\\
COBRA141155.2+341510\tablenotemark{a} & 141.5$\pm$4.5 & 198.6$^{+4.2}_{-4.2}$ & 190.71 & 362.63$^{+0.09}_{-0.09}$ & 18.12 & \nodata\\
COBRA222729.1+000522\tablenotemark{a} & 155.2$\pm$6.6 & 216.4$^{+3.4}_{-3.4}$ & 153.61 & 185.84$^{+0.04}_{-0.04}$ & 17.59 & \nodata\\
\enddata
\tablenotetext{a}{Radio sources that are SDSS quasars}
\tablenotetext{b}{Sources that lack a third radio component}
\tablenotetext{c}{Sources where one of the radio components was not convolved}
\end{deluxetable*}

\end{document}